\documentclass[12pt]{iopart}

\usepackage{iopams}  
\usepackage{graphicx}
\usepackage{epstopdf}
\usepackage{multirow}

\pdfoutput=1

\begin{document}

\title{Neutrino Oscillations with the MINOS, MINOS+, T2K, and NOvA Experiments.}

\author{Tsuyoshi Nakaya}
\address{Kyoto University, Department of Physics\\ Kyoto 606-8502 \\ Japan}
\ead{t.nakaya@scphys.kyoto-u.ac.jp}

\author{Robert K Plunkett}
\address{Fermi National Accelerator Laboratory \\
Batavia, Illinois 60510 \\
USA}
\ead{plunk@fnal.gov}

\begin{abstract}
This paper discusses recent results and near-term prospects of the long-baseline neutrino experiments MINOS, MINOS+, T2K and NOvA. The non-zero value of the third neutrino mixing angle $\theta_{13}$ allows experimental analysis in a manner which explicitly exhibits appearance and disappearance dependencies on additional parameters associated with mass-hierarchy, CP violation, and any non-maximal $\theta_{23}$. These current and near-future experiments begin the era of precision accelerator long-baseline measurements and lay the framework within which future experimental results will be interpreted.
\end{abstract}

\maketitle


\section{Introduction to Long-Baseline Accelerator Experiments}
\subsection{Motivation and 3-Flavor Model}

Beginning with the successful operation of the K2K experiment~\cite{Ahn:2006zza}, the physics community has seen a profound expansion of our knowledge of the mixing of neutrinos, driven by long-baseline accelerator experiments~\cite{Abe:2013hdq, Abe:2014ugx, minos_three_neutrino_prl, minos_nue_PRL110, Agafonova:2014bcr}, experiments studying atmospheric neutrinos~\cite{Fukuda:1998mi, Ashie:2005ik, Abe:2012jj}, solar neutrinos~\cite{Beringer:1900zz}, and, most recently, high-precision experiments with reactor neutrinos~\cite{minos_3_flavor_PRL_reactor}. 
In this article we describe the current generation of  the running long-baseline neutrino experiments T2K and MINOS/MINOS+ , and the status of the NOvA experiment which was commissioned in 2014. Each of these experiments was designed with primary and secondary goals. For example, MINOS had as its principle justification the measurement, via disappearance of $\nu_\mu$,  of the mixing parameters $\sin^2{\theta_{23}}$ and $\Delta m^2_{32}$, with particular emphasis on $\Delta m^2_{32}$. 
T2K and NOvA were primarily designed to elucidate the structure of the neutrino sector 
%
by studies of $\nu_e$ appearance. However they are making and will continue to make very significant contributions to the study of  $\nu_\mu$ disappearance as well. Similarly, MINOS has measured $\nu_e$ appearance and the angle $\theta_{13}$.  
\\

This situation leads us to a point we will emphasize throughout this article, namely that the traditional distinction between various modes of study of mixing, matter effects, and CP violation is rapidly giving way to a more integrated approach which utilizes both major types of signals to gain maximal information about the somewhat complicated 3-neutrino sector. We will first discuss the relevant formalism and the major measurements required to test it. Next we will provide a technical overview of the powerful neutrino beams required for the measurements, and then proceed to discuss the experiments together with their current measurements and expected sensitivities. Finally, we conclude the article with a discussion of the near-term future for these experiments.
\\

It is our goal to familiarize the reader with the surprisingly rich information already available in studies of this sector of physics, the only one currently not well-handled by the Standard Model. In addition, we will provide context for future discussions of progress to be provided by these experiments and by the exciting future world of very large experiments, and very long baselines.

\subsection{Core measurements}

In the three-neutrino model there is a close relationship among the disappearance and appearance modes of oscillation study, going back to their origin in the PMNS matrix. Following reference~\cite{Parke_2005}, it is possible to write the disappearance possibility for muon neutrinos in vacuum, as
\\

\begin{equation}
 P(\nu_\mu\rightarrow \nu_\mu) = 1 -4 \sin^2\theta_{23}\cos^2\theta_{13}(1 - \sin^2\theta_{23}\cos^2\theta_{13})\sin^2\frac{ \Delta m^2_{eff} L}{4E} 
\label{eq:int:numu1}
\end{equation}
 where $\Delta m^2_{eff}$ incorporates effective leading dependences on the additional PMNS parameters $\Delta m^2_{21}$, $\theta_{13}$, and $\delta_{CP}$ as
 \\
 \begin{equation*}
 \Delta m^2_{eff} = \Delta m^2_{32} + \Delta m^2_{21} \sin^2\theta_{12} + \Delta m^2_{21} \cos\delta_{CP}\sin\theta_{13}
 \tan\theta_{23}\sin 2\theta_{12}
 \end{equation*}
 
 Equation~\ref{eq:int:numu1} may be simply manipulated to yield a form appropriate for the baseline of the T2K experiment, namely
 \\
 \begin{equation}
 P(\nu_\mu\rightarrow \nu_\mu) \simeq 1 - (\cos^4\theta_{13}\sin^2 2\theta_{23} +  \sin^2 2\theta_{13}\sin^2 \theta_{23})\sin^2\frac{ \Delta m^2_{31} L}{4E}
\label{eq:int:numu2}
\end{equation}
 
 Here we see the vital role of the mixing angle $\theta_{13}$, which couples in the PMNS matrix to the CP-violating phase $\delta_{CP}$. It is now well-known that this angle is relatively large, approximately 9~degrees~\cite{Beringer:1900zz}. 
T2K, MINOS/MINOS+, and future NOvA measurements of this angle and its consequences are discussed in sections~\ref{sec:nue}, \ref{sec:numu+} and \ref{sec:future}.  
 
Because the earth between the beam creation point and the detector location forms an essential part of any long-baseline experiment, its effects on the measurements must be considered. This creates both problems and opportunities - problems because of the introduction of degeneracies between matter effects and CP violation, and opportunities because of the possibilities to exploit the differences between neutrino and antineutrino interactions, and from the two mass hierarchies.

 \subsubsection{Appearance measurements and $\sin^2 2\theta_{13}$}\
 \\
 
 A large value of $\theta_{13}$ is key to allowing an integrated approach to oscillation studies.  MINOS, T2K, and soon NOvA use the appearance channel for $\nu_e$  with a muon neutrino beam to probe $\theta_{13}$ directly. 
In appropriate approximation for a muon neutrino
with energy $E_\nu$ of $O(1)$~GeV traveling a distance of $O(100)$~km, 
the leading order equation governing the appearance probability is:
\begin{equation}
 P(\nu_\mu\rightarrow \nu_e) \approx \sin^2\theta_{23} \sin^2 2\theta_{13}\sin^2 \frac{ \Delta m^2_{32} L}{4 E} \label{eq:int:nue1}
\end{equation}
Equation~\ref{eq:int:nue1} is applicable for both neutrino and anti-neutrino oscillations.

The difference between neutrinos and anti-neutrinos in oscillation appears as a sub-leading effect, including 
the solar parameters $\theta_{12}$ and $\Delta m^2_{21}$ and CP violation phase $\delta_{\mathrm{CP}}$.
The probability is expressed~\cite{Cervera:2000kp, Freund:2001pn} as
\begin{equation}
P(\nu_\mu \to \nu_e) \simeq \sin^2 2\theta_{13} T_1 - \alpha  \sin 2\theta_{13} T_2 + \alpha  \sin 2\theta_{13} T_3 + \alpha^4 T_4 \label{eq:int:nue2}
\end{equation}
where $\alpha \equiv \frac{\Delta m^2_{21}}{\Delta m^2_{31}}$ is the small ($\sim 1/30$) ratio between the solar and atmospheric squared-mass splittings, and
\begin{eqnarray*}
T_1 & = & \sin^2\theta_{23} \frac{\sin^2 [(1-A) \Delta]}{(1-A^2)}, \\
T_2 & = & \sin \delta_{\mathrm{CP}} \sin 2\theta_{12} \sin 2\theta_{23}  \sin \Delta \frac{\sin(A \Delta)}{A} \frac{\sin [(1-A) \Delta]}{1-A}, \\ 
T_3 & = & \cos \delta_{\mathrm{CP}} \sin 2\theta_{12} \sin 2\theta_{23}  \cos \Delta \frac{\sin(A \Delta)}{A} \frac{\sin [(1-A) \Delta]}{1-A}, \\ 
T_4 & = & \cos^2 \theta_{23}  \sin^2 2\theta_{12}  \frac{\sin^2(A \Delta)}{A^2}.
\end{eqnarray*}
Here, $\Delta = \frac{ \Delta m^2_{32} L}{4 E_\nu} $ and $A \equiv 2 \sqrt{2} G_F n_e E_\nu/ \Delta m^2_{32}$, 
where $N_e$ is the electron density of Earth's crust. 
In Equation~\ref{eq:int:nue2}, the sign of the second term changes for anti-neutrinos,  governing CP violation
when all three mixing angles, including $\theta_{13}$, have nonzero values. 
With current best knowledge of oscillation parameters, the CP violation (sub-leading term)  can be as large as $\sim$30\% of the leading term.

The $A$ dependence arises from matter effects (caused by additional terms in the Hamiltonian for the electron component of the neutrino eigenstate), which are coupled with the sign of $\Delta m^2_{32}$.
In this paper, we refer to $\Delta m^2_{32} > 0$ as the normal mass hierarchy and $\Delta m^2_{32} < 0$ as the inverted one.

\subsubsection{Hierarchy, Octants, and CP Violation ($\delta_{CP}$)}\
\\

The subleading terms shown in Equations~\ref{eq:int:numu1} and~\ref{eq:int:nue2} make it possible in principle for oscillation measurements to be sensitive to the octant of $\theta_{23}$ 
{\it ($\theta_{23} < \pi /4$ or $\theta_{23} > \pi /4$), }
in the case it is not maximal (i.e. $= \pi /4$). 
Even for maximal mixing,  the additional dependencies in Equation~\ref{eq:int:nue2}, and further terms added to Equation~\ref{eq:int:numu2} enable the combination of disappearance and appearance results to begin to give clues to the CP and hierarchy puzzles.


\section{Beamlines}
\subsection{NuMI}
\label{subsec:NUMI_Section}

\begin{figure}
\includegraphics[scale=0.75]{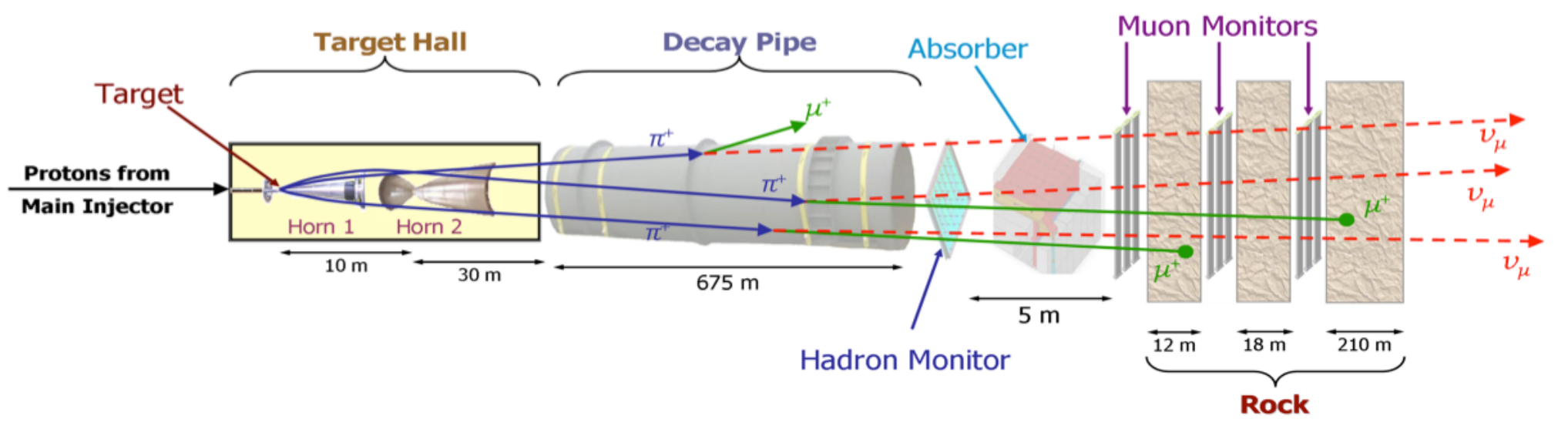}
\caption{ Layout of the NuMI beamline at Fermilab, showing main components of the target, focusing, and decay systems~\cite{NUMI_Beamline}.}
\label{fig:NUMI_beam}
\end{figure}

The conceptual beginning of the NuMI beamline dates to the era of the construction of the Fermilab Main Injector. The beamline, together with its associated tunnels, experimental halls, surface buildings and infrastructure, was built between 1999 and 2004. Datataking with the beam began in March of 2005, and continues to this time. The complex currently consists of a primary beam transport, the NuMI target hall, a 675 meter He-filled decay pipe of one meter radius, a hadron absorber and muon flux monitor area, meters of rock shielding, and two experimental halls - the first housing the MINOS near detector and the MINERvA detector, and the second housing the NOvA near detector. Figure~\ref{fig:NUMI_beam} shows the general configuration of the beamline. 

Beam from the Fermilab Booster is accelerated in the Main Injector to 120 GeV, and then extracted with a system of fast kickers. As part of an upgrade to beam power, commissioning is underway to stack beam in the Fermilab Recycler Ring before transfer to the Main Injector. At the NuMI target the proton beam consists of 6 batches, with a total extraction period of 10 $\mu s$. The time between extractions has varied from 2.2 s to the current 1.3 s. This low duty factor allows the MINOS and NOvA experiments to trigger on a simple timing window, which facilitates the surface location of the NOvA far detector. 
Overlaying multiple injections from the booster in the main injector (slip-stacking) has allowed the beam intensity to reach 375 kW. The NuMI beam for NOvA is anticipated to reach 700 kW, with a similar time structure.

\begin{figure}
\begin{center}
\includegraphics[scale=0.45]{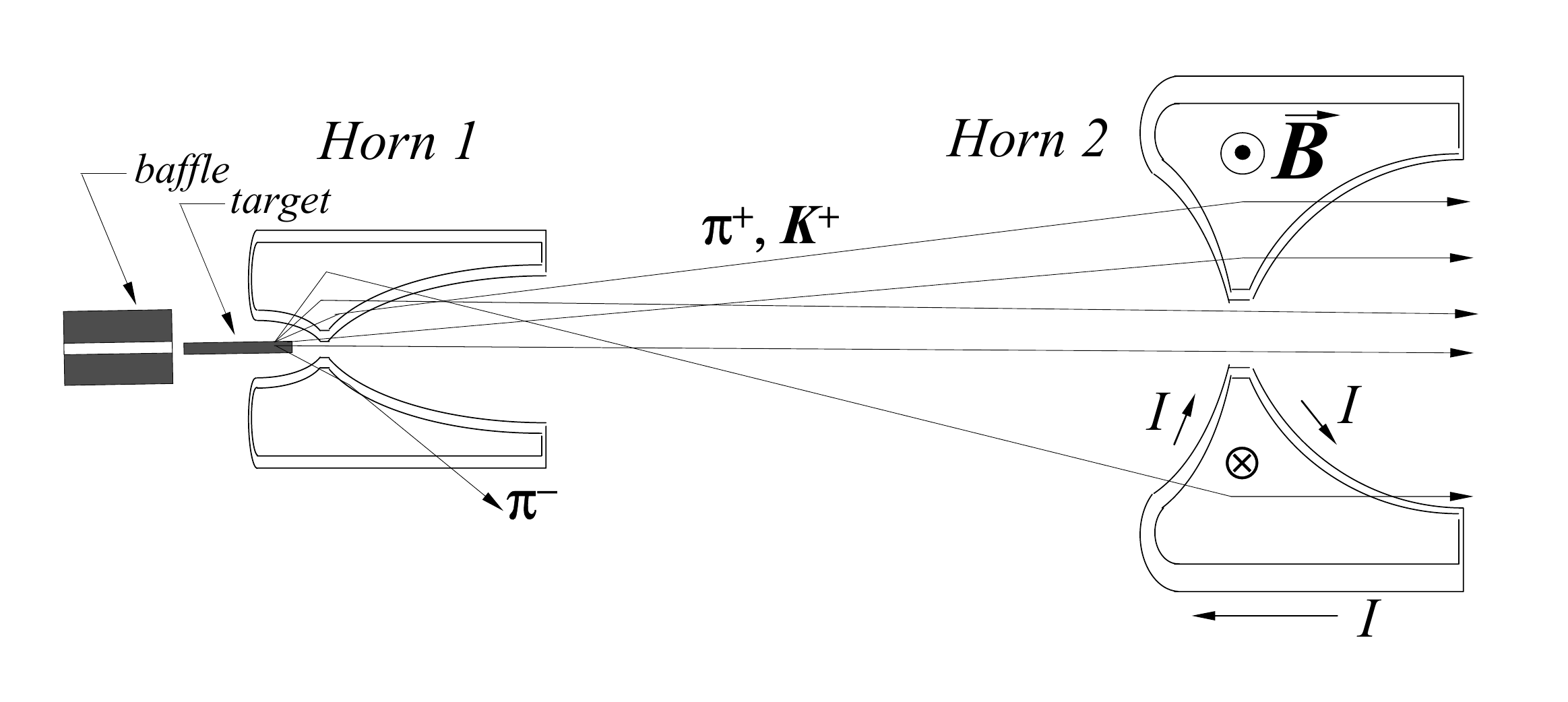}
\caption{ Detail of the magnetic focusing horn system for the NuMI beam line~\cite{minos_CC_PRD}.}
\label{fig:NUMI_horn}
\end{center}
\end{figure}

\begin{figure}
\includegraphics[scale=0.65]{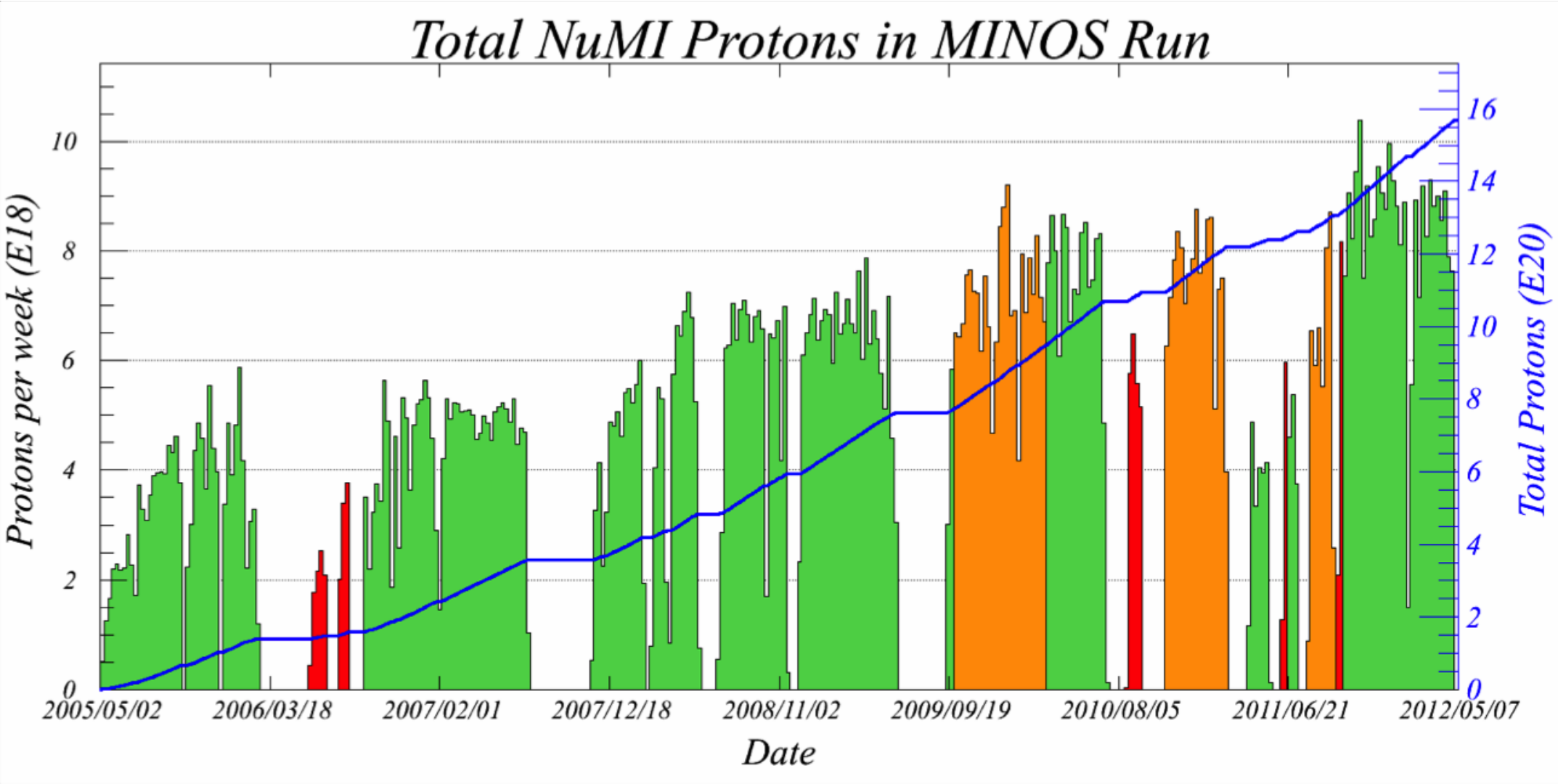}
\caption{Total NuMI beam delivery during MINOS running. The total collected in the principle neutrino configuration was $10.7$ x $10^{20}$ protons. For the antineutrino configuration the total was $3.4$ x $10^{20}$ protons.}
\label{fig:NUMI_protons}
\end{figure}

Secondary hadrons created in the interaction of the extracted proton beam with a 94 cm graphite target are focused by a system of two magnetic horns. Historically, the focused hadron momentum (which translates into neutrino energy) has been adjustable by moving the target w.r.t the first horn. This is more difficult in the NOvA beam configuration, which is optimized for the off-axis application. Figure~\ref{fig:NUMI_horn} shows the focusing schema of the NuMI beam.The majority of the data samples used in MINOS physics analyses have used the NuMI beam line focused at its lowest practical energy configuration, with a peak neutrino energy at approximately 3 GeV. Additional samples at higher neutrino beam energy settings have been used to provide information on the intrinsic $\nu_e$ component of the beam, an important and irreducible background for the measurement of $\sin^2 2\theta_{13}$. Overall beam production for the MINOS running period 2005-2012, is shown in Figure~\ref{fig:NUMI_protons}.

An important feature of horn-focused beamlines like the NuMI beamline is the ability to convert from focusing positive hadrons (primarily $\pi^+$) to negative hadrons (primarily $\pi^-$), which creates a beam heavily enriched in antineutrinos. Using this beam, MINOS has published special studies of the oscillation parameters of $\bar{\nu}_\mu$ \cite{minos_nubar_PRL108}.

Targetry which can withstand the repeated high power proton pulses needed for a neutrino beam represents a technical challenge. The NuMI targets used for MINOS data taking were constructed of 47 segmented graphite fins. A total of 7 targets were used in the period 2005 to 2012, with exchanges usually due to failures in auxiliary cooling systems. In one case the target material experienced significant degradation, visible in the produced neutrino rates. Significant engineering changes have occurred for the targets to be used in the NOvA beam, which must withstand 700 kW operations. These include detailed changes to the graphite fins to allow for an increase in primary beam spot size from 1.1 mm to 1.3 mm, and, importantly, a significant relocation of the water cooling tubing to decrease its vulnerability. With these changes, it is expected that the NuMI targets in the NOvA era will survive a minimum of a year of high power operation before any replacement is needed. 

\subsection{Off axis neutrino beam}
\label{subsec:Off_Axis_Section}

\indent An off axis neutrino beam (OAB) configuration~\cite{Beavis:1995a} is a method to produce a narrow band energy neutrino beam.
In the OAB configuration, the axis of the beam optics is intentionally shifted by a few degrees from the detector 
direction. With a finite decay angle, the neutrino energy becomes almost independent of the parent pion energy due 
to characteristics of the two body decay kinematics of the pion with Lorentz boost. 
The off-axis beam principle can be illustrated with a simple algebraic example. Let us model the beam as consisting of pions which are fully focused in the on-axis directions. The transverse and perpendicular components of the decay neutrino momentum obey the relations:

\begin{equation}
P_T=P^* \sin\theta^* ,  P_L= \gamma P^* (1+\cos \theta^*) \approx E_\nu
\end{equation}

where $P^*$ and $\theta^*$ are the decay momentum and angle in the rest frame of the decaying particle. The fixed off-axis angle condition is $\theta_{lab} = P_T/P_L$.

Near $\theta^*=\pi/2$, we have $\Delta P_T \approx 0$ for variations in $\theta^*$, and therefore 
\begin{equation}
\Delta P_L = \frac{\Delta P_T}{ \theta_{lab}} \approx 0
\end{equation}

\begin{figure}
	\begin{center}
		\includegraphics[width=7cm]{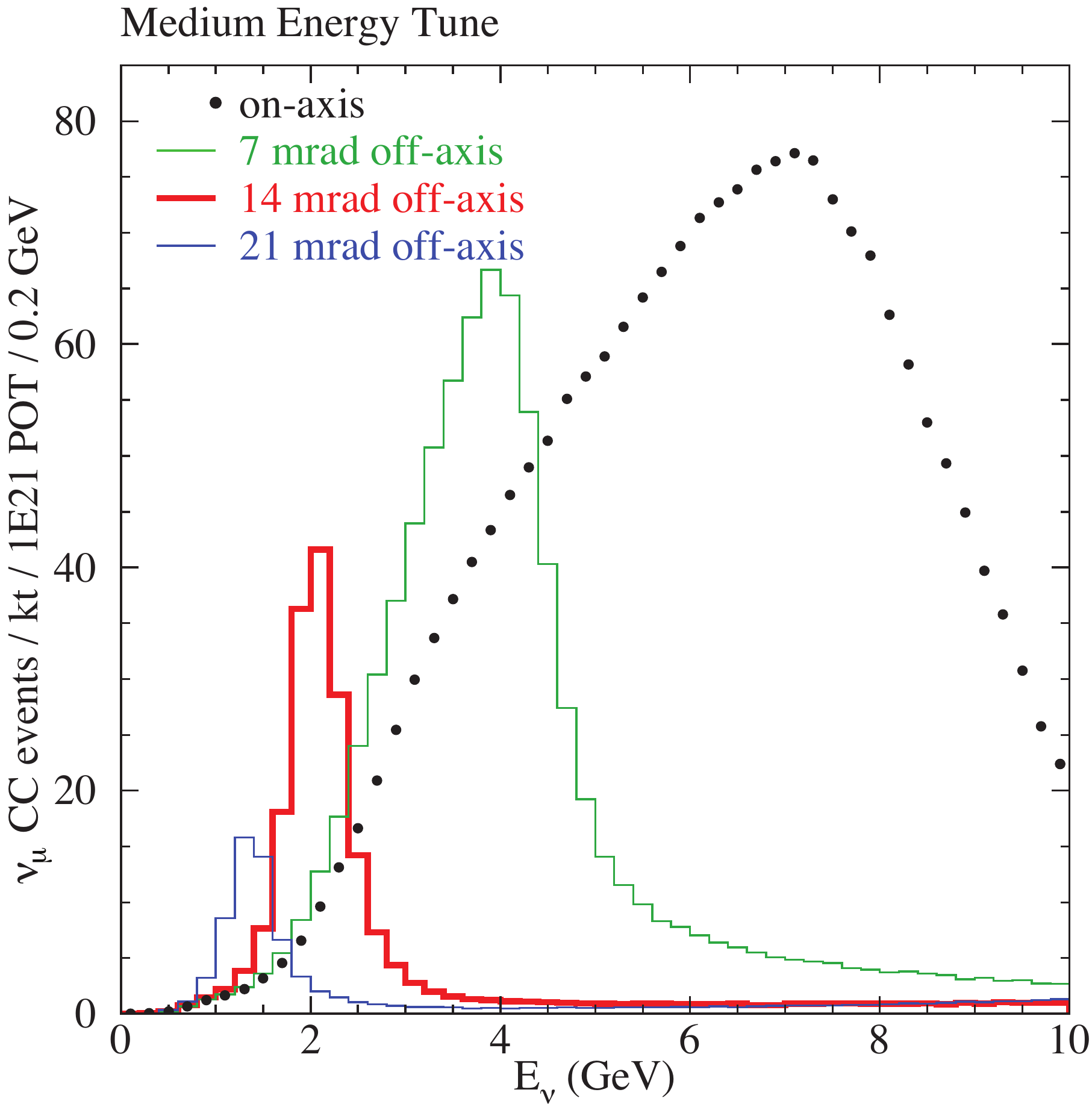}
			\caption{The neutrino spectra (flux times cross-section) for various angles in a medium-energy NuMI beam ~\cite{nova_offaxis_spectra}.}
			\label{fig:NUMI_me_spectra}
	\end{center}
\end{figure}

Physically, the constraint on the angle means that the variation of neutrino energy that normally occurs when $\theta^*$ varies is greatly reduced, and parent particles of many energies contribute to a single peak in neutrino energy.
As a consequence, the peak energy of the neutrino beam depends on the off-axis angle. Figure~\ref{fig:NUMI_me_spectra} illustrates this effect graphically for several off-axis angles in the NuMI configuration.

By changing the off axis angle, it is possible to tune the neutrino beam energy 
to maximize the sensitivity of the oscillation parameters.
As a reference, the off axis angle can be varied from 2.5 
to 3.0 degree in the T2K beamline, which corresponds to a mean energy of neutrinos 
in the range from 0.5 to 0.9 GeV.

\begin{figure}[htbp]
	\begin{center}
		\includegraphics[width=8.2cm]{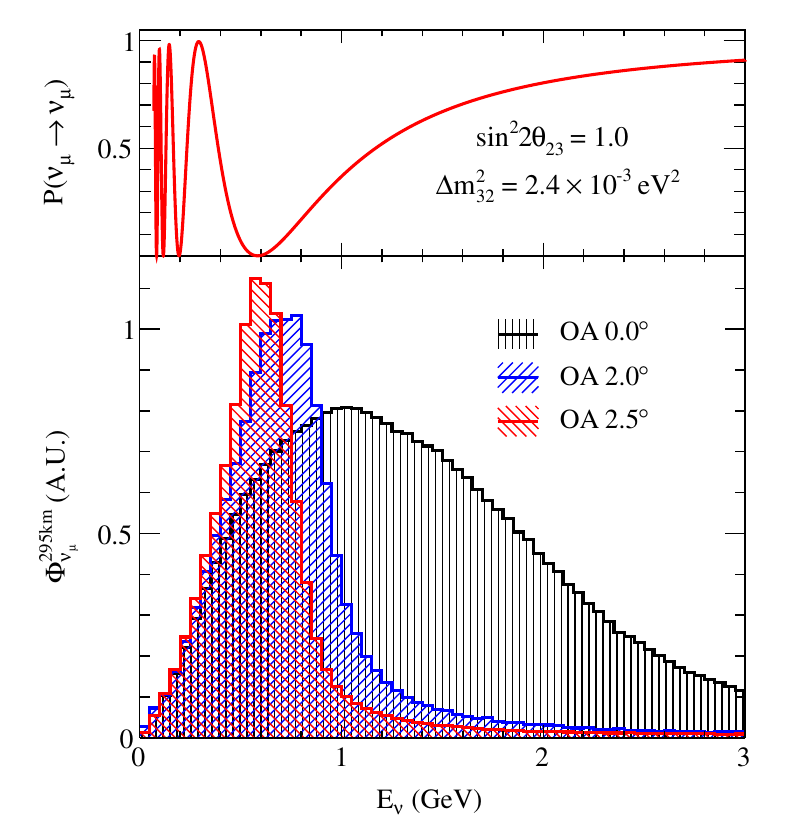}		        
			\caption{The neutrino oscillation probability of $\nu_\mu \to \nu_\mu$ and the neutrino energy spectrum with different off-axis angles in T2K from~\cite{Abe:2012av}}  \label{fig:t2k:offaxis}
	\end{center}
\end{figure}

The neutrino energy spectra at the far detector (Super-Kamiokande) with different off-axis angles in T2K are shown in Figure~\ref{fig:t2k:offaxis}~\cite{Abe:2012av}. In T2K, the off-axis angle is set to 2.5~degrees.

The NOvA experiment is situated at an off-axis angle of 14 mrad (0.8 degrees).  With a higher beam energy focusing than used for the MINOS program, this results in a large flux at the neutrino energy associated with oscillations. At the same time, it reduces backgrounds from neutral current (NC) interactions from higher energies and from intrinsic beam $\nu_e$, which have a wider energy distribution.

\
\\

\subsection{T2K neutrino beam}
\begin{figure}[htbp]
	\begin{center}	\includegraphics[width=7.6cm]{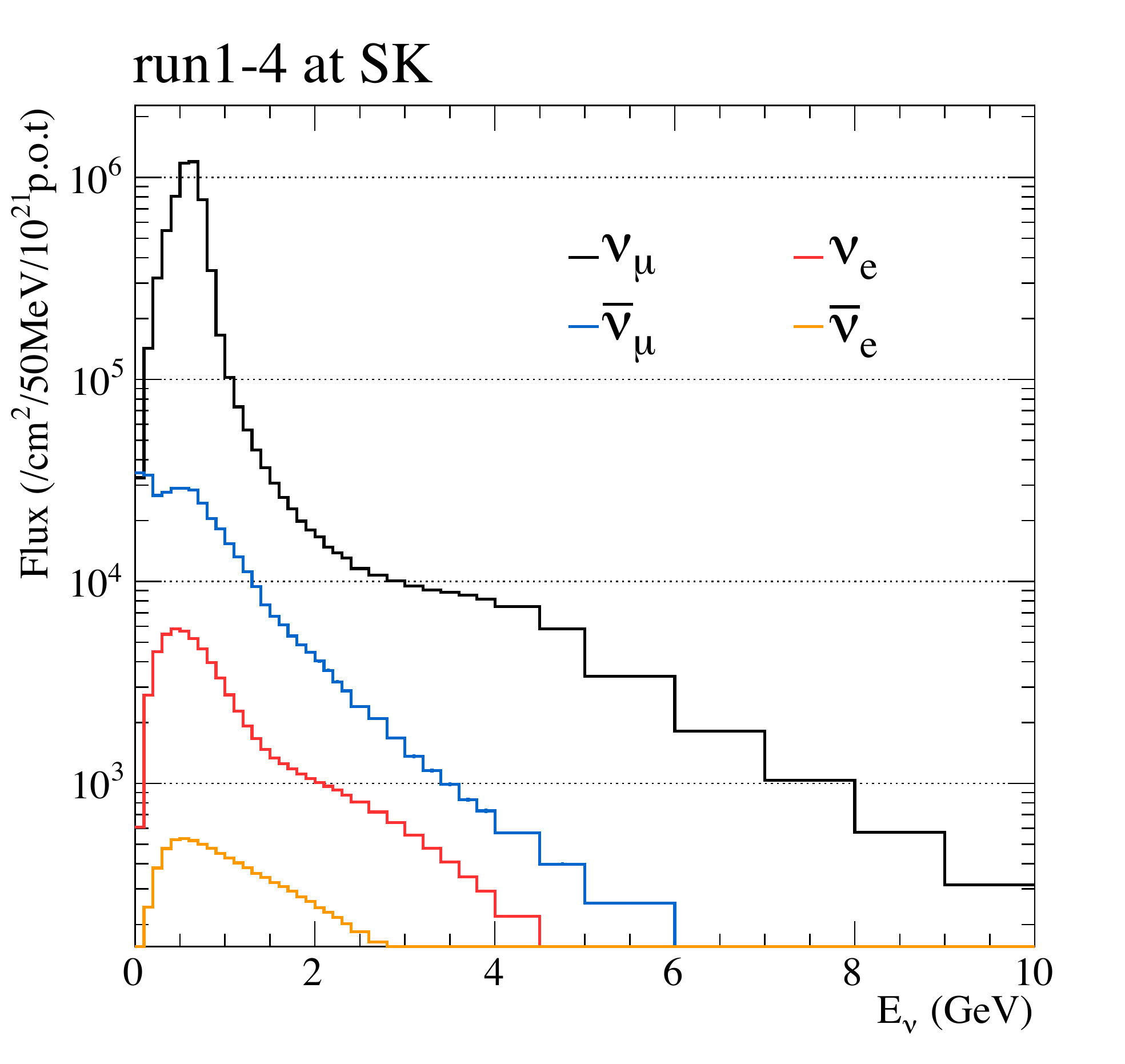}		        
				\includegraphics[width=7.6cm]{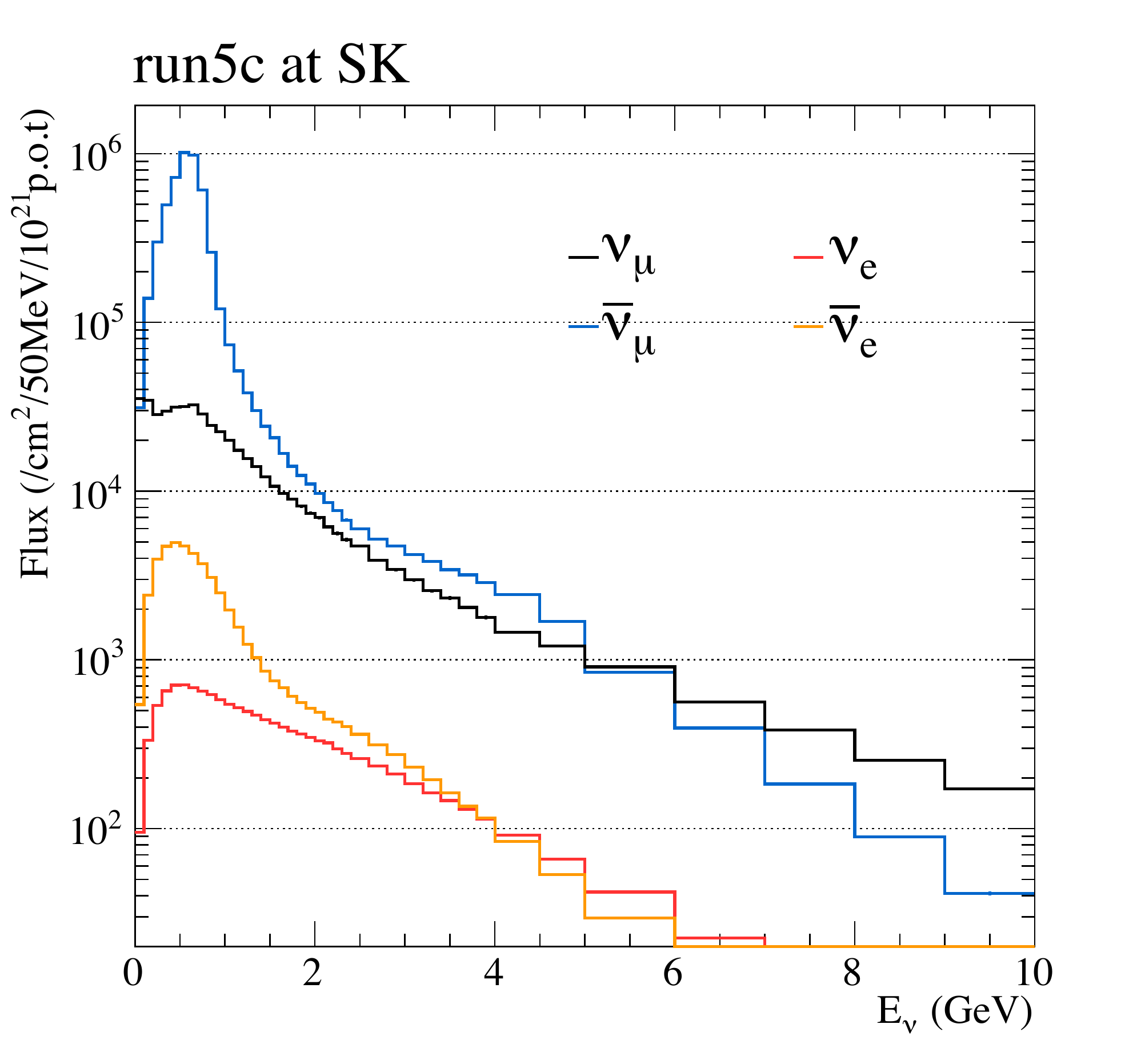}		        
			\caption{Prediction of the T2K beam flux for neutrinos and antineutrinos at Super-K. 
			The left plot is for the neutrino beam mode made by focusing the positively charged particles, 
			and the right is for the anti-neutrino mode made by focusing the negative ones.
			The flux above $E_\nu = 10$~GeV is not shown although the flux is simulated up to $E_\nu = 30$~GeV. }  \label{fig:t2k:flux}
	\end{center}
\end{figure}

J-PARC, the Japan Proton Accelerator Research Complex, is the accelerator complex supplying 30~GeV protons to the T2K experiment.
An intense neutrino beam with  a narrow-band energy spectrum is produced using the  off-axis technique.
The beam energy is tuned to the oscillation maximum ($\sim 600$~MeV for the T2K baseline of 295~km), which also suppresses the high energy component contributing to background generation.
The left plot in Figure~\ref{fig:t2k:flux} shows the prediction of the T2K neutrino beam flux at the far detector, Super-Kamiokande (Super-K). 
The flux is dominated by muon neutrinos with a small fraction (at the level of a few \%) of intrinsic electron neutrinos, referred to as "beam $\nu_e$". The beam $\nu_e$ component is a major background when searching for electron neutrino appearance.

The proton beam is directed onto a graphite target which is designed to accept 750 kW beam power. The target is a graphite rod of 91.4 cm long and 2.6 diameter with $1.8 g/cm^3$ density. The target is helium-cooled. Since the current beam power is still around 350 kW, there is still a margin for safety. The details of the T2K target are found in ref. \cite{Abe:2011ks}. The positively charged particles (mainly pions) produced are focused by three magnetic horns, typically operated at 250~kA. The decay of the charged particles in a 100~m decay volume produces the neutrino beam.
By reversing the direction of the horn current, negatively charged particles are focused to produce the anti-neutrino beam.
The prediction of the anti-neutrino beam flux is shown at the right plot in Figure~\ref{fig:t2k:flux}.
Thanks to the off-axis technique, the signal to noise ratio of the anti-neutrino beam flux is as good as 30~\footnote{The signal to noise ratio of anti-neutrinos to neutrinos is typically much worse than that of neutrinos to anti-neutrinos.} at the flux peak, while the wrong sign component of neutrinos is broadly distributed in energy.

\begin{figure}[htbp]
	\begin{center}
		\includegraphics[width=15cm]{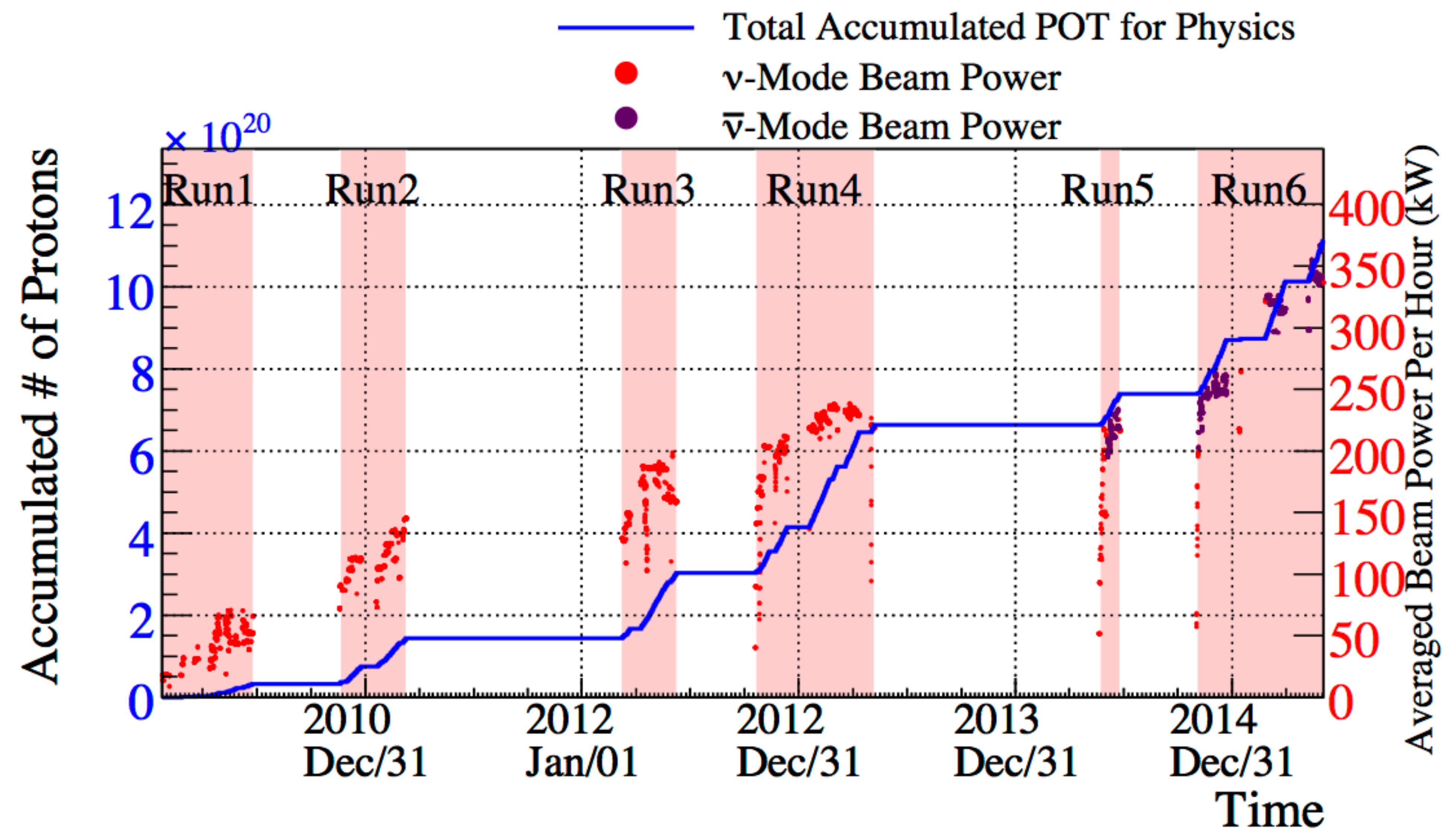}		        
			\caption{The history of the delivered protons to the T2K experiment for analysis.
			The dots show the number of protons per pulse, and the lines show the integrated number of protons.
			The red dots are for the neutrino beam running, and the purple dots are for the anti-neutrinos.}  \label{fig:t2k:pot}
	\end{center}
\end{figure}

T2K started physics data taking in January 2010. Although the data taking was interrupted on March 11, 2011 by the Great East Japan Earthquake, the experiment collected $6.57 \times 10^{20}$~POT (protons on target) for analysis before May 2013~\footnote{J-PARC stopped operation in May 2013 because of the hadron hall accident.}.
 The history of data taking is shown in Figure~\ref{fig:t2k:pot}. 
In 2014, anti-neutrino beam running began. Today, a maximum beam power of 370~kW has been recorded in J-PARC.

\section{Electron Appearance Analysis} \label{sec:nue}

\subsection{MINOS}
\label{subsec:MINOS_Electron_Section}

The MINOS experiment, with its magnetized steel calorimeters, was principally designed to detect and classify the charged current reactions $\nu_\mu+N\rightarrow \mu^-+X$ and  $\bar {\nu}_\mu + N\rightarrow \mu^+ + X$. In order to detect and measure the appearance of $\nu_e$ and $\bar{\nu_e}$, which indicate a non-zero $\theta_{13}$, sophisticated statistical techniques must be used to disentangle the relative contributions of this signal from the similar neutral-current background. To do this, MINOS uses the LEM (Library Event Matching) technique \cite{minos_nue_PRL107}. This procedure uses large ($>10^7$) simulated samples of signal and background events to form event-by-event comparisons of the observed deposited charge in detector channels with the equivalent simulated deposited charge in the library events. The LEM procedure gives a set of output variables which are used as input to a simple artificial neural net, giving a statistical discriminant ( $\alpha_{LEM}$ ), which can be used to identify signal and background components of the data. 

The discriminant  $\alpha_{LEM}$ is formed by the output of a neural net which has been given as inputs the reconstructed event energy and characteristics of the 50 best-matched library events, namely 
i) the fraction of these library events that are nue CC events, 
(ii) their average inelasticity (y) and
 (iii) the average overlapping fraction of charge on strips between the data and the 50 library events. Events with $\alpha_{LEM}>0.6$ are selected for further analysis.

\begin{figure}
\begin{center}
\includegraphics[scale=.5]{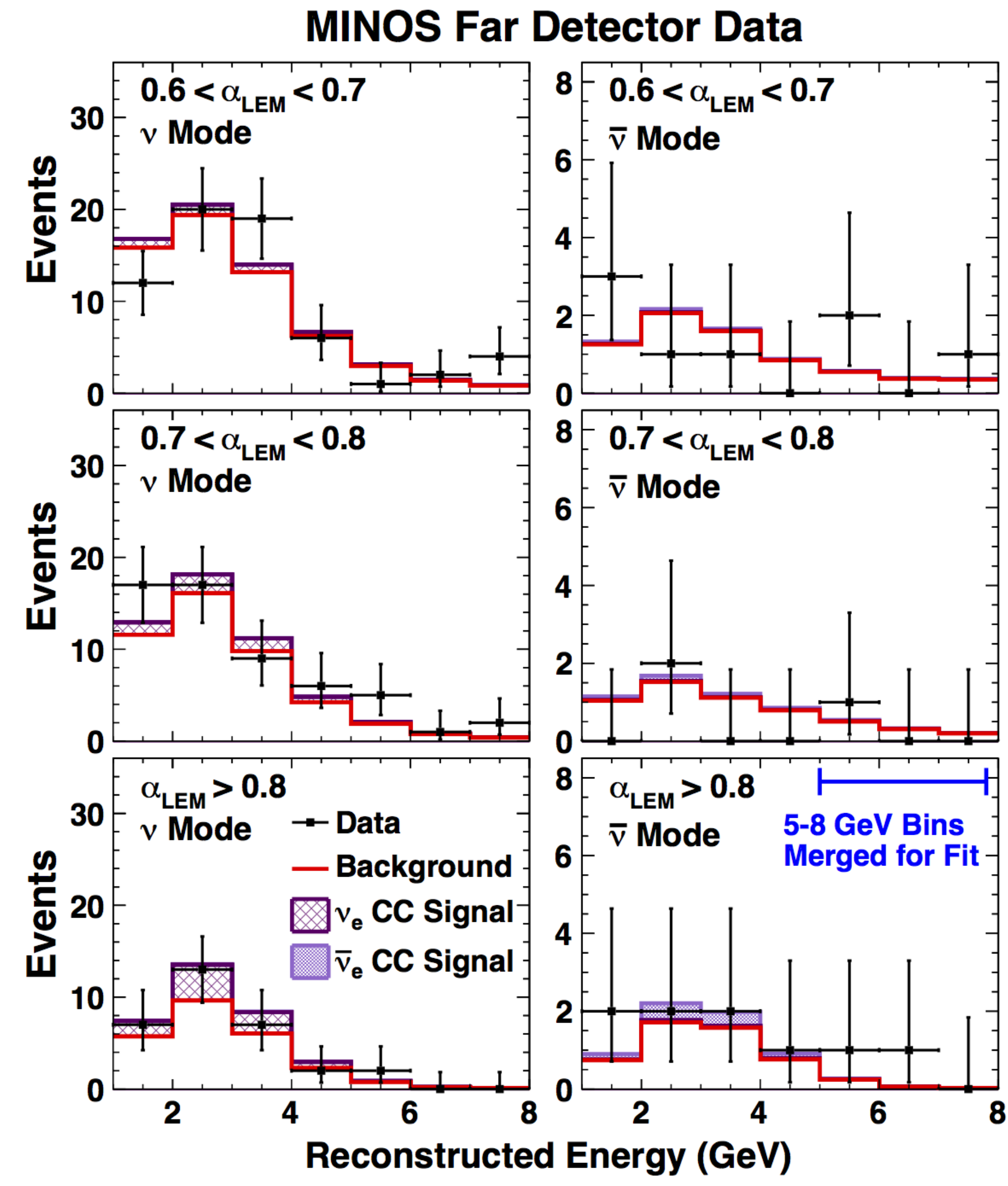}\
\caption{MINOS far detector data with statistical errors \protect\cite{minos_nue_PRL110} used for $\nu_e$ appearance analysis, compared with expectations for sin$^2$(2$\theta_{13}$) = 0.051, $\Delta m^{2}_{32}$ $>$ 0, $\delta = 0$, and $\theta_{23}$ = $\pi/4$}.
\label{fig:MINOS_Electron_Data}
\end{center}
\end{figure}
\begin{figure}
\begin{center}
\includegraphics[scale=.5]{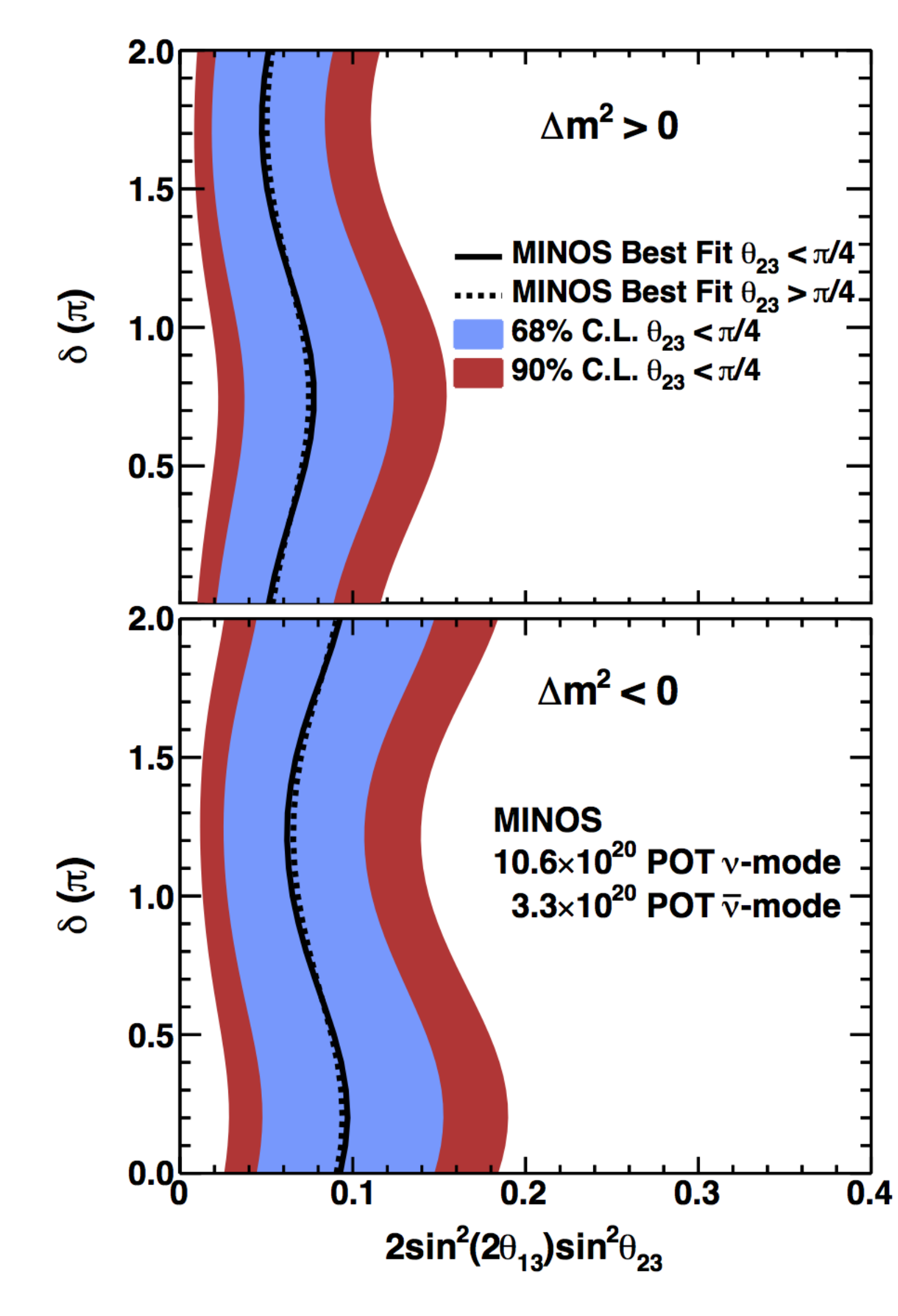}
\caption{MINOS allowed contours at 68\% and 90\% confidence for the measured quantity  $2 \sin^2 2\theta_{13}\sin^2\theta_{23}$, as a function of $\delta_{CP}$. Systematic and statistical uncertainties are included, and results for both assumed hierarchies are displayed.}
\label{fig:MINOS_Electron_Result}
\end{center}
\end{figure}
\begin{figure}
\begin{center}
\includegraphics[scale=.4]{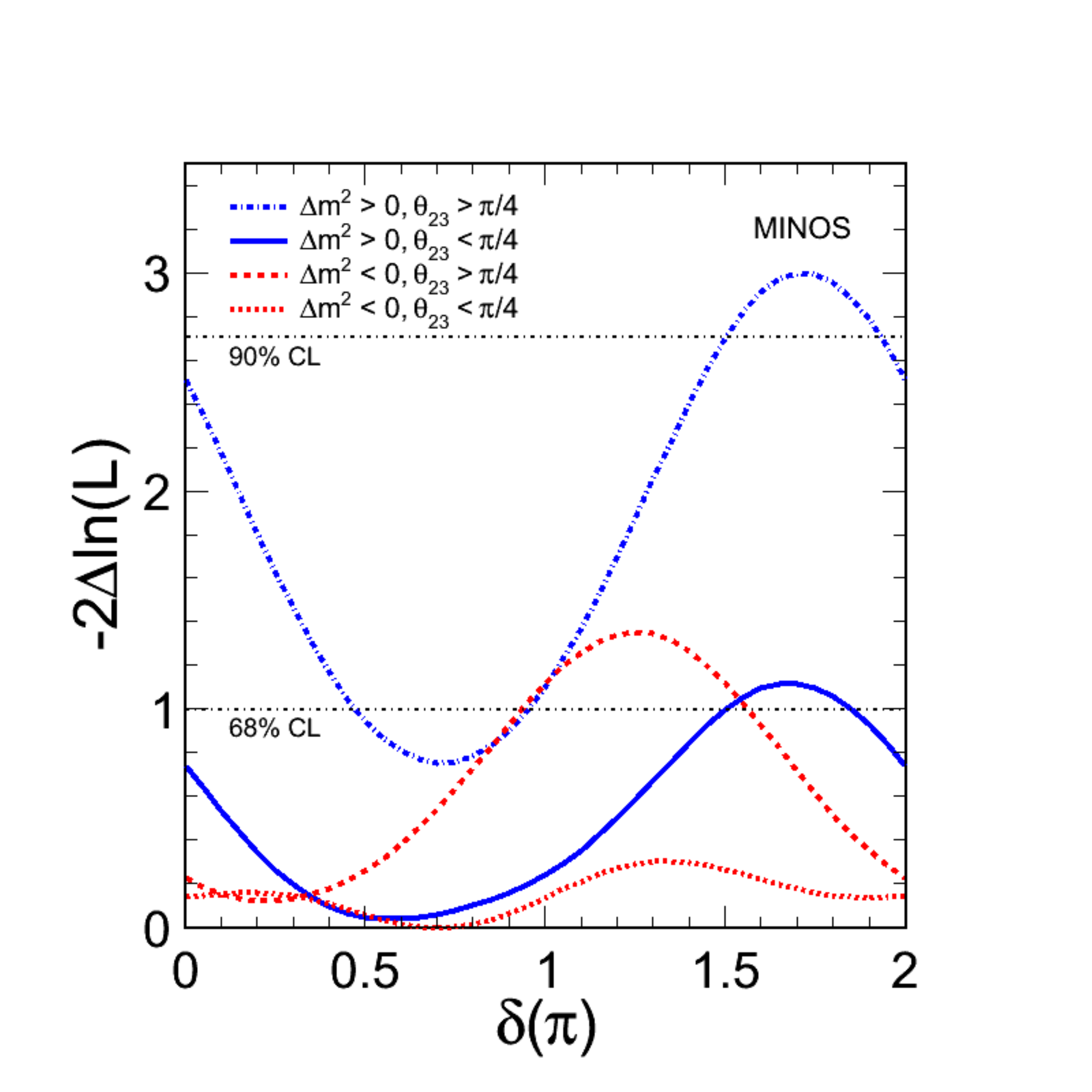}
\caption{Variation of likelihood (compared to best fits) with $\delta_{CP}$, plotted as $-2\Delta ln(L)$, for the observed $\nu_e$ appearance in MINOS, shown for varying combinations of other oscillation parameters. Values above the horizontal dashed lines are disfavored at either 68\% or 90\% C.L.}
\label{fig:MINOS_Electron_CP}
\end{center}
\end{figure}

Next, to search for appearance of $\nu_e$ due to the oscillation phenomenon, the spectra of the varying background components present in the beam ($\nu_\mu$-CC, NC, and residual beam $\nu_e$-CC) need to be estimated using data from the near detector. This is done in MINOS by comparing samples obtained from different beam focusing configurations for decaying secondaries, as discussed in ref. \cite{minos_nue_PRD82}. Figure \ref{fig:MINOS_Electron_Data}, from ref. \cite{minos_nue_PRL110} shows the far detector MINOS data and the expected backgrounds, for various bins of $\alpha_{LEM}$.\\
\\
The final elements required to produce an appearance measurement are extrapolation of the background (and oscillated signal) estimates between near and far detectors, and an estimate of the signal efficiency. The first is done by comparison of the background measured in the ND with its simulated value, giving a correction factor that can be applied to the equivalent simulation of the far detector. 
The technique is simpler than that used for MINOS CC appearance measurement;
however the essential equivalence of the methods has been demonstrated in ref. \cite{minos_CC_PRD}.
In order to estimate the signal efficiency, hybrid events were created by substitution of a simulated electron shower in shower-subtracted, well-identified CC events. The efficiencies obtained were $>55\%$ in both beam configurations.\\
\\
The principal systematic errors affect the result are uncertainties in the background estimation and in the signal efficiency. They are 3.8\% (4.8\%) and 2.8\% (3.1\%), respectively, for the $\nu$ ($\bar{\nu}$) modes. The measurement is dominated by statistical errors, affecting both the signal and the background estimation. MINOS systematic errors are discussed further in section \ref{subsec:MINOS_Sys_Section}.\\
\\
\\
After establishment of these techniques and their systematic errors, MINOS can now use the near detector spectrum to extrapolate the expectation of signal and background for hypothesized values of the the physical parameters $\theta_{13}$, $\delta_{CP}$, and mass hierarchy, to determine statistically allowed and disallowed regions. The overall background estimation for the $\nu$ beam configuration is  127.7 background events. For parameter values of  $\sin^2 2\theta_{13}=0.1$,  $\delta_{CP}=0$, $\theta_{23}=\pi/4$ and normal hierarchy, $33.7\pm{1.9}$ appearance events are expected, giving a signal/background ratio of $S/B = 0.26$. A total of 152 events are observed. Contributions to the analysis from the $\bar{\nu}$ beam configuration are small, totaling only 21.4 expected and 20 observed events. The result from ref. \cite{minos_nue_PRL110} is shown in Figure~\ref{fig:MINOS_Electron_Result} for the full MINOS dataset, consisting of $10.6 \times 10^{20}$ POT for the $\nu$ beam configuration and $3.3 \times 10^{20}$ POT for the $\bar{\nu}$ beam configuration. This analysis does not distinguish between neutrinos and antineutrinos. The result shows the characteristic features of periodic variation with $\delta_{CP}$, as well as shape inversion with hierarchy choice. MINOS cites best-fit values for $2 \sin^2 2\theta_{13}\sin^2\theta_{23}$ of $0.051^{+0.038}_{-0.030}$ under the normal hierarchy assumption, and $0.093^{+0.052}_{-0.049}$ for the inverted hierarchy, with 90\% confidence ranges of 0.01 - 0.12 and 0.03 - 0.18 , respectively. The best fits are all computed for $\delta_{CP} = 0$, and $\theta_{23}<\pi /4$.\\
\\
It is of interest to examine the parameter space probed by this appearance analysis more closely. The MINOS collaboration has computed the change in likelihood for excursions of the CP-violating phase $\delta_{CP}$ for four combinations of the hierarchy and $\theta_{23}$ octant parameters. This result from ref. \cite{minos_nue_PRL110} is shown in Figure \ref{fig:MINOS_Electron_CP}. The experiment disfavors 31\% of the total 3-parameter space ($\delta_{CP}$, hierarchy, octant) at 68\%C.L. and shows a suggestive, but statistically limited preference for the inverted mass hierarchy scenario.
\\
\\
\\
\\

\subsection{T2K} \label{sec:t2k:nue}
The first evidence of non-zero $\theta_{13}$ was reported~\cite{Abe:2011sj, minos_nue_PRL107}  in the $\nu_\mu \to \nu_e$ appearance channel in 2011.
Today, with more data collected in T2K,  the $\nu_\mu \to \nu_e$ transition is well established~\cite{Abe:2013hdq}.
Twenty-eight electron candidate events in T2K have been observed in the T2K far detector (Super-K)  by requiring one Cherenkov "ring", identified as an electron type with visible energy greater than 100~MeV. In addition, a newly developed algorithm was applied to suppress background events with a $\pi^0 \to 2 \gamma$,  where one of the photons is missed in reconstruction. The details of event selection are found in~\cite{Abe:2013xua, Abe:2013hdq}.
The number of observed events, compared with the expectations is shown in Table~\ref{tab:t2k:nue}.
\begin{table}[th]
\begin{center}
\caption{The number of observed events with the MC expectations and efficiencies. The oscillation parameters are assumed to be
$\textrm{sin}^{2}2\theta_{13}=0.1$, $\sin^2 \theta_{23}=0.5$, $|\Delta m^2_{32}|=2.4\times10^{-3}$ eV$^{2}$, $\delta_{\mathrm{CP}}=0$, and $\Delta m^2_{32}>0$} \label{tab:t2k:nue}
\begin{tabular}{lc|c|ccccc}
\hline
\hline
   & \multirow{2}{*}{Data}   & Total  & Signal $\nu_\mu \to \nu_e$    & $\nu_\mu + \overline{\nu}_\mu$ & Beam $\nu_e + \overline{\nu}_e$ & \multirow{2}{*}{NC}   \\ 
                          &      &   MC  & CC & CC    & CC         &                       \\ 
\hline                                                                                                  %
$\nu_e$ events                      & 28   &    21.6 & 17.3    & 0.1   & 3.2       & 1.0          \\ 
\hline
Efficiency [\%]                         & -     & - &  61.2   & 0.0   & 19.1      & 0.4    \\ 
\hline
\hline
\end{tabular}
\end{center}
\end{table}
The observed number of events, 28, is significantly larger than the expected number, $4.92 \pm 0.55$, with $\theta_{13}=0$, but is consistent with the expectation of $21.6$ with $\sin^2 2\theta_{13} = 0.10$ and $\delta_{\mathrm{CP}}=0$.

\begin{figure}
\begin{center}
\includegraphics[scale=1.5]{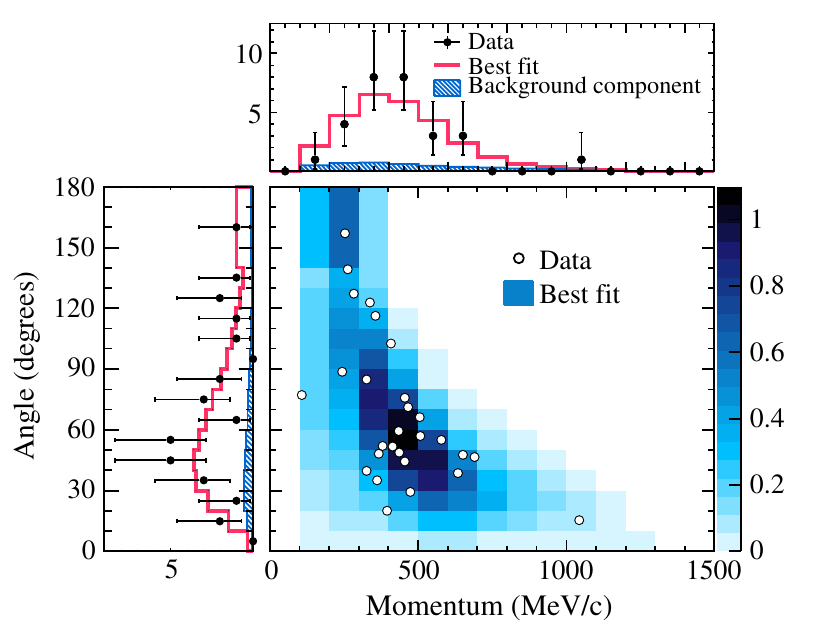}
\caption{The T2K electron momentum versus angle distribution for 28 single-ring electron events, together with the MC expectation in~\cite{Abe:2013hdq}. 
The best fit value of $\sin^2 2 \theta_{13} = 0.140$ in the normal hierarchy case is used for the expectation.} \label{fig:t2k:nue1}
\end{center}
\end{figure}

The best fit value of $\theta_{13}$ has been evaluated to be $\sin^2 2 \theta_{13} = 0.140 \pm 0.038 (0.170 \pm 0.045)$ with a 68~\% confidence level (C.L.), by fixing the other oscillation parameters:  $\sin^2 2 \theta_{12} = 0.306$, $\Delta m_{21}= 7.6 \times 10^{-5}$~$\rm eV^2$, $\sin^2 \theta_{23}=0.5$, $|\Delta m^2_{32}|=2.4\times10^{-3}$ eV$^{2}$, and $\delta_{\mathrm{CP}}=0$.
Figure~\ref{fig:t2k:nue1} shows the electron momentum versus angle distribution (sensitive to the oscillation), which is used to extract the oscillation parameters $\sin^2 2 \theta_{13} $ and $\delta_{\mathrm{CP}}$ which give the best fit values.
The significance for a nonzero $\theta_{13}$ is calculated to be $7.3$~$\sigma$.

\begin{figure}
\begin{center}
\includegraphics[scale=1.5]{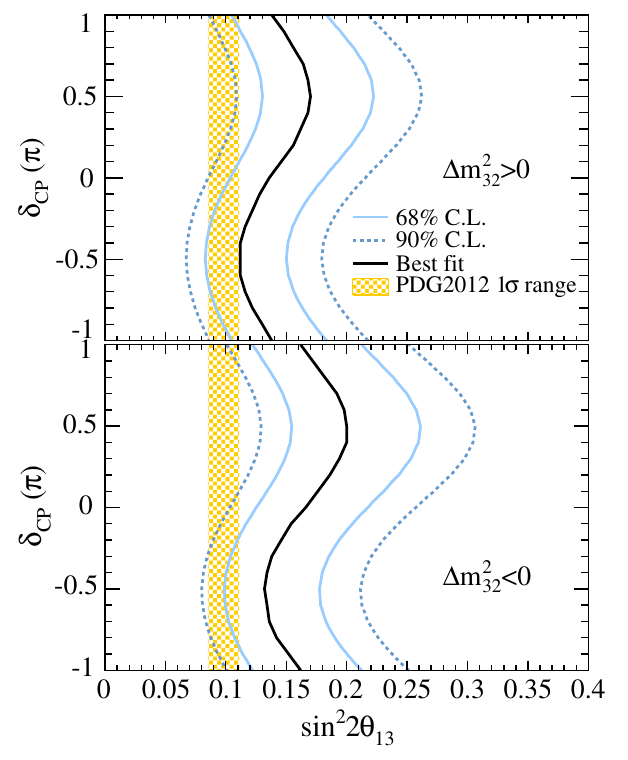}
\caption{The allowed regions for $\sin^2 2 \theta_{13}$ as a function of $\delta_{\mathrm{CP}}$ for the normal mass hierarchy case (top) and the inverted one (bottom) from~\cite{Abe:2013hdq}. The value of $\theta_{13}$ from reactor experiments in PDG2012 is shown as the shaded region } \label{fig:t2k:nue2}
\end{center}
\end{figure}

\begin{figure}
\begin{center}
\includegraphics[scale=1.5]{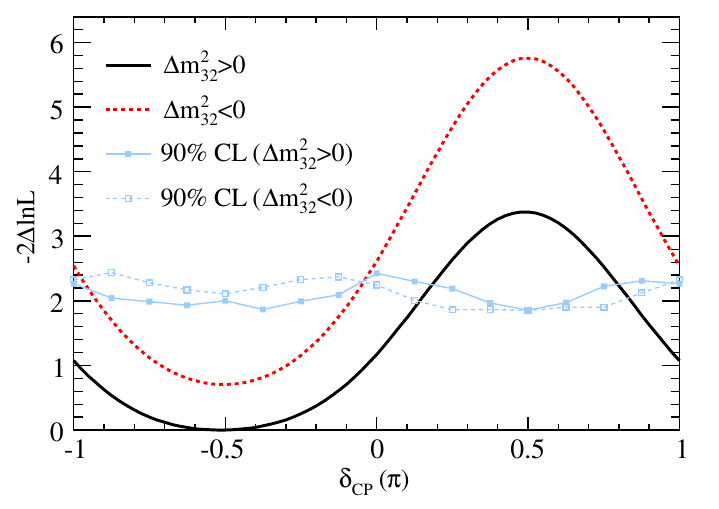}
\caption{The $-2 \Delta \ln L$  value as a function of $\delta_{\mathrm{CP}}$ with the reactor $\theta_{13}$ constraints  
for normal and inverted hierarchies from~\cite{Abe:2013hdq}. The likelihood is marginalized over  $\theta_{13}$, $\theta_{23}$ and $\Delta m^2_{32}$.
The 90~\% C.L. is evaluated by using the Feldman-Cousins method. The $\delta_{\mathrm{CP}}$ regions with values
above the 90~\% C.L. lines are excluded.} \label{fig:t2k:nue3}
\end{center}
\end{figure}

Allowed regions for $\sin^2 2 \theta_{13}$ as a function of $\delta_{\mathrm{CP}}$ are evaluated as shown in Figure~\ref{fig:t2k:nue2}, where the 
values of $\sin^2 \theta_{23}$ and $\Delta m^2_{32}$ are varied in the fit with additional constraints from~\cite{Abe:2013fuq}.
In order to be sensitive to $\delta_{\mathrm{CP}}$, T2K uses the value of $\theta_{13}$, $0.098 \pm 0.013$, from reactor experiments in PDG2012~\cite{Beringer:1900zz}.
The $-2 \Delta \ln L$ in the fit as a function of $\delta_{\mathrm{CP}}$ is extracted,  and is shown in Figure~\ref{fig:t2k:nue3}.
The T2K measurement, together with the reactor $\theta_{13}$ value prefers $\delta_{\mathrm{CP}} = - \pi/2$ with an exclusion of $0.19 \pi < \delta_{\mathrm{CP}} < 0.80 \pi (-\pi < \delta_{\mathrm{CP}} <- 0.97 \pi \mbox{ and } -0.04 \pi < \delta_{\mathrm{CP}} < \pi)$ with normal (inverted) hierarchy at 90~\% C.L. This may be a hint of CP violation in neutrinos.
%

\section{Precise measurements of oscillation parameters}  \label{sec:numu+}

\subsection{T2K} \label{sec:t2k:numu}
\begin{figure}
\begin{center}
\includegraphics[scale=1.3]{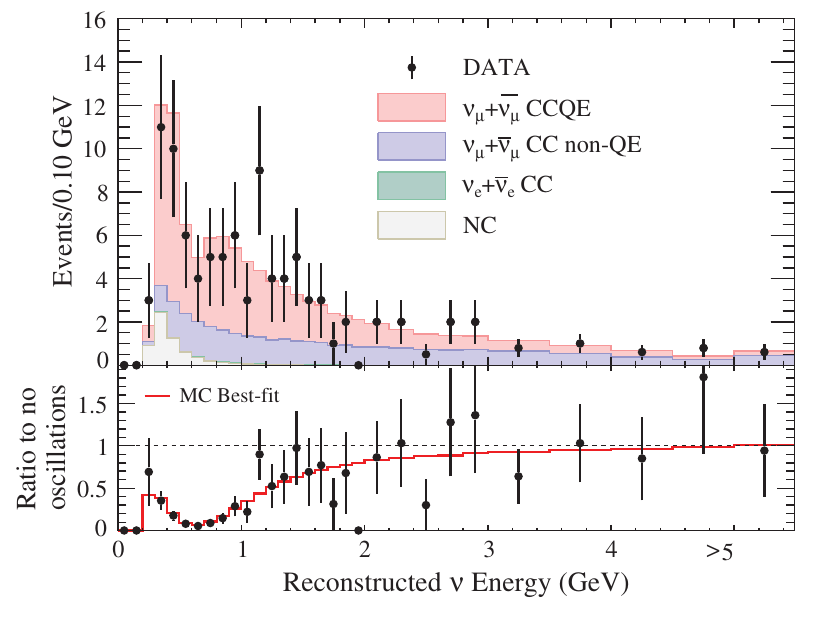}
\caption{The neutrino energy spectrum for single-ring muon events in~\cite{Abe:2014ugx}. (Top) The observed and expected spectra with the event categories in the simulation. (Bottom) The ratio of the observed spectrum to the no-oscillation hypothesis with the best-fit case.} \label{fig:t2k:Enumu}
\end{center}
\end{figure}

The most precise measurement of $\theta_{23}$ has been carried out~\cite{Abe:2014ugx} by T2K based on the data set of $6.57 \times 10^{20}$~POT.
First, a single-ring muon sample\footnote{The number of rings corresponds to the number of observed particles in Super-K} is selected by requiring one muon-type Cherenkov ring with momentum greater than 200~$\rm MeV/c$ in Super-K.
The details of the event selection are found in~\cite{Abe:2014ugx}.
One hundred twenty events are selected while the expectation without neutrino oscillation is $446.0 \pm 22.5$~(syst.).
The neutrino energy for each event is calculated under the quasi-elastic (QE) assumption using the expression
\begin{equation}
E^{rec}_\nu = \frac{m^2_p - (m_n-E_b)^2-m_l^2+2(m_n-E_b)E_l}{2(m_n-E_b-El+p_l \cos \theta_l)}, \label{eq:enu}
\end{equation}
where $m_p$ is the proton mass, $m_n$ the neutron mass, $m_l$ the lepton mass, $E_l$ the lepton energy, and $E_b=27$~MeV the binding energy of a nucleon inside a  $^{16} O$ nucleus.
Figure~\ref{fig:t2k:Enumu} shows the neutrino energy  of the observed 120 events with the MC expectations for neutrino oscillations.

\begin{figure}
\begin{center}
\includegraphics[scale=1.3]{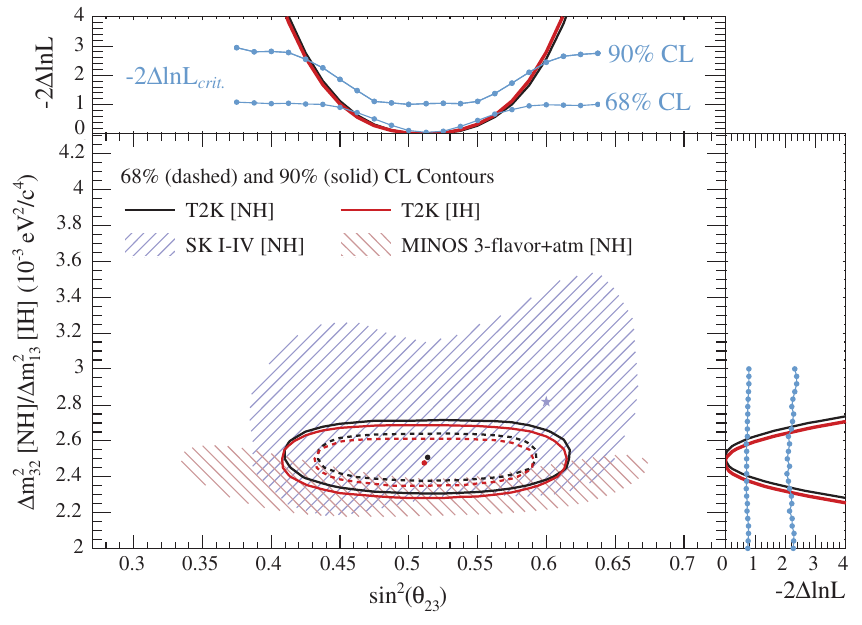}
\caption{T2K Contours of oscillation parameters $\sin^2 \theta_{23}$ versus $|\Delta m^2_{32(13)}|$ for 68~\% and 90~\% C.L. region.
The one-dimensional profile likelihoods are also shown for each oscillation parameter in the top and right windows.
The 1D $-2 \Delta \ln L_{\rm critical}$ values for the normal mass hierarchy are shown in the windows.
The Super-Kamiokande~\cite{Himmel:2013jva} and MINOS~\cite{minos_three_neutrino_prl} results are also shown for comparison.} \label{fig:t2k:theta23}
\end{center}
\end{figure}

Using the number of events and the neutrino energy spectrum,  
the oscillation parameters $(\sin^2 \theta_{23}, \Delta m^2_{32(13)})$ are estimated with an un-binned maximum likelihood fit for the normal (inverted) mass hierarchies. Details of the method are found in~\cite{Abe:2014ugx}. The result is shown in Figure~\ref{fig:t2k:theta23}.

The best fit value with the 1D 68~\% confidence intervals are $\sin^2 \theta_{23} = 0.514^{+0.055}_{-0.056}$ $(0.511 \pm 0.055)$ and 
$\Delta m^2_{32} = 2.51 \pm 0.10$ $(\Delta m^2_{13} = 2.48 \pm 0.10) \times 10^{-3} \rm eV^2$
for the normal (inverted) hierarchy case. The result is consistent with the maximal possible disappearance probability and is more precise than previous measurements, especially for $\sin^2 \theta_{23}$.

\subsection{Joint analysis of $\nu_\mu$ disappearance and $\nu_e$ appearance samples in MINOS}
\label{subsec:MINOS_Joint_Section}

\begin{figure}
\begin{center}
\includegraphics[scale=0.6]{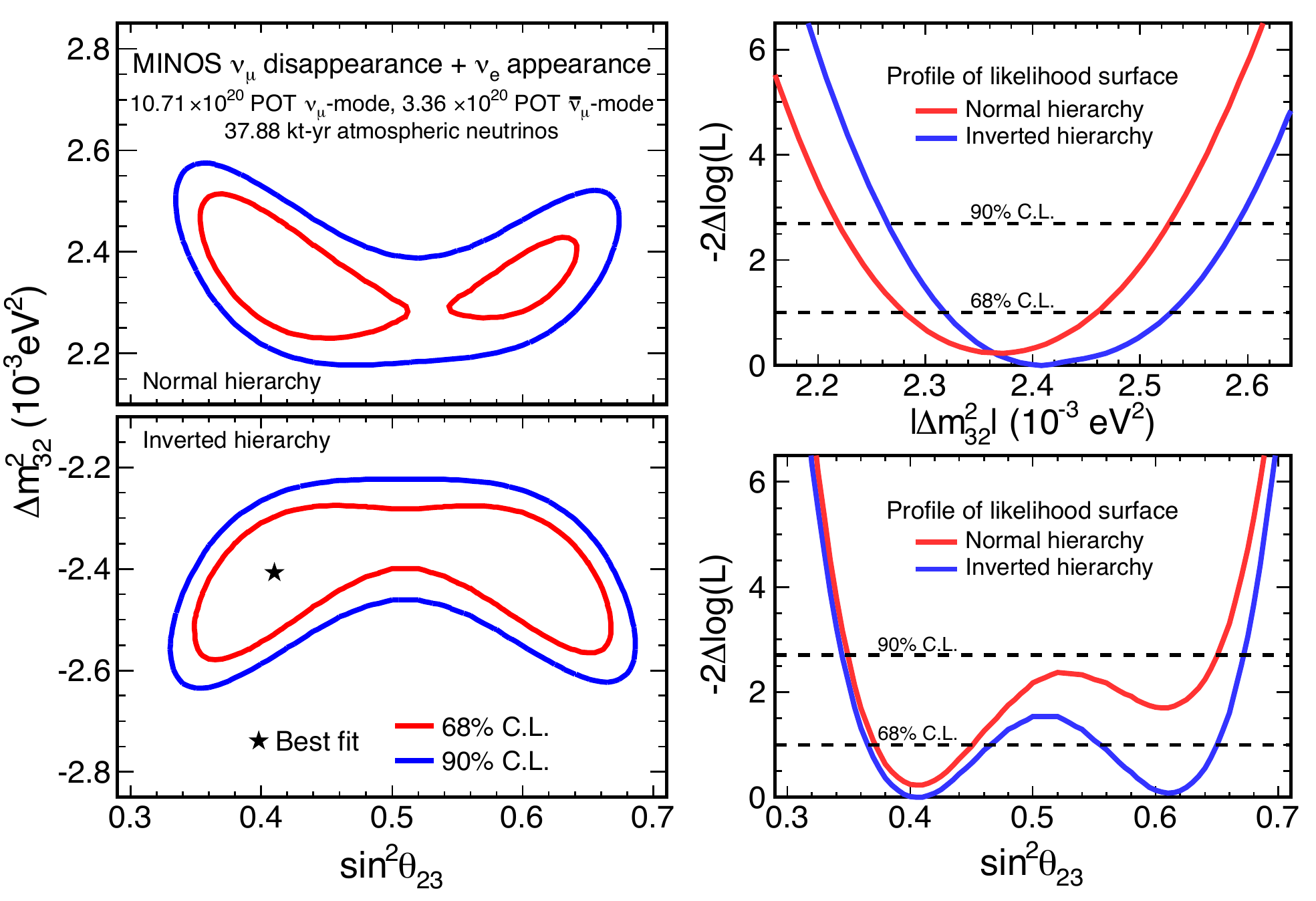}
\caption{MINOS 3-flavor oscillation parameters using both disappearance and appearance data. Left: Confidence level contours for assumed normal and inverted hierarchies, computed using $-2\Delta{ln(L)}$ w.r.t. the overall best fit point (star). Right: One-dimensional likelihood profiles for the parameters. All results from \cite{minos_three_neutrino_prl}.  }
\label{fig:MINOS_3_flavor_result}
\end{center}
\end{figure}

Beginning in 2005 (2003 for collection of atmospheric data), the MINOS experiment has provided precision measurements of the oscillation parameters $\Delta m^2$ and $\sin^2 (2\theta)$ for effective definitions of these parameters in a two-neutrino approximation.  Most recently, MINOS has quoted
 $2.28$ x $10^{-3}<|\Delta m^2_{32}| < 2.46$ x $10^{-3}$ e$V^2$ (68\% confidence) 
 and  a 90\% C.L. range for $\theta_{23}$ of  $0.37<\sin^2 \theta_{13}<0.64$ (both normal mass hierarchy),
using a complete 3-neutrino description of the data \cite{minos_three_neutrino_prl}. Atmospheric data and appearance data are included in the combined fits. In particular, the $\nu_e$ appearance data and atmospheric data provide, in principle, sensitivity to additional information concerning mass hierarchy and CP phase. As an example, the atmospheric data sample, divided into neutrino and antineutrino samples for up-going multi-GeV events, shows different matter effects for normal and inverted hierarchies. In the current sample, these additional sensitivities are limited, as shown by the presentation in 
Figure~\ref{fig:MINOS_3_flavor_result}, and the fitted values of $|\Delta m^2_{32}|$ and $\sin^2\theta_{23}$ are consistent \cite{minos_three_neutrino_prl}. The fit results are obtained using constraints from external data. In particular, a value of
$\sin^2 \theta_{13} =0.0242\pm{0.0025}$ has been taken from a weighted average of reactor experiment values \cite{minos_3_flavor_PRL_reactor}, and solar oscillation parameters are taken from \cite{minos_3_flavor_PRL_fogli}.

\
\\

\subsection{Joint analysis of $\nu_\mu$ disappearance and $\nu_e$ appearance samples in T2K}

The oscillation probability of $\nu_\mu \to \nu_e$ depends on many oscillation parameters: $\sin^2 \theta_{13}$, $\sin^2 \theta_{23}$, $\Delta m^2_{32}$ and $\delta_{\mathrm{CP}}$;  that of $\nu_\mu \to \nu_\mu$ depends mainly on $\sin^2 \theta_{23}$ and $\Delta m^2_{32}$.
Therefore, all oscillation parameters can be efficiently extracted by fitting two data samples of $\nu_e$  and $\nu_\mu$ simultaneously.
For this purpose, the T2K collaboration developed analysis techniques to fit both $\nu_e$ and $\nu_\mu$ samples.
One method is based on the $\Delta \log L$ method, and the other is based on the Markov Chain Monte Carlo (MCMC) method.
The $\theta_{13}$ constraint from PDG2012~\cite{Beringer:1900zz} is applied in the analysis.

\begin{figure}
\begin{center}
\includegraphics[width=15cm]{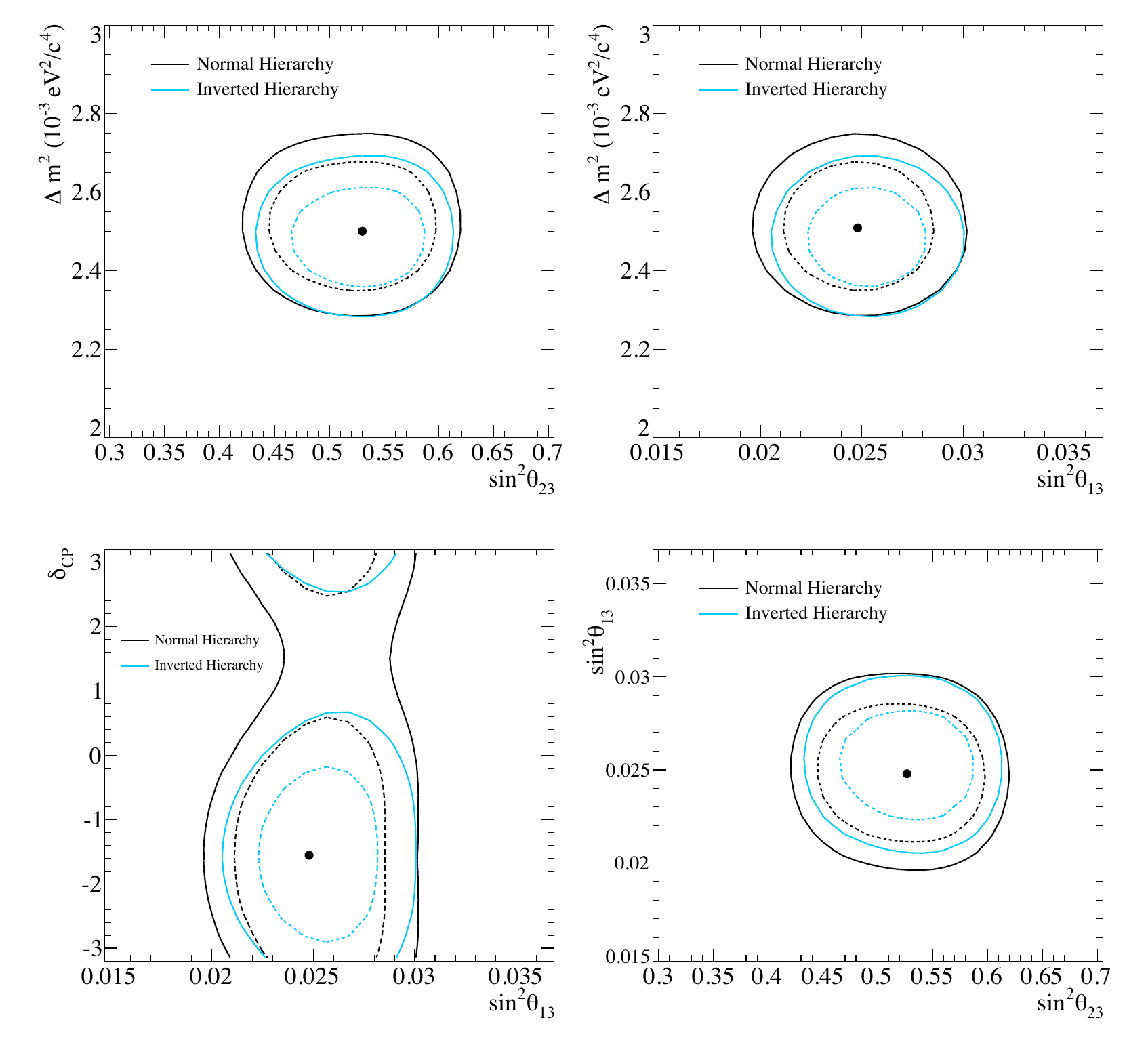}
\caption{T2K Contours of oscillation parameters calculated with the $\Delta \log L$ method for both normal and inverted hierarchy cases~\cite{Abe:2015awa}: (top-left) $\sin^2 \theta_{23}$ versus $|\Delta m^2_{32(13)}|$,  (top-right)  $\sin^2  \theta_{13}$ versus $|\Delta m^2_{32(13)}|$, (bottom-left)  $\sin^2 \theta_{13}$ versus $\delta_{\mathrm{CP}}$, and (bottom-right) $\sin^2 \theta_{23}$ versus $\sin^2 \theta_{13}$. The 90~\% (68~\%) C.L. are shown in the solid (dashed) lines with the best fit point shown by the mark.}
\label{fig:t2k:join1}
\end{center}
\end{figure}

With the $\Delta \log L$ method, T2K measures $\sin^2 2 \theta_{13}$, $\sin^2 \theta_{23}$, $\Delta m^2_{32(13)}$ and $\delta_{\mathrm{CP}}$ as shown in Figure~\ref{fig:t2k:join1}. The results are consistent with those shown in Section~\ref{sec:t2k:nue} and ~\ref{sec:t2k:numu}, and the correlations between parameters are properly treated.

\begin{figure}
\begin{center}
\includegraphics[width=7.5cm]{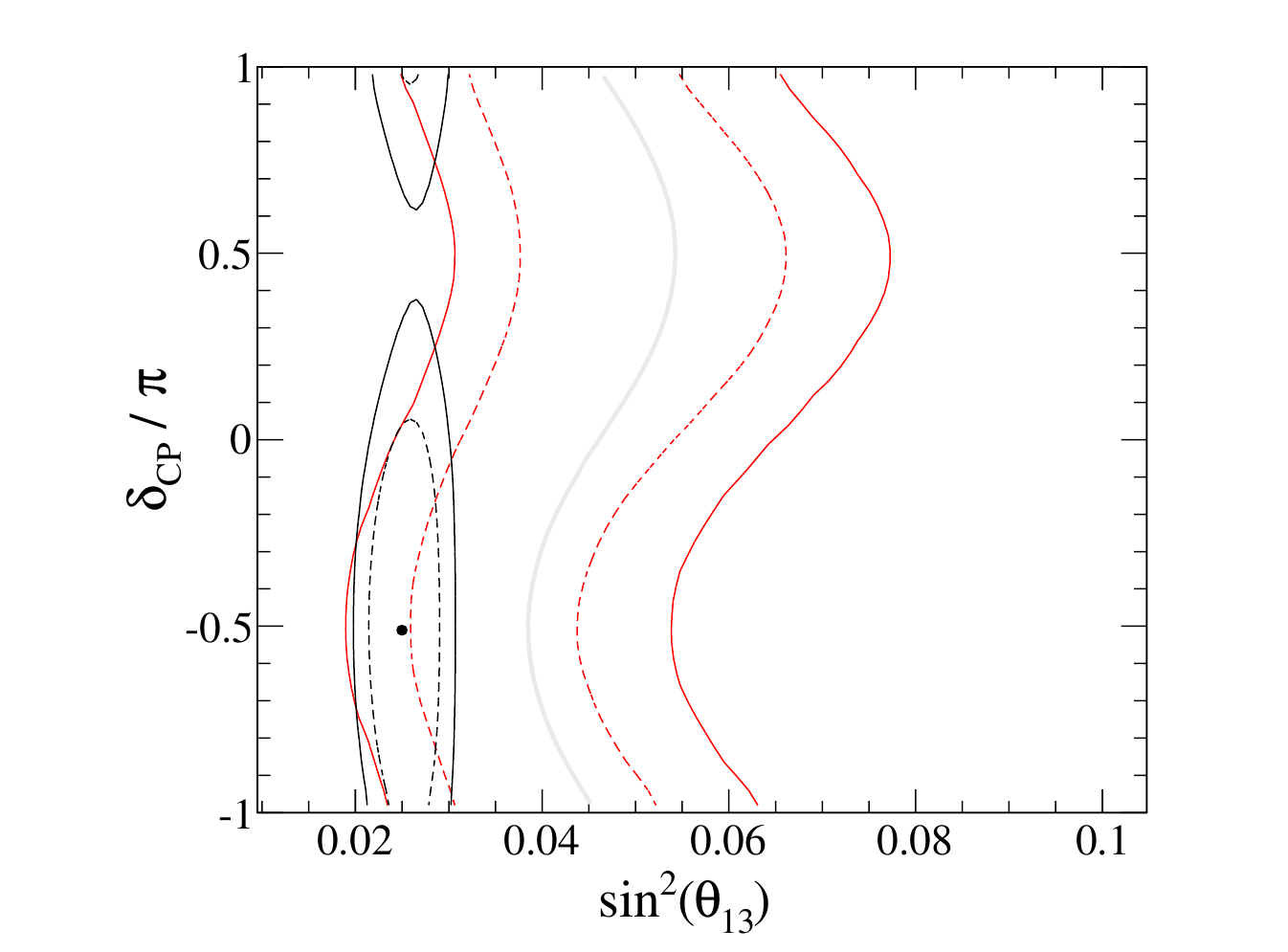}
\includegraphics[width=8cm]{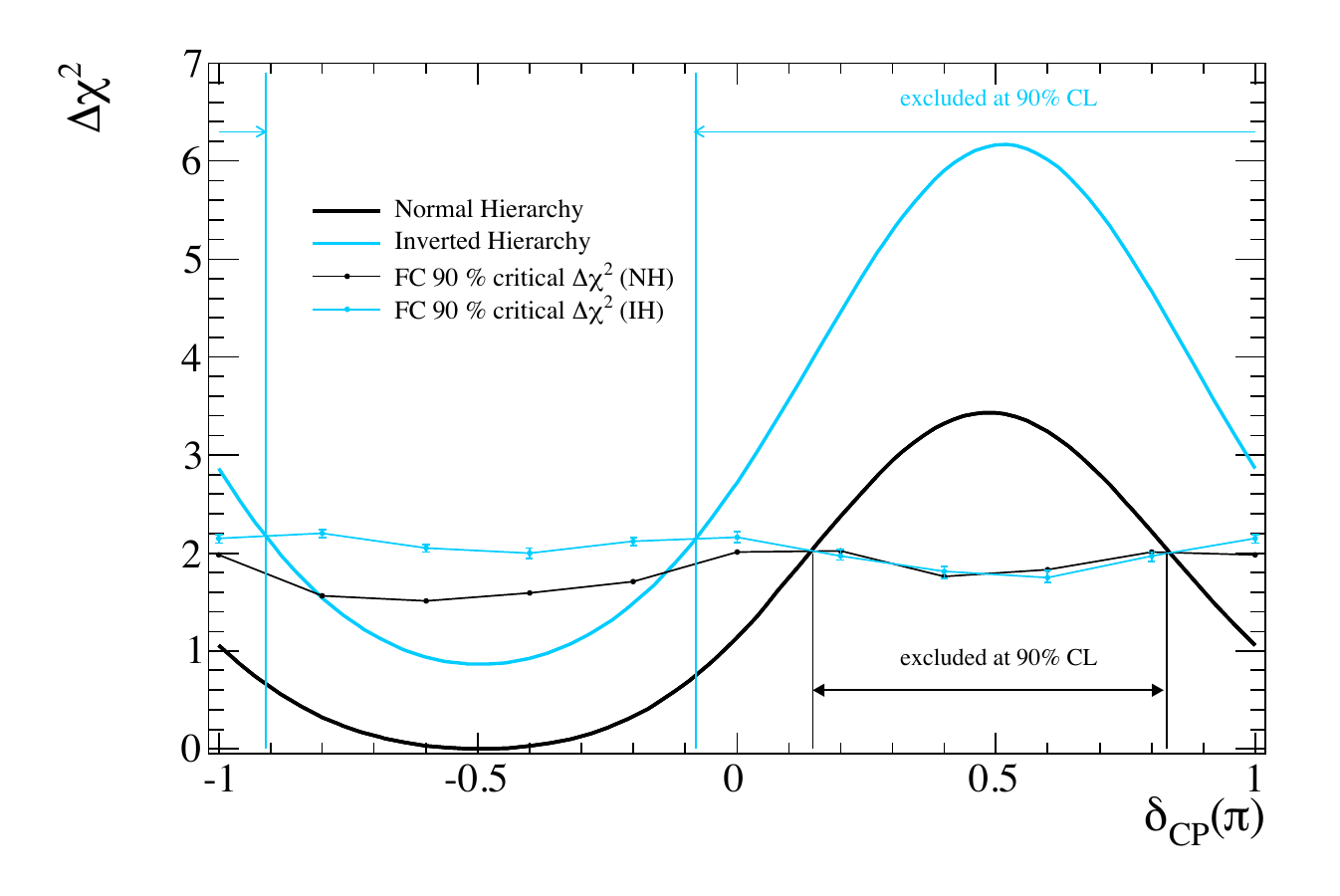}
\caption{(Left) The T2K credible intervals in the $\sin^2 \theta_{13}$ versus $\delta_{\mathrm{CP}}$ plane calculated in the MCMC method~\cite{Abe:2015awa}.
The 68~\% intervals are show by the dashed lines and the 90~\% intervals are by the solid lines.
These are constructed by marginalizing over both mass hierarchies. 
The red lines are extracted only from the T2K data set, and the black line and the point are extracted from both the T2K data set and the reactor $\theta_{13}$ constraint. 
(Right) The $-2 \Delta \ln L$  value as a function of $\delta_{\mathrm{CP}}$ in the MCMC method with both $\nu_e$  and $\nu_\mu$ samples and reactor $\theta_{13}$ constraint~\cite{Abe:2015awa}. The $\delta_{\mathrm{CP}}$ regions with values
above the 90~\% C.L. lines are excluded.} 
\label{fig:t2k:join2}
\end{center}
\end{figure}

With the MCMC method, the quantities $-2 \Delta \ln L(\equiv \ln L(\delta_{\mathrm{CP}}) - \ln L(\mbox{best fit values})) $ in the fit, as a function of $\delta_{\mathrm{CP}}$, are evaluated as shown in Figure~\ref{fig:t2k:join2}. The best fit value and the preferred regions at 90~\% C.L. are consistent with the result of the $\nu_e$ only sample shown in Figure~\ref{fig:t2k:nue3}.

\begin{figure}
\begin{center}
\includegraphics[width=9cm]{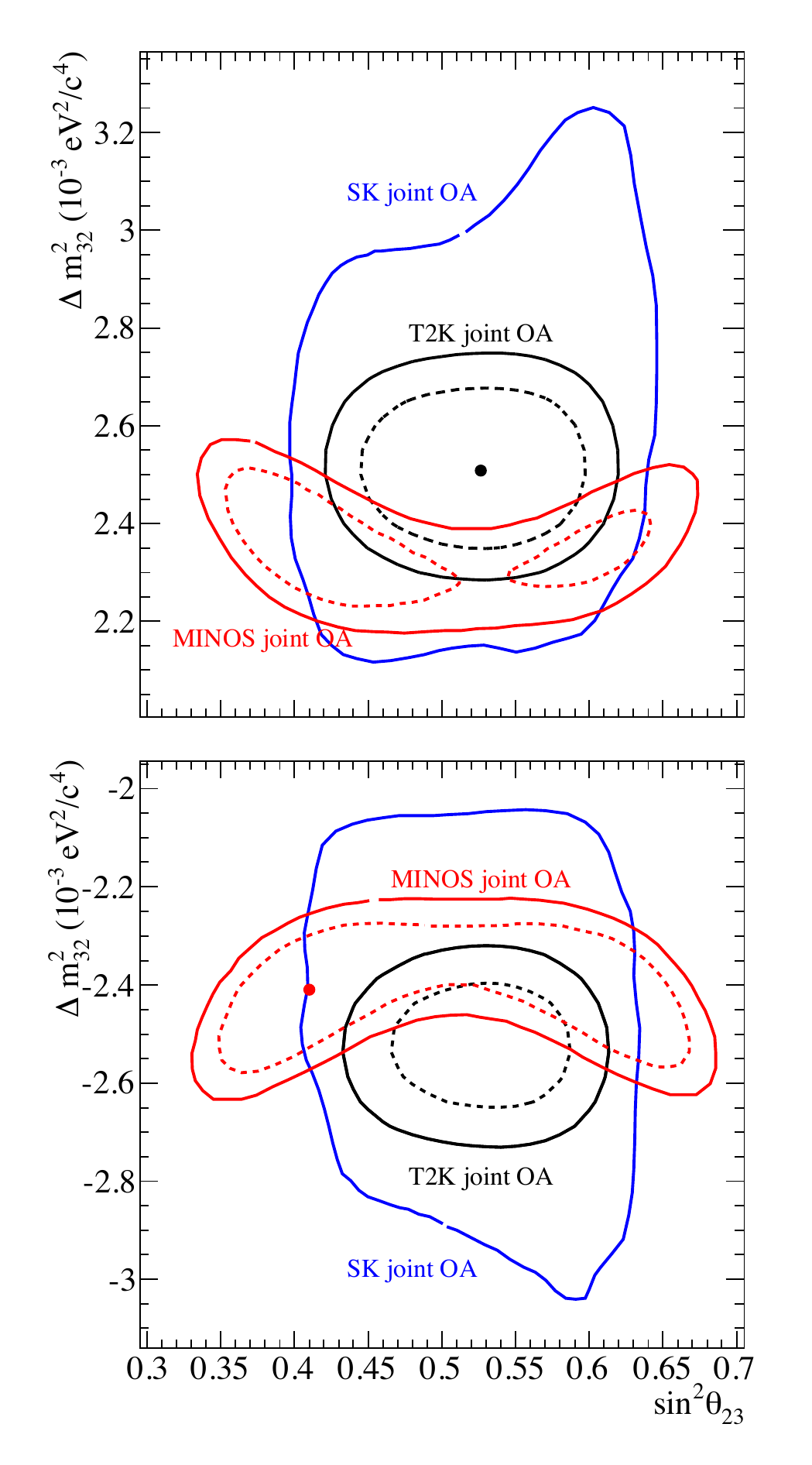}
\caption{T2K 68~\% (dashed) and 90~\% (solid) CL regions contours of oscillation parameters $\sin^2 \theta_{23}$ versus $\Delta m^2_{32}$ for normal (top) and inverted (bottom) mass hierarchy~\cite{Abe:2015awa}. The best fit point is shown as the black mark in the normal mass hierarchy. The Super-Kamiokande~\cite{Himmel:2013jva} and MINOS~\cite{minos_three_neutrino_prl} results are also shown for comparison.} 
\label{fig:t2k:join3}
\end{center}
\end{figure}

In Figure~\ref{fig:t2k:join3}, the credible intervals calculated in the MCMC method are shown in the $\sin^2 \theta_{23}$ versus $|\Delta m^2_{32}|$ plane, for both normal and inverted mass hierarchy cases. The results are compared with other measurements by Super-Kamiokande~\cite{Himmel:2013jva} and MINOS~\cite{minos_three_neutrino_prl}. The T2K best fit point is found to lie in the normal mass hierarchy as shown in Figure~\ref{fig:t2k:join3}.
%

\section{Systematic Uncertainties}

\subsection{T2K Beam} \label{sec:sys:beam}
In accelerator neutrino beam experiments, understanding of the properties of the neutrino beam is very important.
An experiment is usually designed to cancel first-order uncertainties of the neutrino beam by adopting the "two detectors" technique, in which one detector, located near the beam production point, is used to monitor the beam and the other, far detector, studies neutrino oscillations.
By normalizing the neutrino events with the near detector measurement, the systematic uncertainties of the beam and the neutrino cross sections are largely canceled. Even with the cancellation, for a precision measurement, understanding of the beam itself is essential.

\begin{figure}
\begin{center}
\includegraphics[width=76mm]{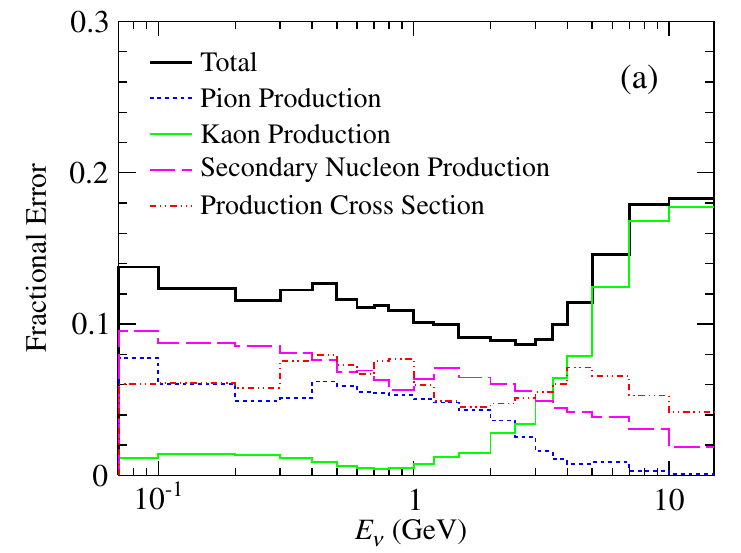}
\includegraphics[width=76mm]{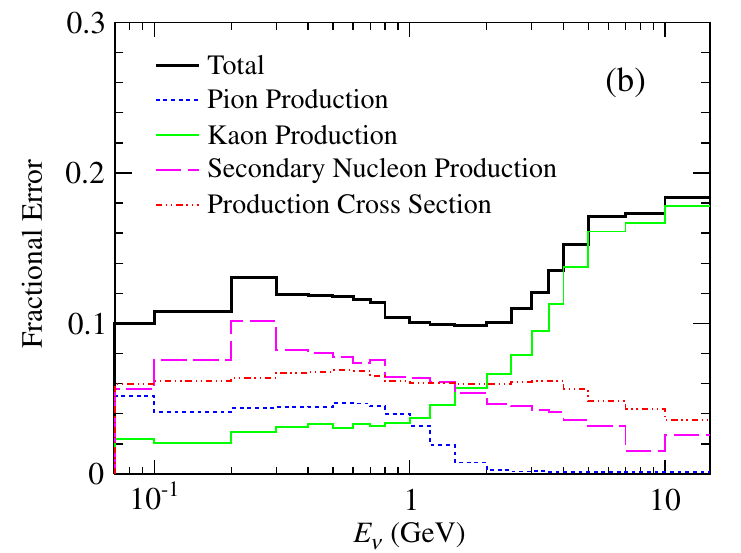}
\caption{Fractional uncertainty of the T2K neutrino beam flux: (left) muon neutrino component and (right) electron neutrino component~\cite{Abe:2013xua}.} \label{fig:t2k:beam sys}
\end{center}
\end{figure}

In the T2K experiment, the neutrino beam is simulated by incorporating real measurements of hadron production, among which large contributions come from the CERN NA61 experiment~\cite{Abgrall:2011ae, Abgrall:2011ts}. 
The uncertainties of hadron production decaying into neutrinos are directly related to the systematic uncertainty in the neutrino beam.
The uncertainty of the neutrino beam flux at the far detector in the T2K experiment is shown in Figure~\ref{fig:t2k:beam sys}.
The fractional uncertainty of the beam is at the $10 \sim 15$~\% level, of which the largest component is still the uncertainty in hadronic interactions.
The second largest component is due to the combined uncertainties of proton beam parameters, alignment of the beam line components and off-axis angle.
In T2K, the beam stability and the off-axis angle are directly monitored using the neutrino beam monitor: INGRID~\cite{Abe:2011xv}.
As shown in Section~\ref{sec:sys:xsec}, the relative beam flux uncertainty between the near and far detectors can be reduced to the 3~\% level including the uncertainty in cross sections constrained by the near detector.

\subsection{Constraints by the T2K near detector measurements} \label{sec:sys:xsec}
The uncertainty of neutrino cross sections is not small, especially in the $\sim$GeV energy region.
Although the uncertainty is typically at the $\sim$20~\% level, the first order uncertainties of the beam and cross sections can be canceled by adopting the two detector technique, as described in Section~\ref{sec:sys:beam}.
For this purpose, MINOS, MINOS+, NOvA and T2K have sophisticated near detectors which collect large amounts of neutrino data to measure the neutrino beam flux and neutrino cross sections with high precision.

\begin{figure}
\begin{center}
\includegraphics[width=52mm]{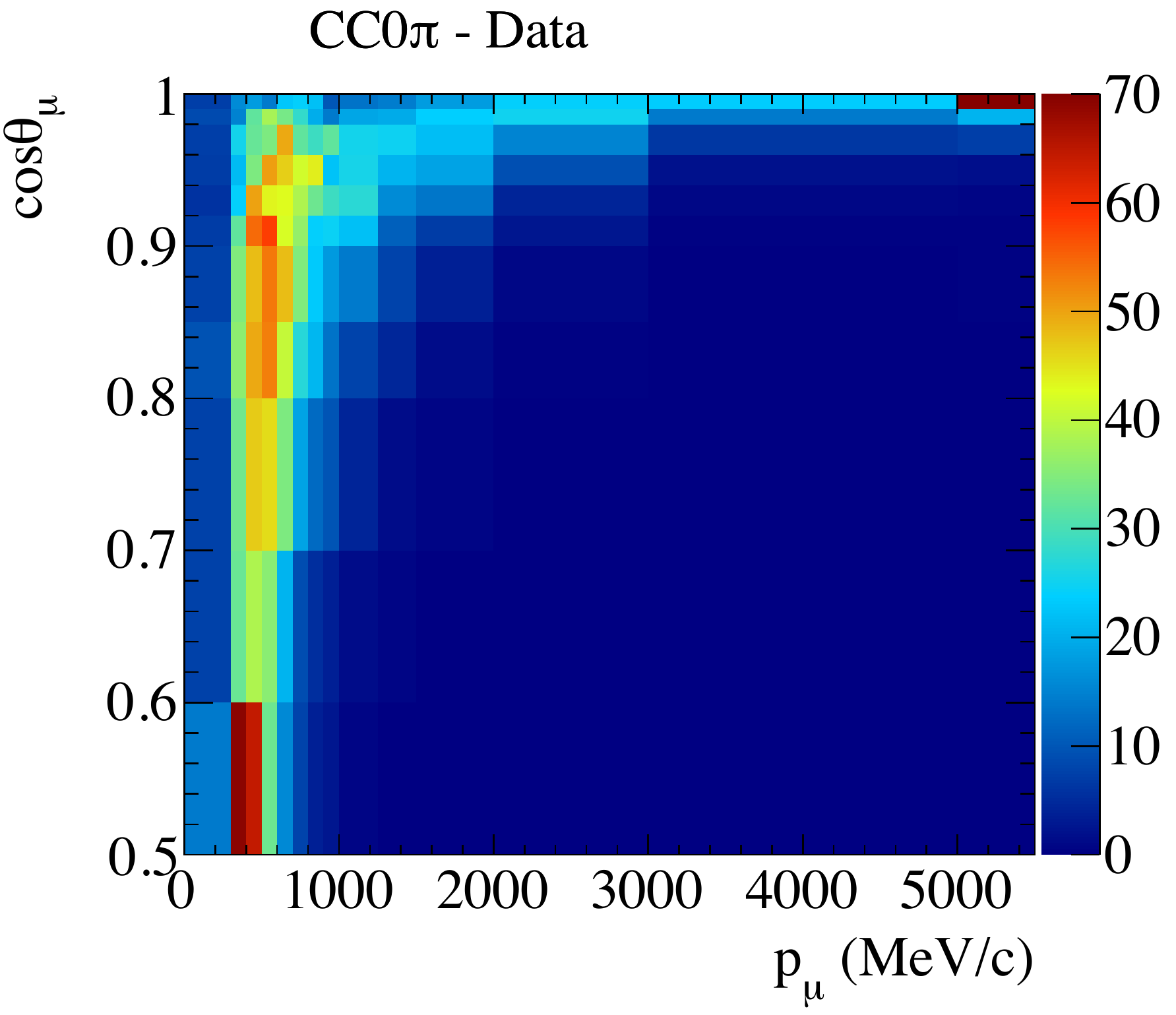}
\includegraphics[width=48mm]{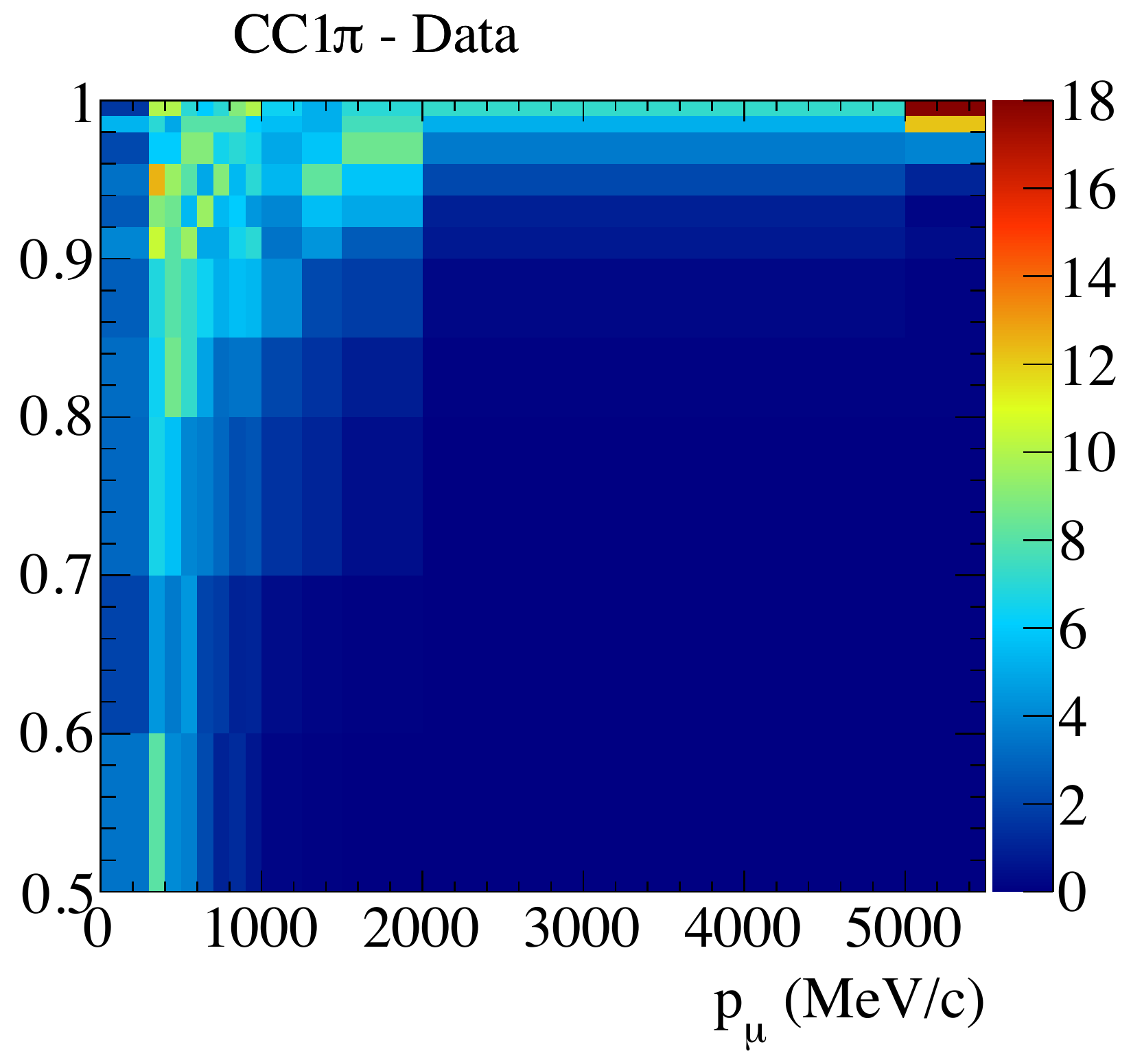}
\includegraphics[width=51mm]{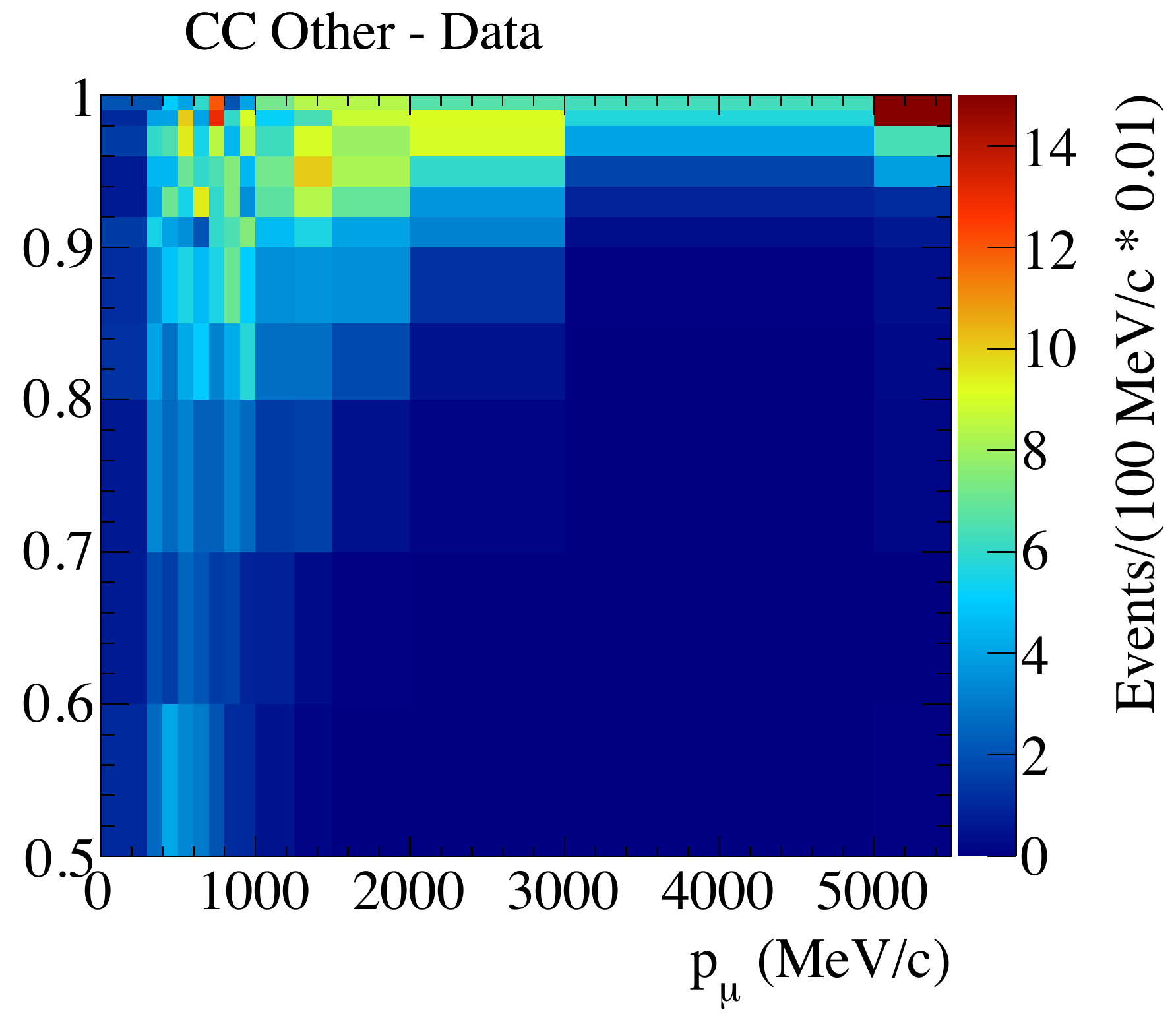}
\caption{The muon momentum versus the scattering angle for $\rm CC 0 \pi$ (left), $\rm CC 1 \pi$ (middle) and CC~Other (right) samples in T2K data} \label{fig:t2k:nd280-1}
\end{center} 
\end{figure}

In the T2K experiment, the near detector called ND280~\cite{Abe:2011ks} is located at 280~m from the beam production target, in the same direction as the far detector, at 2.5 degrees off-axis.  
ND280 consists of two Fine Grained Detectors (FGD), three Time Projection Chambers (TPC), an Electromagnetic Calorimeter system (ECAL), Side Muon Range Detectors (SMRD) and a $\pi^0$ detector (P0D).
Except for the SMRD, the detectors are located inside a dipole magnet of 0.2~T magnetic field.
For neutrino energies around 1~GeV, the dominant neutrino interaction is charged-current (CC) quasi-elastic (QE) scattering, and the second dominant one is CC 1 $\pi$ production. In the higher energy region, deep-inelastic scattering (DIS) becomes dominant. 
In ND280, the following three event categories have been measured: $\rm CC 0 \pi$, $\rm CC 1 \pi$ and CC~Other where the $\rm CC 0 \pi$ sample is for CC QE events, $\rm CC 1 \pi$ for CC 1 $\pi$ production and CC~Other for DIS.
The distributions of muon momentum and scattering angle relative to the neutrino beam are shown in Figure~\ref{fig:t2k:nd280-1} for data.
The neutrino interaction models and the neutrino beam flux are tuned to match the observed distributions in Figure~\ref{fig:t2k:nd280-1}.
After tuning, the uncertainties of the neutrino event rates are summarized in Table~\ref{tab:t2k:sys}. 
The uncertainties of $\pi$ hadronic interactions in the far detector and the detector systematic error in the far detector are also shown.
\begin{table}[th]
\begin{center}
\caption{Fractional uncertainties (\%) of the number of neutrino events in the T2K far detector ~\cite{Abe:2015awa}.
The uncertainties of cross sections are categorized into two parts: One is constrained by the ND280 measurement, and the other is independent of ND280. } \label{tab:t2k:sys}
\begin{tabular}{lcc}
\hline
\hline
Sources  &  $\nu_e$ candidates  & $\nu_\mu$ candidates \\
\hline
flux + cross sections (ND280 constrained) &  3.2  & 2.7 \\
cross sections (ND280 independent)         & 4.7   & 5.0 \\
$\pi$ interactions in the far detector                & 2.5   & 3.0 \\
Far detector systematic 	                          &  2.7  &  4.0 \\
\hline
Total                                                           & 6.8   &  7.7 \\
\hline
\hline
\end{tabular}
\end{center}
\end{table}

\begin{figure}
\begin{center}
\includegraphics[width=70mm]{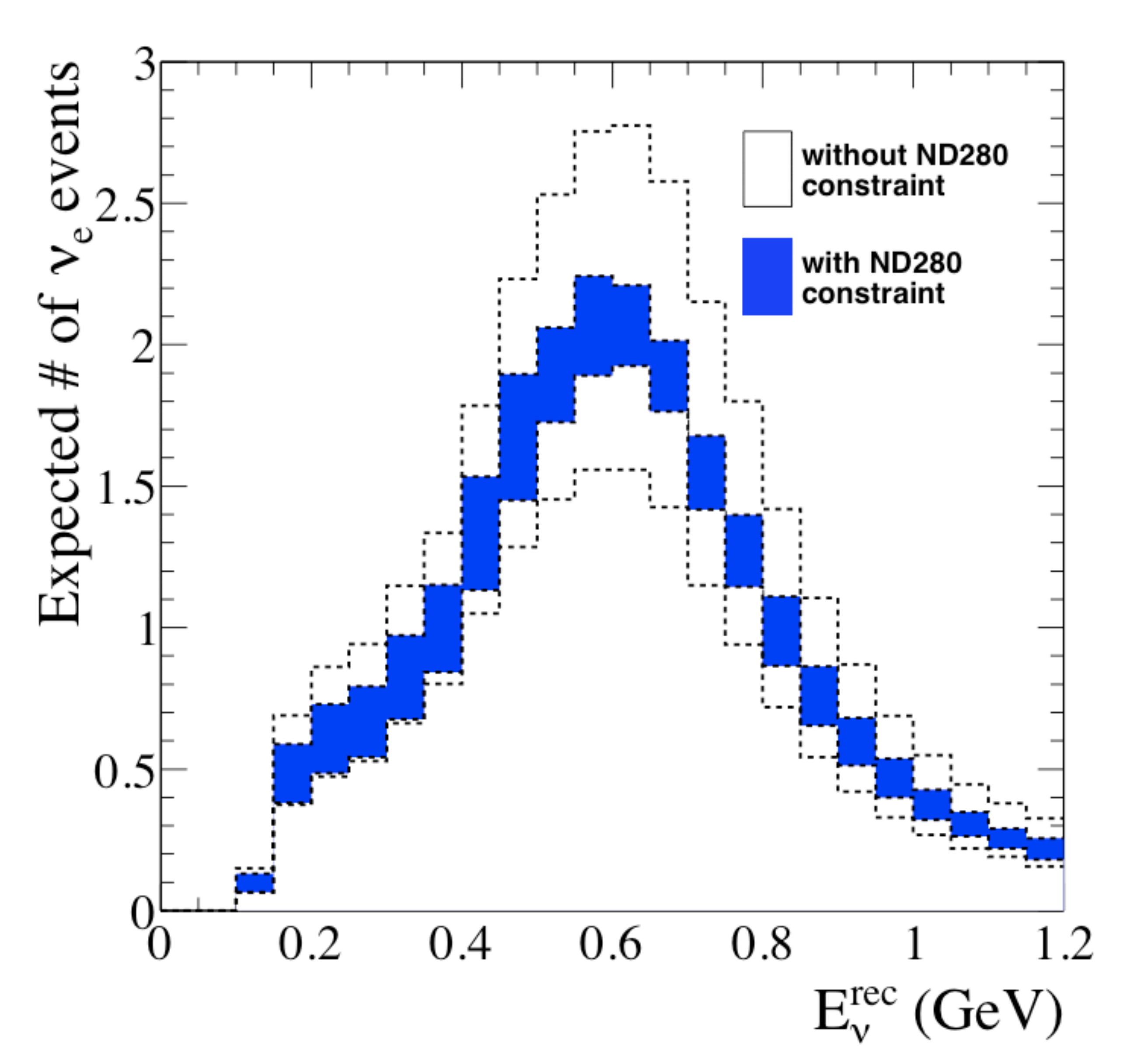}
\includegraphics[width=70mm]{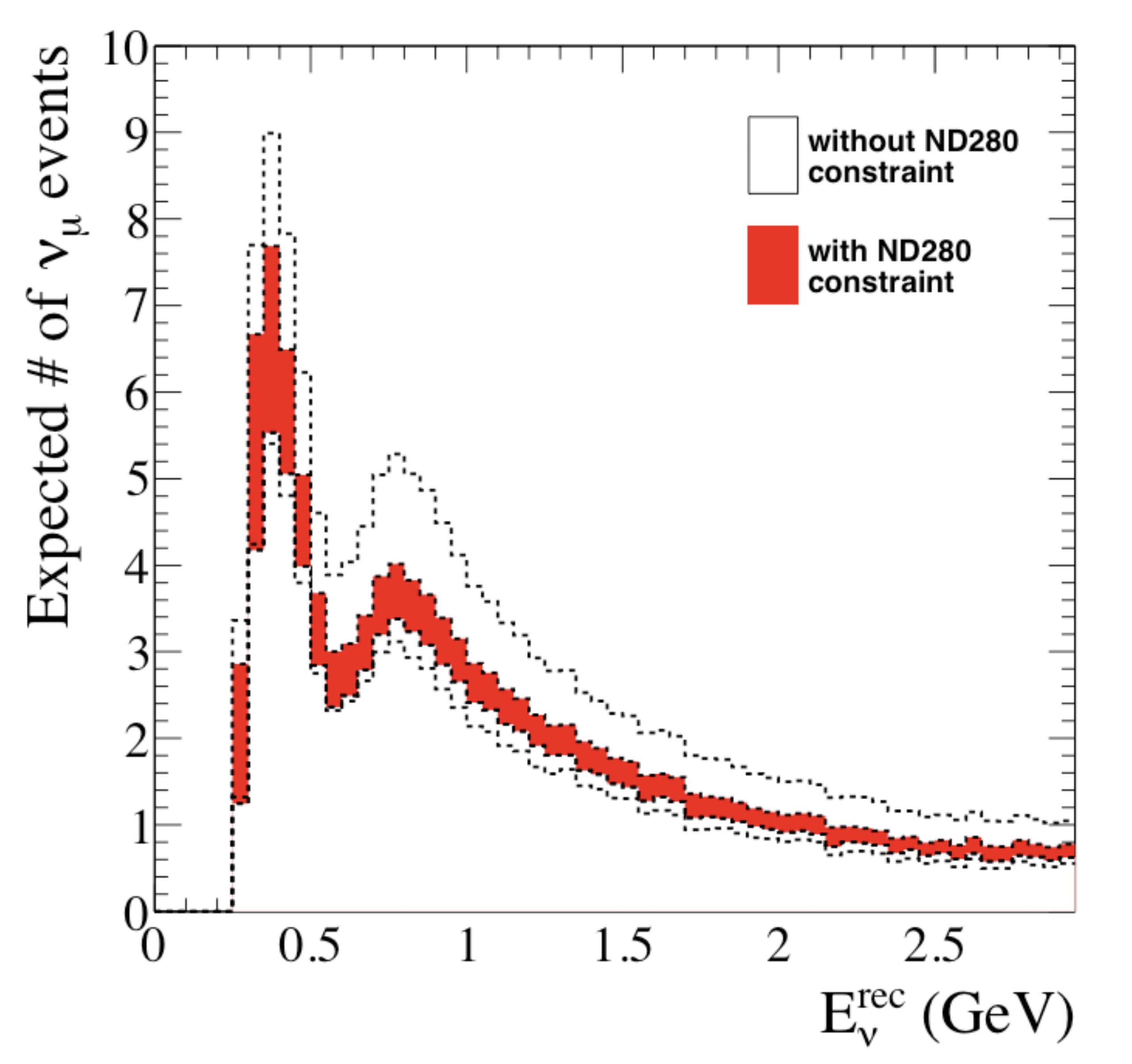}
\caption{The uncertainty of the neutrino events in the T2K far detector as a function of neutrino energy~\cite{Abe:2015awa}. (Left) $\nu_e$ candidate events and (right) $\nu_\mu$ candidate events.} \label{fig:t2k:sys}
\end{center} 
\end{figure}

In T2K, the number of observed $\nu_e$ ($\nu_\mu$) events is 28 (120) with a systematic uncertainty of 6.8 (7.7)~\%.
In Figure~\ref{fig:t2k:sys}, the uncertainties on the expected energy distributions of both $\nu_e$ and $\nu_\mu$ events are shown before and after
constraint by the ND280 measurement.
Today, the T2K sensitivity is not very limited by systematic errors which will also rapidly improve for the future CP violation measurement.
In the near future, T2K expects that the total systematic uncertainty can be reduced down to 5~\% or less.

\subsection{MINOS Systematics}
\label{subsec:MINOS_Sys_Section}

MINOS measurements have significant statistical error. For example, in the combined beam and atmospheric analysis a total of 3117 beam-generated contained-vertex charged current events, distributed across the entire neutrino energy spectrum, are used ~\cite{minos_2_flavor_prl}. The appearance analysis finds 172 events. Notwithstanding these small samples, MINOS has performed complete and detailed analysis of the systematics of the parameter measurements. The sensitivity of the results to systematics also benefits from the great similarity between the MINOS near and far detectors. Here we summarize the most important beam-related systematics and their effects.
\\

The MINOS 3-neutrino combined $\nu_\mu$ disappearance and $\nu_e$ appearance paper \cite{minos_three_neutrino_prl} describes the use of 32 systematic effects as nuisance parameters in the final fits. Of these 13 concern the atmospheric neutrinos; these are not discussed further in this section. There are 4 dominant systematic effects in $\nu_\mu$ disappearance in MINOS. These are, with representative values (from \cite{minos_CC_PRL106}):

\begin{enumerate}
\item Hadronic shower energy (7\% in the oscillation maximum region),
\item $\mu$ energy (2 -3\%, depending on technique),
\item Relative normailzation (1.6\%), and
 \item Residual neutral current (NC) contamination (20\%)
 \end{enumerate}
 
 The systematic knowledge of the muon energy includes measurements by range (in the MINOS steel, 2\%), and by momentum extracted from curvature in the MINOS magnetic field (3\%). The relative normalization error of 1.6\% is derived from knowledge of fiducial masses and relative reconstruction efficiencies \cite{minos_CC_PRL106}.

 Additional systematics are taken into account for $\nu_e$ appearance, affecting both signal and background predictions which are compared with the observed data. Many of the large number of systematic checks have small or negligible effects on the measured parameters; nevertheless they are incorporated in the final fitting procedures as discussed above.
 
 Illustrative values of the systematic effects are quoted in \cite{minos_nue_PRL110}, where the effect of the relevant uncertainties on the far detector background prediction for $\nu_e$ appearance in the 
 $\nu_\mu$ beam are given as:
 
 \begin{enumerate}
 
 \item{Energy scale:} This includes both relative energy scale differences between the near and far detectors and the absolute energy scale. The former affects the $\nu_e$ background prediction by directly impacting the extrapolation from data, and is the most important single systematic (2\%). The latter enters via its effects on the event selection process, and is less important. The combined effect on the background is 2.7\%
 \item{Normalization:} This term refers to effects relating to the relative fiducial masses and exposures of the two detectors. It is quoted as 1.9\%
 \item{$\nu_\tau$ cross section:} A poorly known pseudo-scalar form factor \cite{minos_nutau_cross_section} causes a background uncertainty of 1.7\%.
 \item{All others:} Small effects due to, for example, neutrino fluxes and cross-sections (which largely cancel due to the functional identity of the near and far detectors) and hadronic shower modeling in neutrino interactions.
 \cite{minos_nue_thesis}.
 The sum of these small effects is $<1\%$, showing once more the effectiveness of extrapolating from near detector data.
 \end{enumerate}
 
 The final uncertainty on the $\nu_e$ background is ~4\%, to be compared with its statistical uncertainty of 8.8\%. The numbers cited here apply to the $\nu_\mu$ beam mode only, with similar, but slightly higher values for the $\bar{\nu}_\mu$ beam mode. In addition, there is a systematic error of approximately 5\% on the appearance signal selection efficiency, studied with CC events in which the muon data has been replaced with a simulated electron.

\section{Additional Measurements}

\subsection{Additional Measurements in MINOS}
\label{subsec:MINOS_Additional_Section}

In addition to the primary mission of MINOS and MINOS+, which is the understanding of the three-neutrino oscillation sector, the MINOS detectors and NuMI beam line are capable of a wide variety of additional measurements which enrich our understanding of the physics of neutrinos, and other areas. Measurements published by the MINOS collaboration include:

\begin{enumerate}

\item Searches for additional sterile neutrinos using charged and neutral current interactions \cite{minos_old_nc} \cite{minos_nc},
\item Measurement of neutrino cross-sections \cite{minos_cross_sections},
\item Tests of fundamental symmetries and searches for non-standard interactions. \cite{minos_lv_etc},
\item Studies of cosmic rays at both near and far detectors ~\cite{minos_cosmic}.

\end{enumerate}

In this section we discuss briefly the first item, searches for sterile neutrinos, and present an example of a fundamental symmetry test.

 Anomalies seen in short baseline experiments \cite{miniboone_contours} and others have generated great interest in the possibility of a fourth neutrino which would not have Standard Model interactions. Neutral currents in the MINOS detectors are visible as hadronic showers without an accompanying lepton. All active flavors of neutrinos produce neutral currents equivalently, so that the three-neutrino oscillation phenomenon should not cause any depletion with respect to expectations in the observed far detector spectrum. This is indeed seen to be the case, as documented in \cite{minos_nc} which measures a limit on the fraction of neutrinos which can have oscillated to sterile neutrinos, $f_s$, of $f_s <$ 22\%. More recently, preliminary analysis of further data has generated limits on the sterile mixing angle $\theta_{24}$ which extend the range of previous experiments ~\cite{minos_neutrino_2014}.

\begin{figure}
\begin{center}
\includegraphics[scale=0.35]{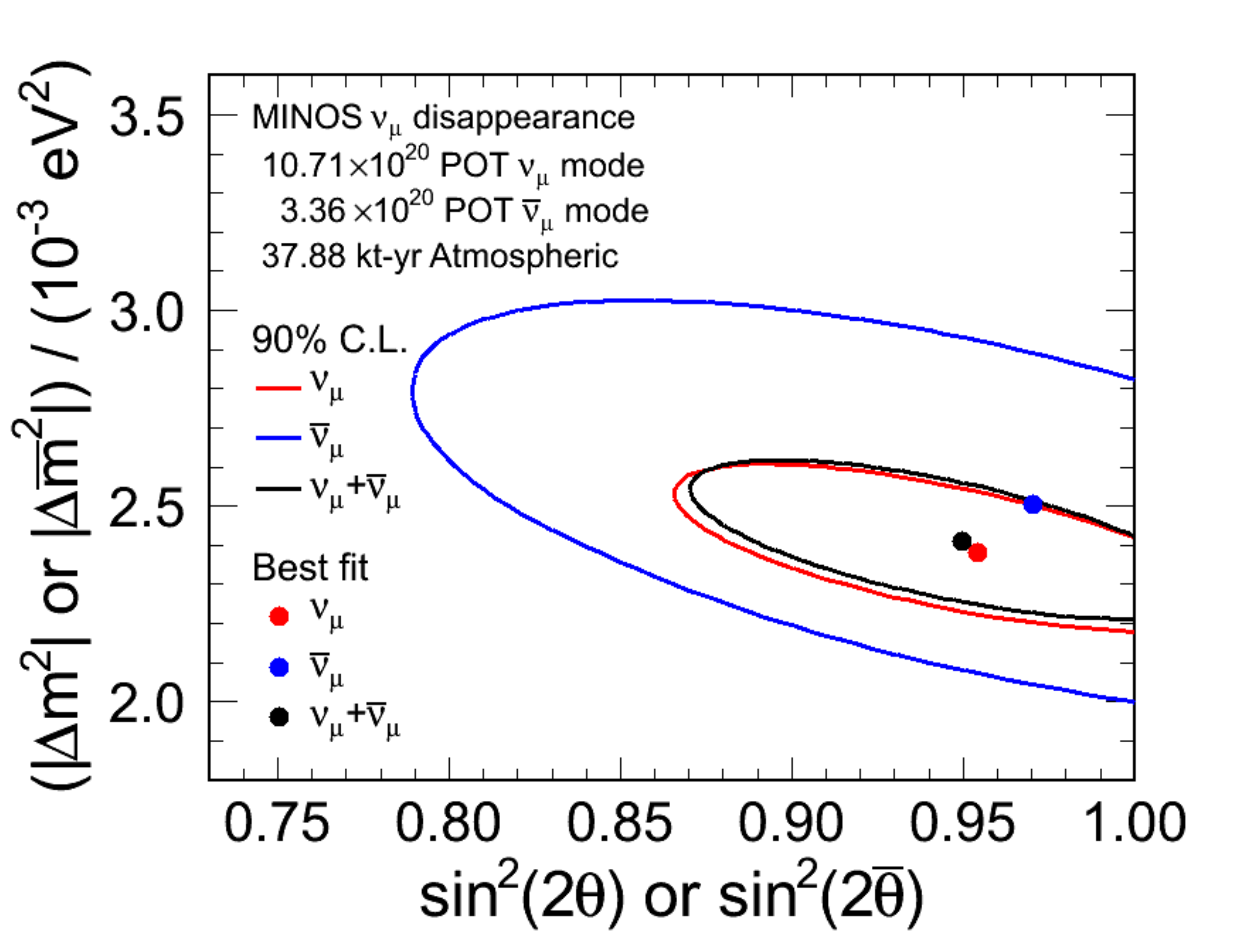}
\caption{MINOS comparison of $\nu_\mu$ and $\bar{\nu_\mu}$ oscillations from \cite{minos_2_flavor_prl}. The 90\% confidence level allowed regions and best fit values are shown for $\nu_\mu$ and $\bar{\nu_\mu}$ oscillations, and for a fit in which the parameters are assumed to be identical.}
\label{fig:MINOS_4param}
\end{center}
\end{figure}

A fundamental test of CPT symmetry is the equivalence of oscillation parameters obtained from $\nu_\mu$ and $\bar{\nu_\mu}$. The magnetized MINOS detector can perform an event-by-event comparison of these parameters enabling an accurate test of this prediction. The resulting allowed regions (from \cite{minos_2_flavor_prl}) are displayed in figure \ref{fig:MINOS_4param}. The difference in $|\Delta m^2|$ obtained, in a two-flavor model, is $| \Delta \bar{m}^2 | - |\Delta m^2| = (0.12^{+0.24}_{-0.26}) \times{10^{-3}} eV^2$.

\subsection{Additional Measurements in T2K}
In addition to neutrino oscillation studies, T2K conducts various measurements on neutrino-nucleus cross sections.
As described in Section~\ref{sec:sys:xsec}, the understanding of neutrino cross sections is important to reduce systematic uncertainties of neutrino oscillation measurements, which could improve the sensitivity to neutrino oscillations.

The cross sections measured in T2K are summarized in Table~\ref{tab:t2k:xsec}.
As the first step, T2K measures the muon neutrino charged current (CC) inclusive cross sections with the T2K off-axis near detector (ND280)~\cite{Abe:2013jth} and the on-axis near detector (INGRID)~\cite{Abe:2014nox}. In future, these analyses will be more sophisticated to measure exclusive channels, such as CC-QE, CC 1$\pi$ production, and CC-coherent $\pi$, as energy dependent differential cross sections.
With INGRID, there are two types of neutrino detector with different target materials. One has an iron target, and the other has a plastic (CH) target.
By using two targets, the CC inclusive cross sections on iron and plastic, and the ratio of cross sections are measured.
The CC-QE cross sections are also measured with the CH target~\cite{Abe:2015oar}.
With ND280, the electron neutrino CC inclusive cross sections can also be measured~\cite{Abe:2014agb, Abe:2014usb}, using the powerful particle identification performance of TPC and ECAL. In the analysis~\cite{Abe:2014usb}, the electron neutrino contamination in the beam is measured relative to the prediction in the simulation. T2K also has divided the measurement into two contributions: One is the electron neutrino from kaon decay and the other is from muon decay.
The neutral current (NC) gamma production cross section in neutrino-oxygen interaction has also been measured~\cite{Abe:2014usb} by using the far detector, Super-Kamiokande.
All the results are consistent with the predictions in the neutrino interaction generator libraries NEUT~\cite{Hayato:2009zz} and GENIE~\cite{Andreopoulos:2009rq}. Finally, with ND280, the electron neutrino disappearance sample was searched to investigate neutrino oscillations to sterile neutrinos~\cite{Abe:2014nuo}.
\begin{table}[htbp]
\begin{center}
\caption{Neutrino Cross Sections measurements in T2K for charged current (CC) and neutral current (NC) inclusive processes. 
The measurements of cross sections are given per nucleon. The ratio of cross sections is also shown in some measurements. } \label{tab:t2k:xsec}
\begin{footnotesize}
\begin{tabular}{lccc}
\hline
\hline
{\normalsize Mode}  & {\normalsize Results ($\frac{cm^2}{\rm nucleon}$ or the ratio)} & {\normalsize $<E_\nu>$~(GeV)} & {\normalsize Reference} \\
\hline
$\nu_\mu$ CC inclusive & $ (6.91 \pm 0.13 (stat) \pm 0.84 (syst)) \times 10^{-39} $ & 0.85 & \cite{Abe:2013jth} \\
$\nu_e$ CC inclusive & $ (1.11 \pm 0.09 (stat) \pm 0.18 (syst)) \times 10^{-38} $ & $\sim$1.3 & \cite{Abe:2014agb} \\
$\nu_\mu$ CC inclusive on $Fe$& $ (1.444 \pm 0.002 (stat) ^{+0.189}_{-0.157} (syst)) \times 10^{-38} $ & 1.51 & \cite{Abe:2014nox} \\
$\nu_\mu$ CC inclusive on $CH$& $ (1.379 \pm 0.009 (stat) ^{+0.178}_{-0.147} (syst)) \times 10^{-38} $ & 1.51  & \cite{Abe:2014nox} \\
$\nu_\mu$ CC ratio of $Fe/CH$ & $  1.047 \pm 0.007 (stat) \pm 0.035 (syst) $ & 1.51  & \cite{Abe:2014nox} \\
$\nu_\mu$ CC-QE on $CH$ & $  (10.64 \pm 0.37 (stat) ^{+2.03}_{-1.65} (syst)) \times 10^{-39} $ & 0.93  & \cite{Abe:2015oar} \\
$\nu_\mu$ CC-QE on $CH$ & $  (11.95 \pm 0.19 (stat) ^{+1.82}_{-1.47} (syst)) \times 10^{-39} $ & 1.94  & \cite{Abe:2015oar} \\
NC $\gamma$ & $  (1.55 ^{+0.65}_{-0.33}) \times 10^{-38} $ & 0.63  & \cite{Abe:2014dyd} \\
$\nu_e$ CC ratio  & $ \sigma_{\nu_e CC}/\sigma_{\rm prediction} = 1.01 \pm 0.10 $ & $\sim$1.3 & \cite{Abe:2014usb} \\
$\nu_e$ (K decay) CC ratio  & $ \sigma_{\nu_e CC}/\sigma_{\rm prediction} = 0.68 \pm 0.30 $ &  ---  & \cite{Abe:2014usb} \\
$\nu_e$ ($\mu$ decay) CC ratio  & $ \sigma_{ \nu_e CC}/\sigma_{\rm prediction} = 1.10 \pm 0.14 $ & --- & \cite{Abe:2014usb} \\
\hline
\hline
\end{tabular}
\end{footnotesize}
\end{center}
\end{table}
%

\section{The Frontier and future sensitivity} \label{sec:future}

\subsection{NOvA}
\label{subsec:NOVA_Section}

The NOvA experiment is the principal appearance-mode long-baseline experiment at Fermilab. It makes use of the off-axis NuMI beam as discussed in previous sections. The far and near detectors minimize passive mass, moving toward the ideal of a totally active fiducial volume which will be realized with future detectors, such as e.g. DUNE liquid argon TPC's. A totally active detector typically has better performance because all charged particles in neutrino interactions are reconstructed properly with good efficiency. In the case of NOvA the active medium is liquid scintillator mixed into oil. The basic segmentation of the detector is planes of 3.6 cm x 5.6 cm PVC tubes, separated by thin walls.  Light is collected in each tube by a looped wavelength shifting fiber that is directed onto a single pixel of an avalanche photodiode (APD). The APD's are cooled by a hybrid thermoelectric/water system to a temperature of $-15^{\circ}$ C.

\begin{figure}[htbp]
\begin{center}
\includegraphics[scale=0.6]{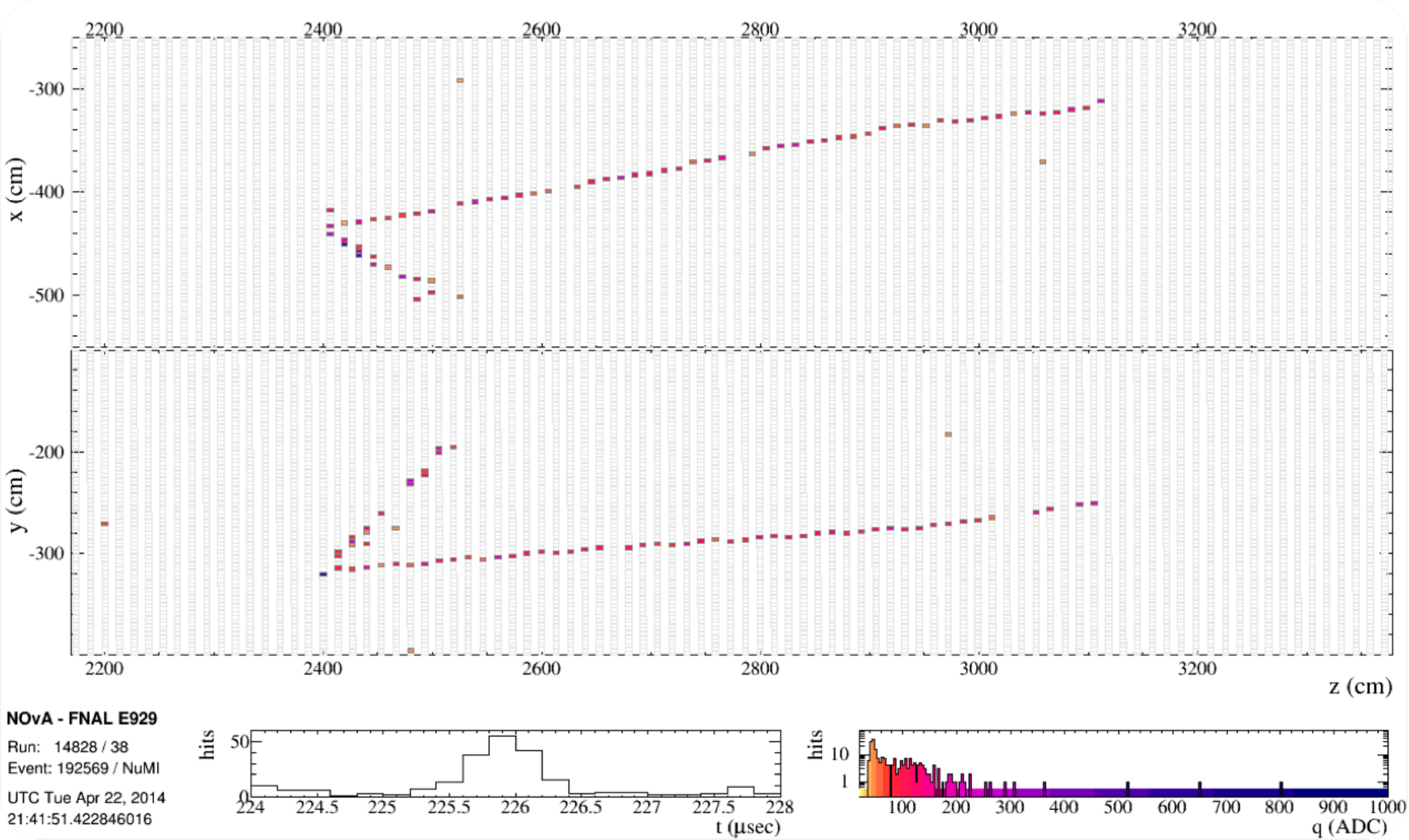}
\label{fig:NOVA_event}
\caption{Charged-current neutrino interaction observed in the NOvA Far Detector (Courtesy NOvA Collaboration).}
\end{center}
\end{figure}

There are a total of 344,064 tubes arranged in an interchanging pattern of horizontal and vertical planes, giving an active structure 15.6 m x 15.6 m, and 60 m long. The active scintillator mass is 8.7 ktonnes with 5.3 ktonnes devoted to the PVC support structure, fiber readout, and other structural components, giving a total far detector mass of 14.0 ktonnes. Figure~\ref{fig:NOVA_event} shows a typical NOvA far detector event.\
\\
In addition to the far detector, a functionally equivalent near detector is located at Fermilab. This detector consists of horizontal planes 4.0 m x 4.0 m, together with a steel muon catcher to assist in measurement of the spectrum of CC events. The expected average occupancy of the near detector is $\approx 6$ events per spill at full intensity, which will be separated by their time of occurrence, as is already successfully done in MINOS and MINERvA. 

\begin{figure}
\begin{center}
\includegraphics[scale=0.5]{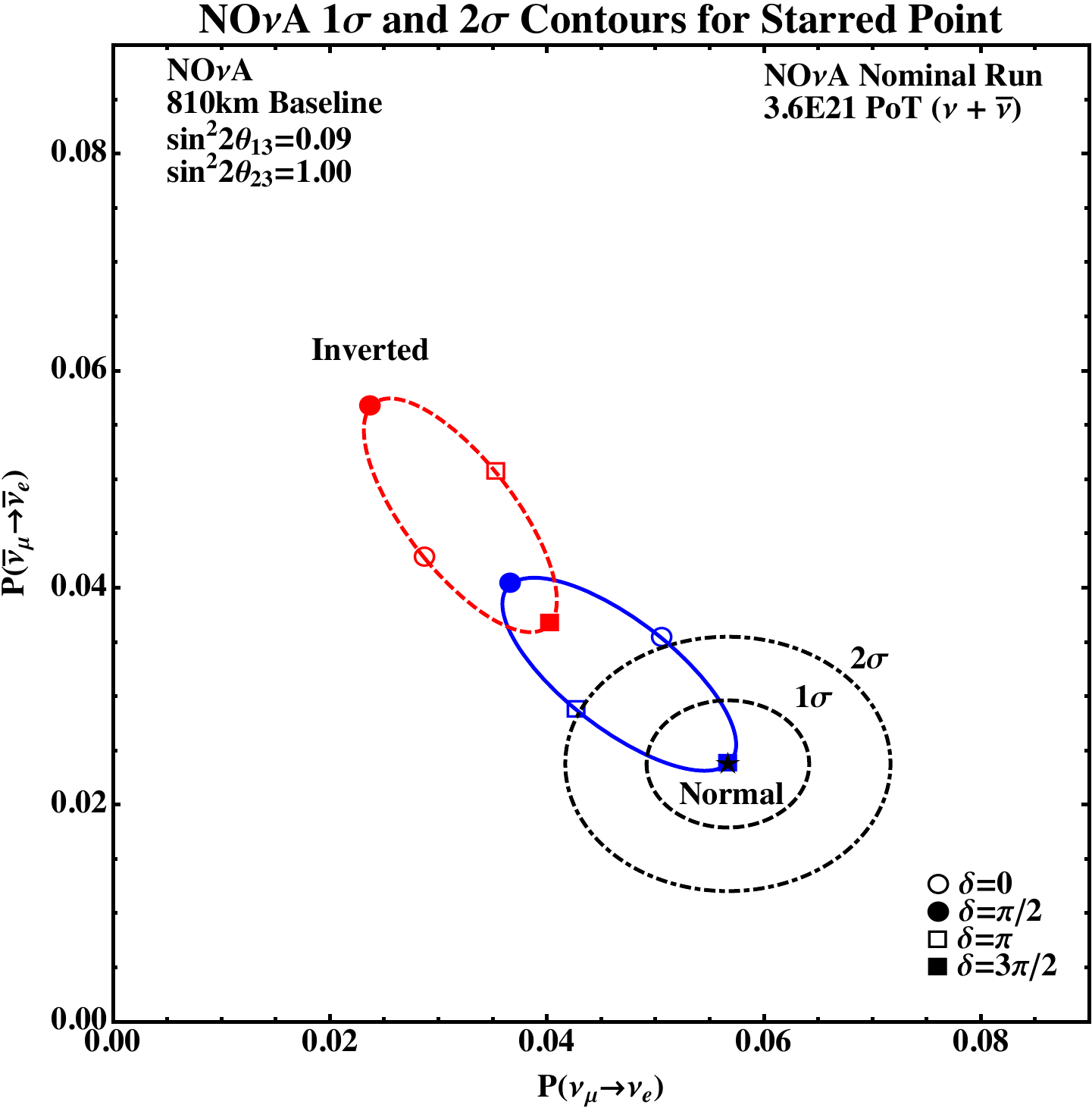}
\includegraphics[scale=0.5]{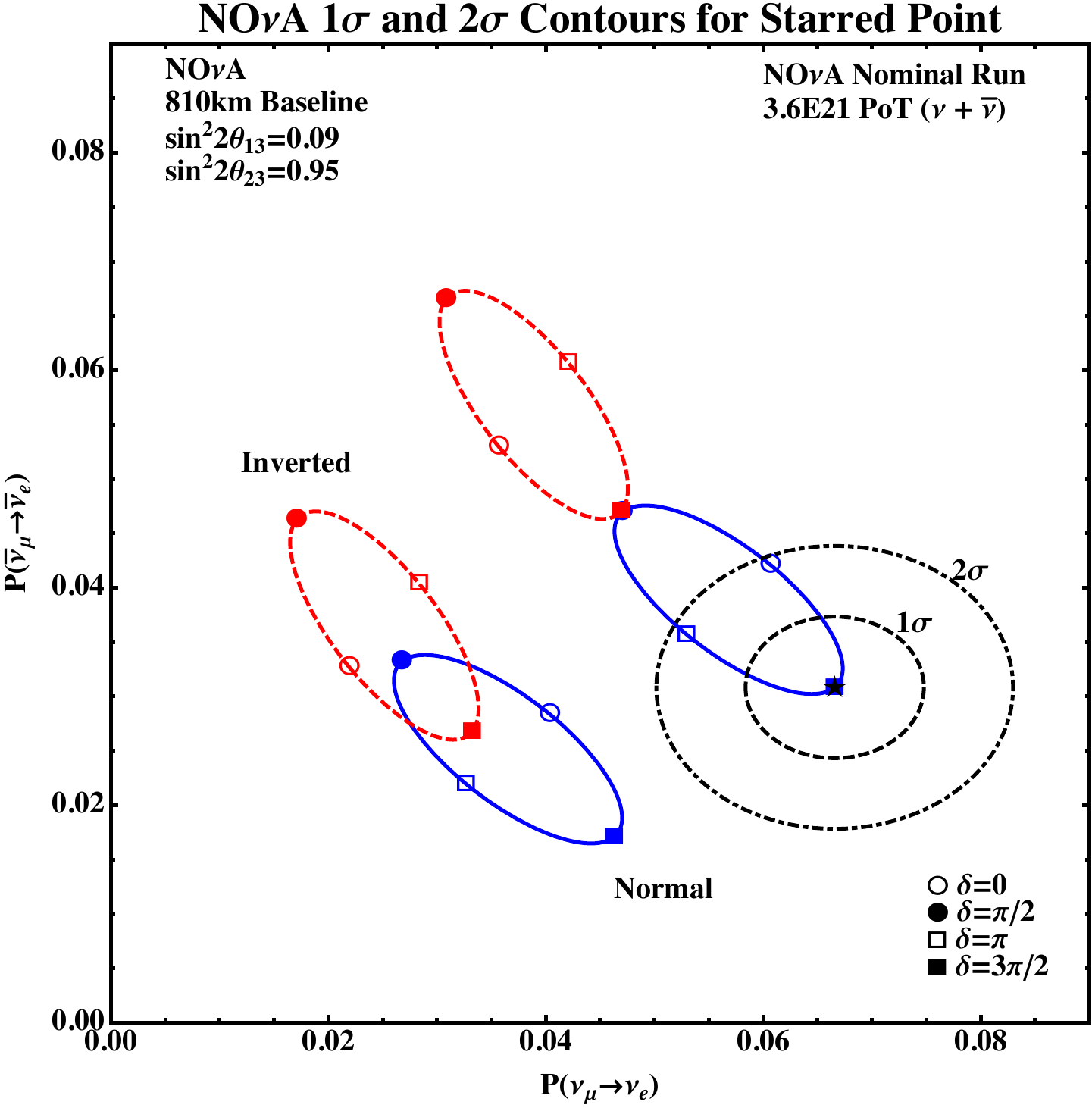}
\caption{Biprobability plot showing the effect of mass hierarchy, $\delta_{CP}$, and $\nu_\mu$ vs $\bar\nu_\mu$ exposure. The left panel shows the expected appearance probabilities for $\nu_e$ ($\bar\nu_e$) when the mixing angle $\theta_{23}=\pi/4$ (maximal mixing.) The right panel shows the same for assumed non-maximal mixing with $\theta_{23}<\pi/4$ and $\theta_{23}>\pi/4$. (Courtesy NOvA Collaboration).}
\label{fig:NOVA_biprob}
\end{center}
\end{figure}

\begin{figure}
\begin{center}
\includegraphics[scale=0.35]{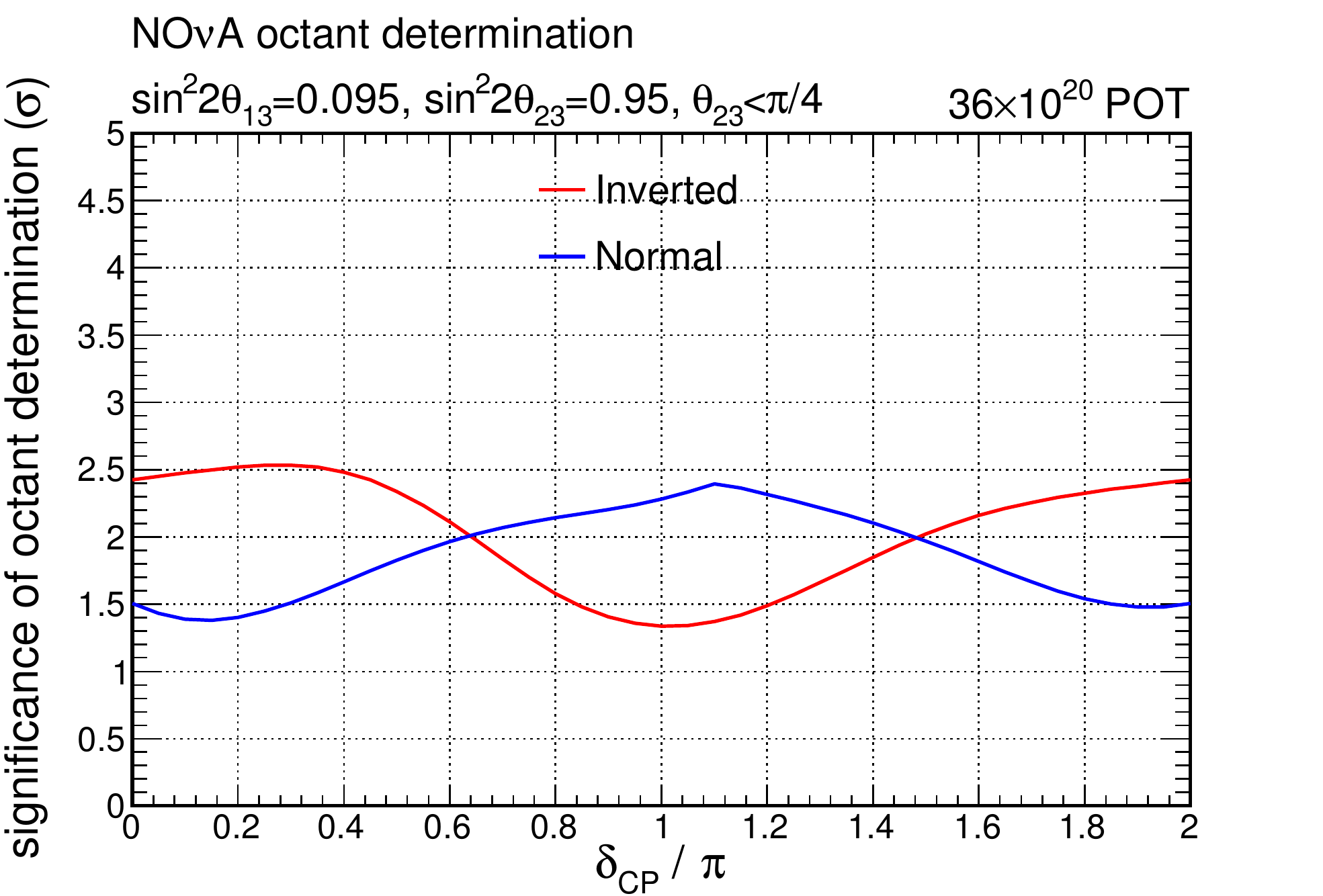}
\includegraphics[scale=0.35]{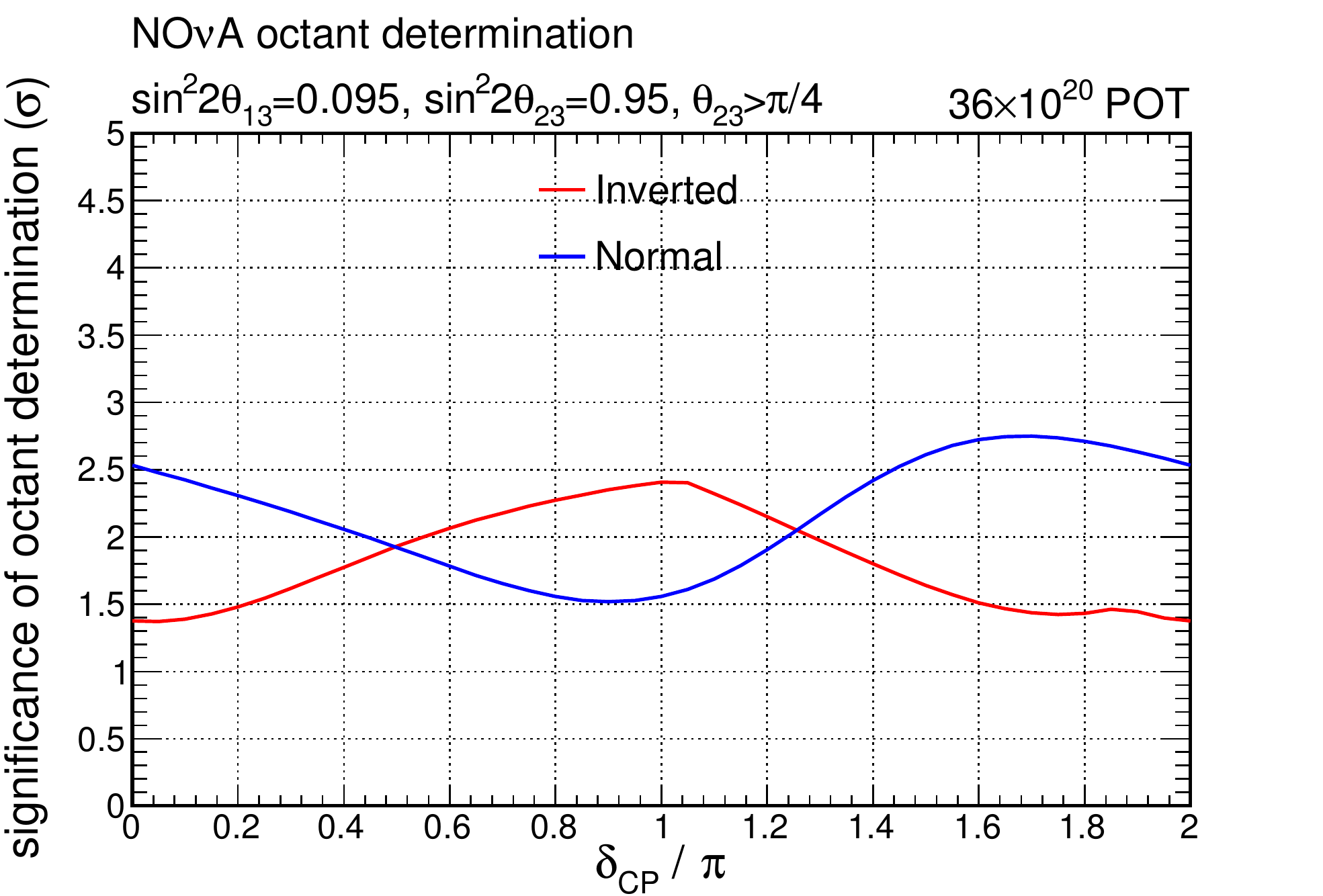}
\caption{NOvA sensitivities for resolution of the octant of non-maximal $\theta_{23}$, assumed to be at $\sin^2 2\theta_{23}=0.95$. Left (right) panel shows the sensitivity for $\theta_{23}<\pi/4$ ($\theta_{23}>\pi/4$). (Courtesy NOvA Collaboration).}
\label{fig:NOVA_octant}
\end{center}
\end{figure}

The relatively large value of $\theta_{13}$ provides the opportunity for a rich harvest of physics results for NOvA via both disappearance and appearance, with both neutrino and antineutrino beams. Here we focus on the appearance channel, with its rich information about $\theta_{13}$, mass hierarchy, and $\delta_{CP}$. Disappearance measurements are also highly sensitive and give excellent information on the octant of $\theta_{23}$. During early running, combination with MINOS+ will be exploited, as discussed separately below.
\\
As discussed in earlier sections, the measurements of $\nu_e$ appearance of NOvA are significantly affected by matter effects, potential CP violation, mass hierarchy, and the distinctions between neutrinos and anti-neutrinos. A useful tool for understanding these dependencies is the biprobability plot, in which one axis displays the appearance probability for $\nu_e$ and the other the probability for $\bar{\nu_e}$. Figure~\ref{fig:NOVA_biprob} shows this situation graphically, for the case of maximal mixing. The results are clearly separated for many values of $\delta_{CP}$, with areas of overlap for the regions around $\delta_{CP}=\pi/2$  ($\delta_{CP}=3\pi/2$) for normal (inverted) mass hierarchy. Moreover, the subleading effects cause a separation in the appearance probabilities depending on the octant of $\theta_{23}$. 
The sensitivity of NOvA to the octant depends on the value of $\theta_{23}$, and somewhat on both the hierarchy and the value of $\delta_{CP}$. For a value of $\sin^2 2\theta_{23}$ of 0.95, there is considerable sensitivity for all values of those parameters, exceeding 95\% CL for significant portions of the parameter space with the nominal exposure, shown in Figure~\ref{fig:NOVA_octant}. \\
 
\begin{figure}
\begin{center}
\includegraphics[scale=0.35]{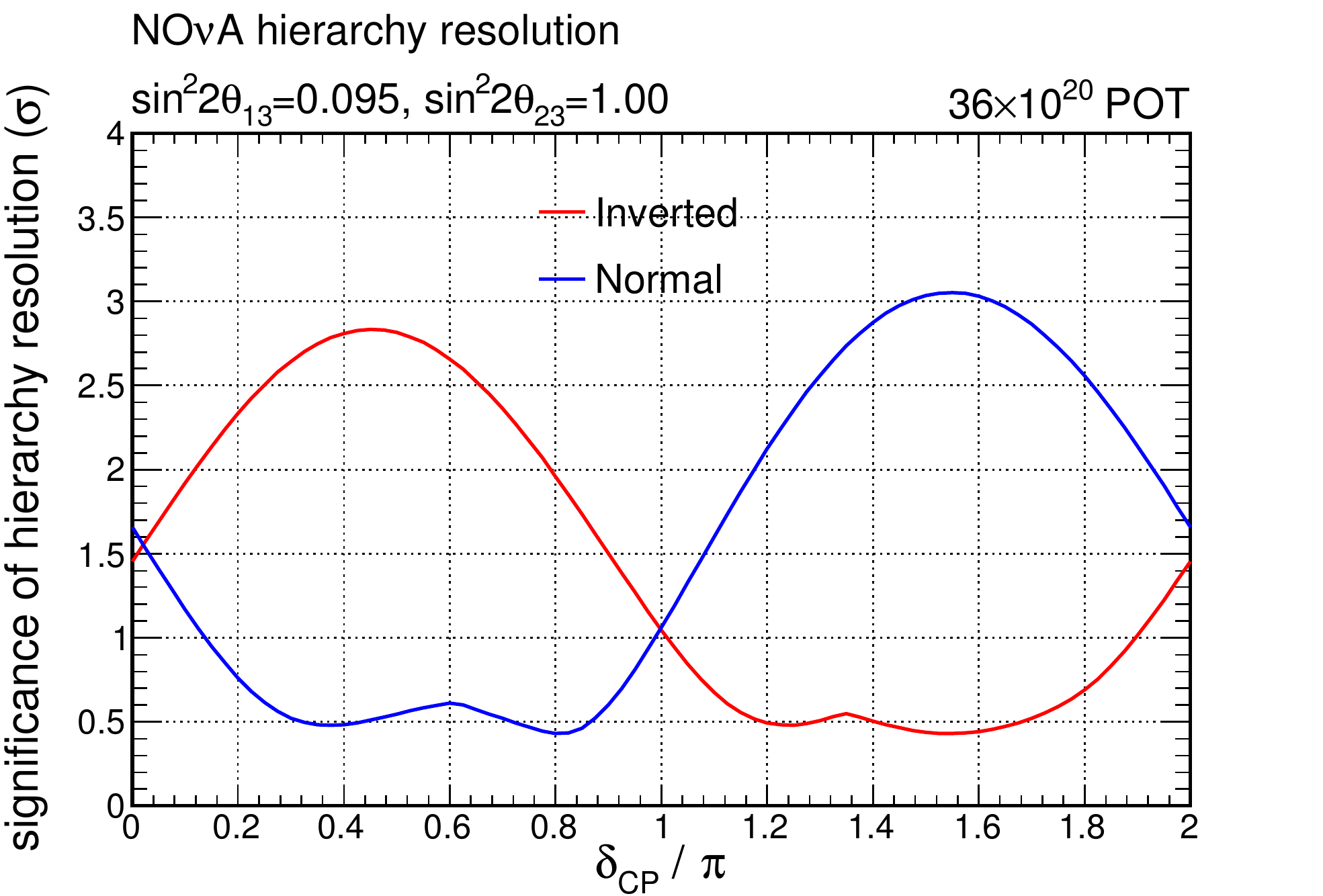}
\includegraphics[scale=0.35]{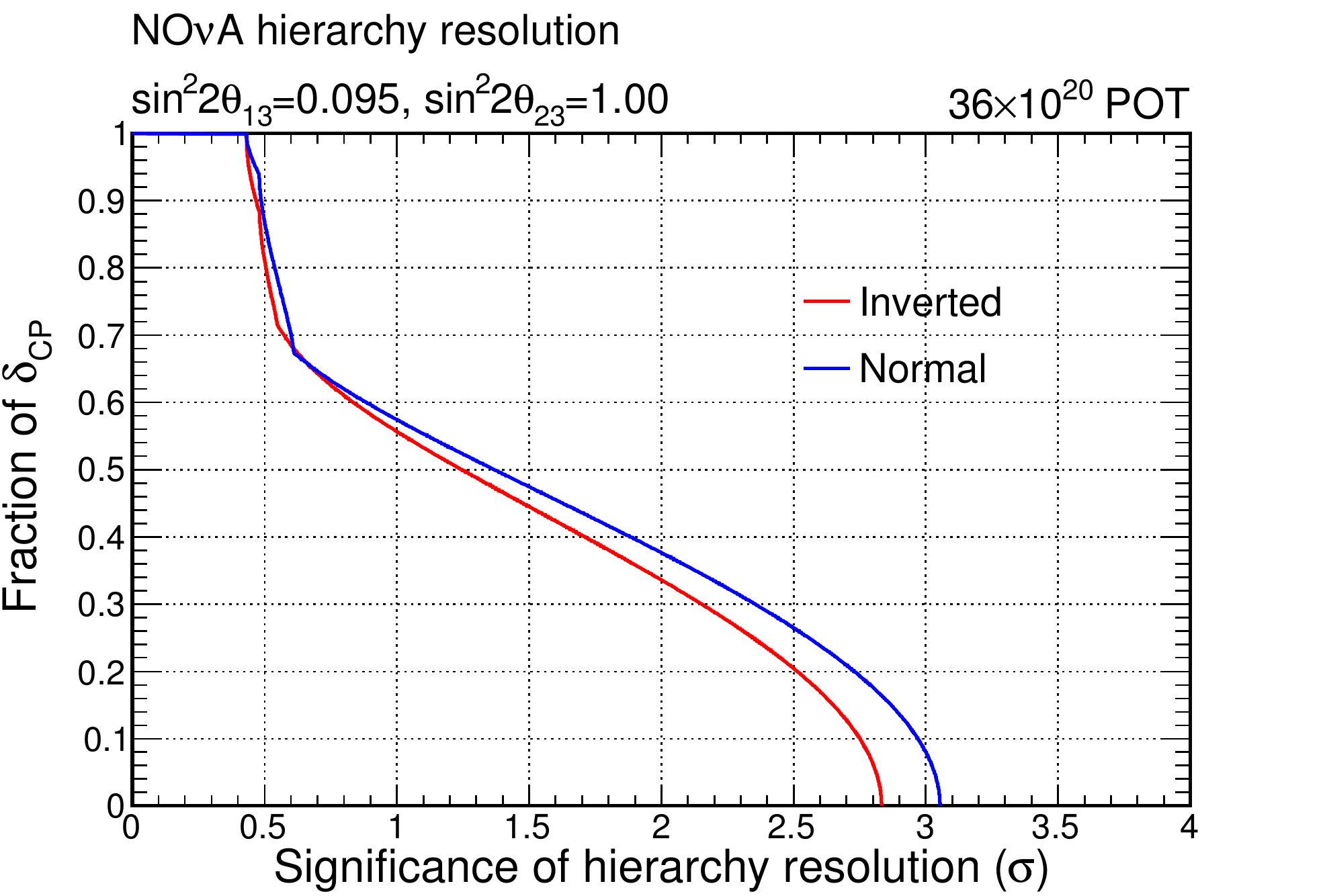}
\caption{Sensitivity of NOvA to the neutrino mass hierarchy. The left panel shows the sensitivity as a function of $\delta_{CP}/\pi$. The right panel shows the fraction of $\delta_{CP}$ space covered for various degrees of statistical significance $\sigma$. Both plots are for maximal $\theta_{23}$. Approximate one third of the $\delta_{CP}$ range gives a hierarchy determination at 95\% C.L. (Courtesy NOvA Collaboration).}
\label{fig:NOVA_hierarchy}
\end{center}
\end{figure}

\begin{figure}
\begin{center}
\includegraphics[scale=0.35]{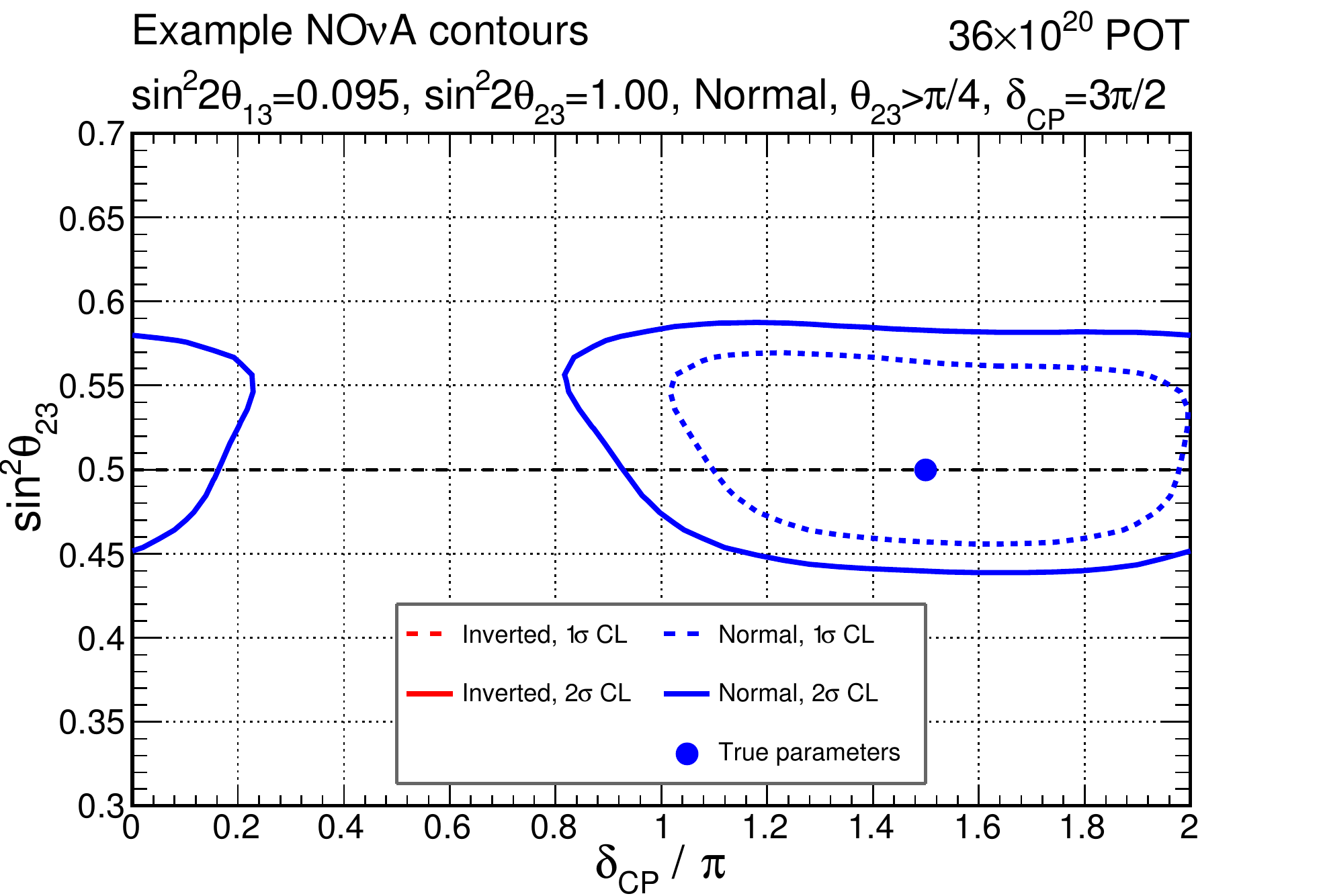}
\includegraphics[scale=0.35]{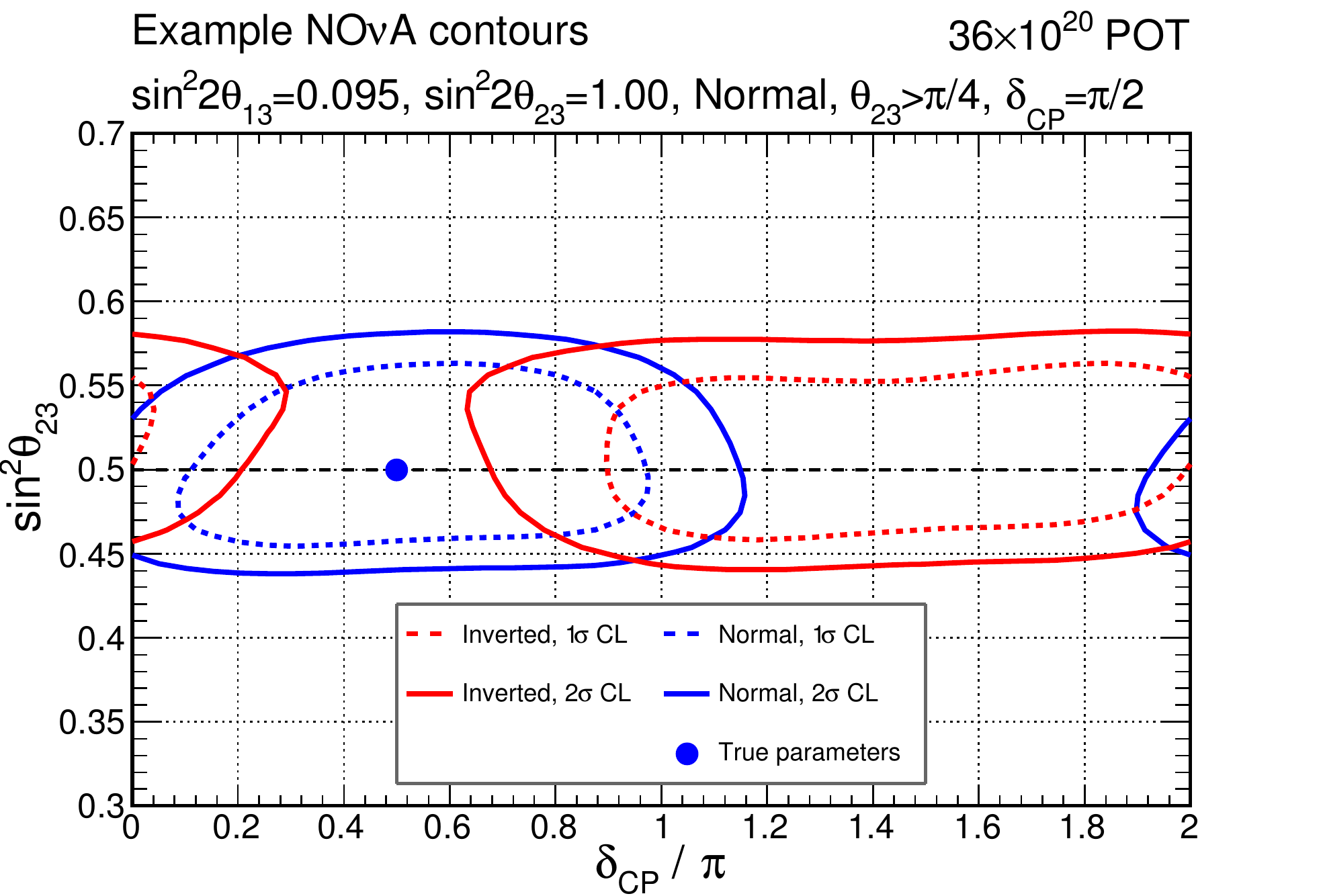}
\caption{Examples of joint NOvA sensitivity contours for $\sin^2\theta_{23}$ and $\delta_{CP}$, for a nominal run of $36\times10^{20}$ protons on target in the 700 kW NuMI beam. In the left panel $\theta_{23}$ is maximal, and $\delta_{CP}$ is chosen for maximum separation between the hierarchies. The inverted hierarchy and some values of $\delta_{CP}$ are disfavored. A less-favorable case is shown in the right panel, where $\delta_{CP}$ is chosen to illustrate the difficulty of distinguishing the hierarchies. In both-cases $\theta_{13}$ is taken as an external input (Courtesy NOvA Collaboration).}
\label{fig:NOVA_2par}
\end{center}
\end{figure}

As suggested by Figure~\ref{fig:NOVA_biprob}, a principle goal of NOvA is to gain information about the mass hierarchy of the neutrino eigenstates. We can see the sensitivity 
of a "standard" expected exposure in the NuMI beam in Figure~\ref{fig:NOVA_hierarchy}. Hierachy and CP violation information are coupled, leading to the right panel of the figure, in which the fractional coverage of "CP-space" $0-2\pi$ at which the hierarchies can be separated is shown as a function of the significance of the separation.
\\
\\
Because of parameter ambiguities, the study of CP violation in neutrino oscillations is particularly challenging. Full resolution of the problem may require the very large detectors of Hyper-K and DUNE, currently under discussion. However, particularly in favorable cases, important information can be obtained. Figure~\ref{fig:NOVA_2par} illustrates the situation by plotting the simultaneous significance of two quantities, $\sin^2\theta_{23}$ and $\delta_{CP}$, for both hierarchies. In confused cases information is obtained about the likely correlations of $\delta_{CP}$ and the hierarchy, and  less ambiguous cases will favor specific regions of $\delta_{CP}$.
\\

\subsection{MINOS+}
\label{subsec:MINOS_Plus_Section}

\begin{figure}
\begin{center}
\includegraphics[scale=0.35]{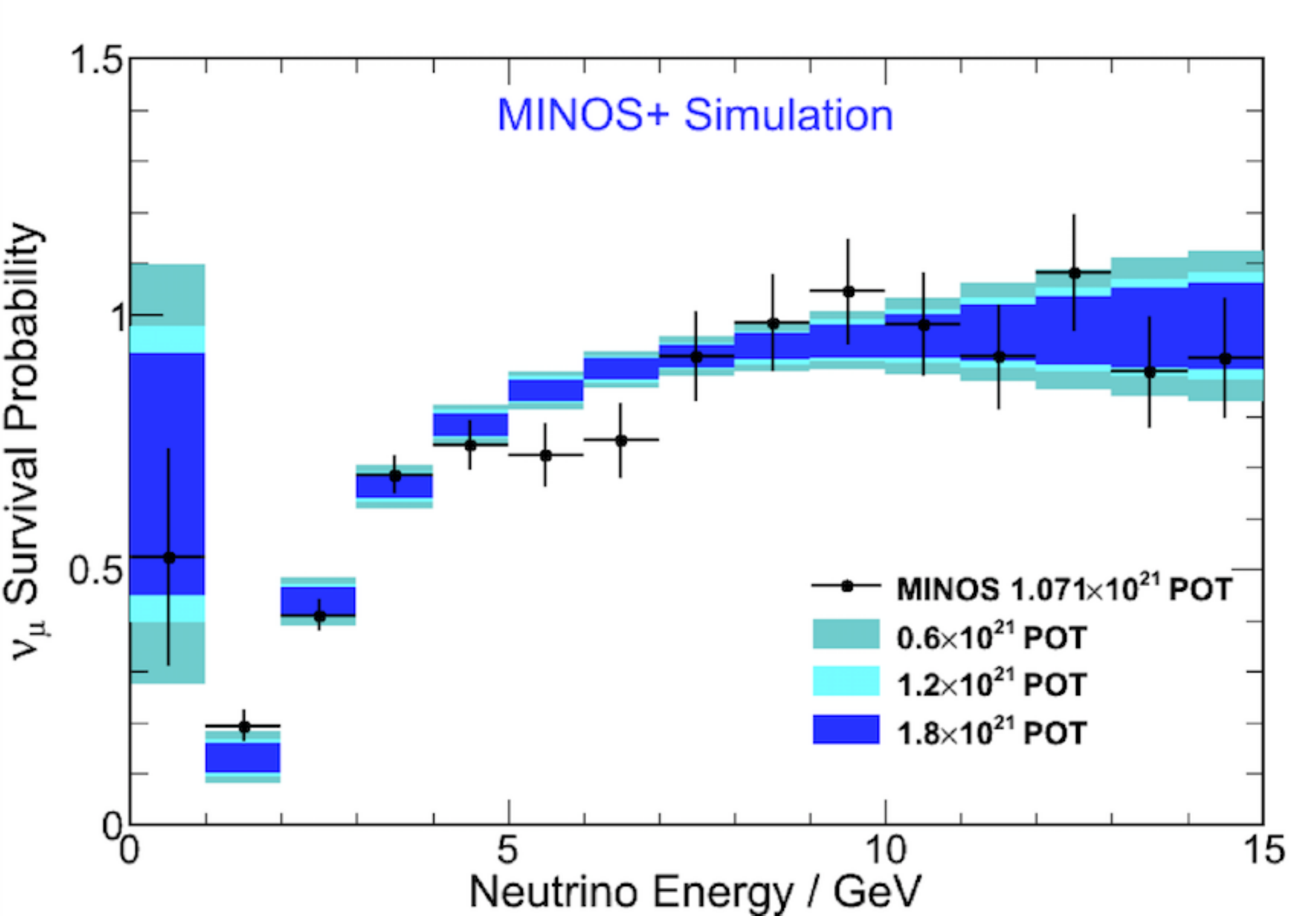}
\caption{Ratio of oscillated to unoscillated predictions for $\nu_\mu$ charged-current events in the MINOS+ experiment, as a function of exposure in terms of protons on target (POT). The statistical precision in the neutrino energy region of 5-7 GeV is much improved by the high rates available from the NuMI complex (Courtesy MINOS+ Collaboration).}
\label{fig:MINOS_Plus_Spectrum}
\end{center}
\end{figure}

The MINOS experiment, as described in this paper, has studied the region in L/E near the oscillation minimum in detail, using primarily low-energy settings of the NuMI beam. As discussed in section \ref{subsec:Off_Axis_Section} the requirements of off-axis kinematics for the NOvA experiment lead to a need for a higher energy on-axis setting of the on-axis beam.
This beam, with its associated larger event rates in both MINOS near and far detectors, is exploited by the MINOS+ experiment. Fig \ref{fig:MINOS_Plus_Spectrum} shows the expected structure of the data spectrum which will be obtained by MINOS+. The experiment, will collect on the order of 3000 CC and 1200 NC events for each exposure of $6.0\times 10^{20}$ protons on the NuMI target (roughly annually). These large event rates allow a varied physics program.
\\

\begin{figure}
\hspace{-2.0 cm}\includegraphics[scale=0.5]{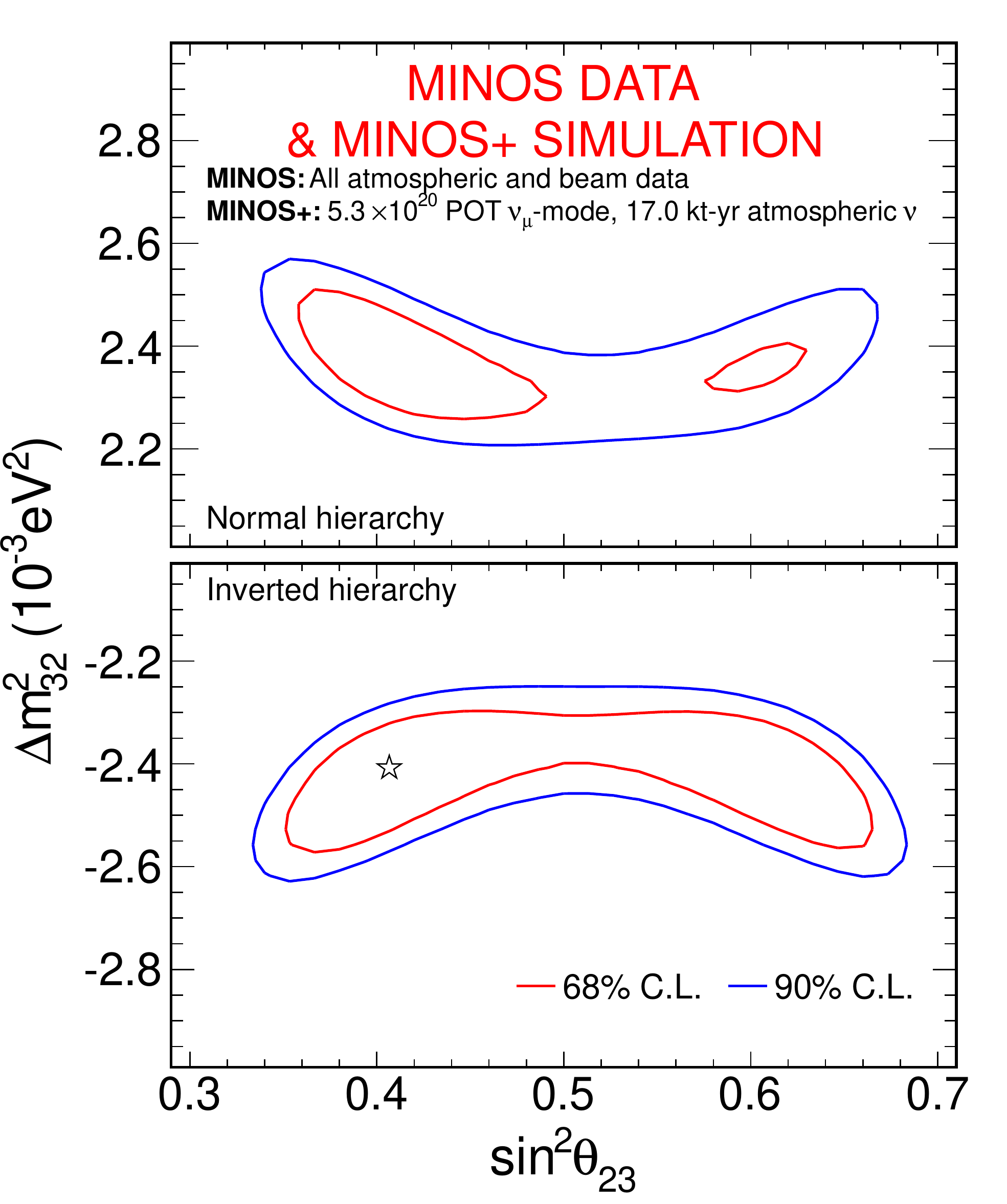}
\includegraphics[scale=0.5]{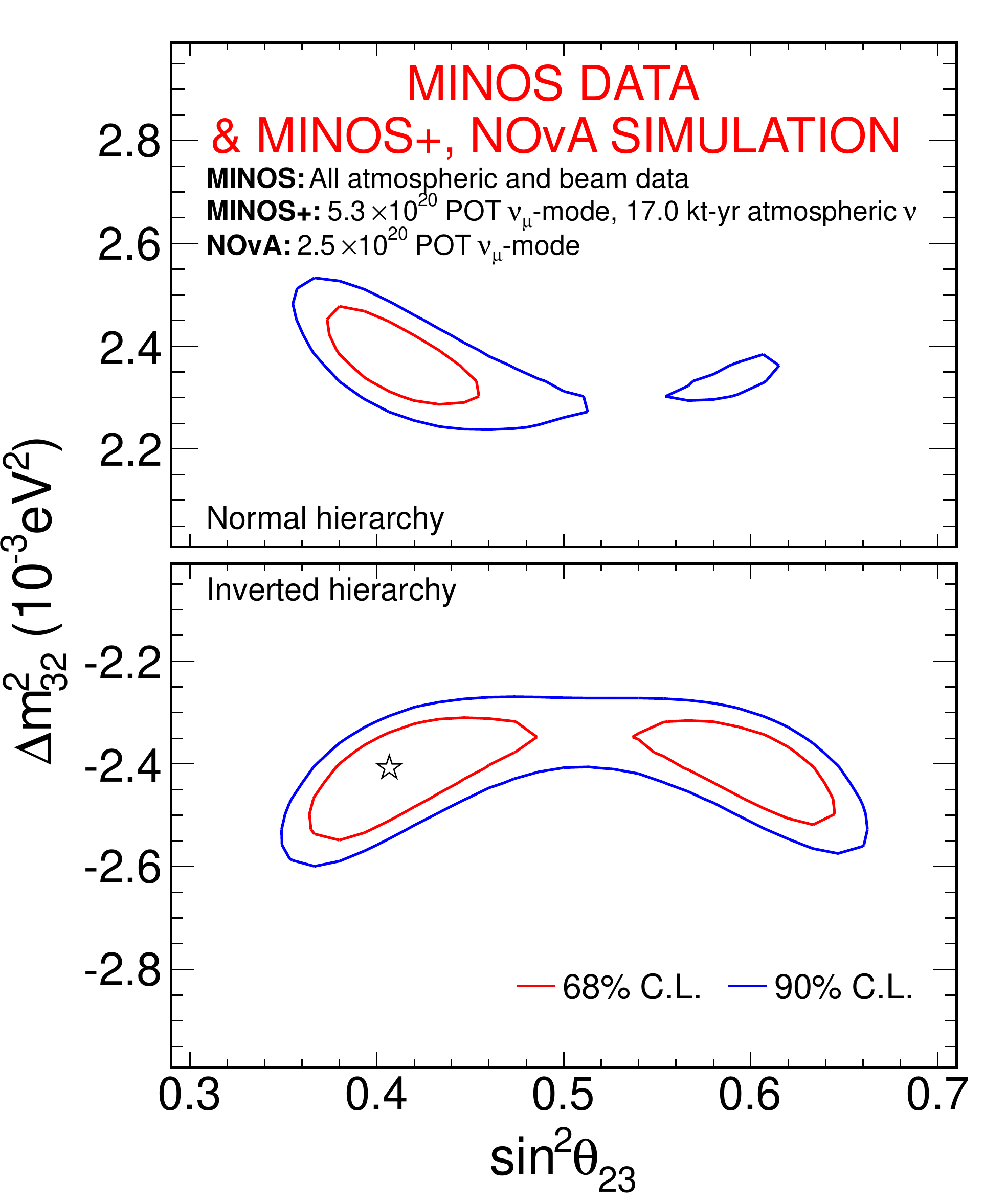}
\caption{Expected near-term precision of determination of the oscillation parameters $\sin^2\theta_{23}$ and $\Delta m^2_{32}$ for the MINOS+ experiment. The left panel shows the results when MINOS and MINOS+ data are combined, and the right panel includes the expected data from the NOvA experiment. The contours are generated using as inputs the best-fit MINOS values from \cite{minos_three_neutrino_prl} (Courtesy MINOS+ Collaboration).}
\label{fig:MINOS+_params}
\end{figure}

Representative physics goals of the MINOS+ experiment include:

\begin{enumerate}
\item Precise verification of the expected spectral shape of the oscillation phenomenon, using the disappearance technique, especially in its transition region in energy between 5 and 7 GeV.
\item Utilization of the precise spectrum together with MINOS and NOvA data to continue improving understanding of the oscillation parameters.
\item Further study of the possibilities of sterile neutrinos, using both CC and NC disappearance.
\item Improved precision on searches for exotic phenomena.
\\
Figure \ref{fig:MINOS+_params} shows an example of (ii), considering all the NuMI program data expected to be obtained in the 2015 time frame. Comparison with Figure~\ref{fig:MINOS_3_flavor_result} shows improved determination of the parameters, especially in the case of the normal mass hierarchy.
\end{enumerate}

\subsection{Future prospects of T2K}
The approved beam for the T2K experiment is $7.8\times 10^{21}$ protons on target (POT).
The results of T2K reported in this paper are based on $6.6 \times 10^{20}$ which is only 8~\% of the original goal~\footnote{T2K collected $1.0\times 10^{21}$~POT on March 26th, 2015.}.
In the near future, the J-PARC accelerator plans to increase the repetition rate of the acceleration cycle by updating the power supply system.
With the upgrade, the beam power of J-PARC will reach 750~kW, and T2K will accumulate the design beam within several years.
In this section, we show the physics sensitivity of T2K with $7.8\times 10^{21}$~POT.
In T2K, there are two beam operation modes: one is neutrino beam and the other is anti-neutrino beam. Since the fraction of the anti-neutrino beam to the neutrino is not fixed yet, we will show both possibilities.

The current goals of T2K with $7.8\times 10^{21}$~POT are
\begin{itemize}
\item Initial measurement of CP violation in neutrinos up to a $2.5 \sigma$ level of significance.
\item Precision measurement of oscillation parameters in the $\nu_\mu$ disappearance with precision of $\delta \Delta m^2_{32} \simeq 10^{-4}$~$\rm eV^2$ and $\delta \sin^2 2\theta_{23} \simeq 0.01$ ; also determination of $\theta_{23}$ octant at 90~\% C.L. if $|\theta_{23} - 45^\circ | > 4^\circ$.
\item Contribution to the determination of the mass hierarchy.
\end{itemize}

\subsubsection{Neutrino events with $7.8\times 10^{21}$ protons on target (POT)}\
\\

Based on the analysis method in~\cite{Abe:2013xua, Abe:2013fuq}, the expected number of $\nu_e$ and $\nu_\mu$ events
\footnote{This study was conducted before T2K developed the special $\pi^0$ rejection algorithm. So, the number of NC background is higher, compared to the results in Section~\ref{sec:t2k:nue}}
 are shown in Table~\ref{ta:t2k:nue} and \ref{ta:t2k:numu} using the following neutrino oscillation parameters: $\textrm{sin}^{2}2\theta_{13}=0.1$, $\sin^2 \theta_{23}=0.5$, $\textrm{sin}^{2}2\theta_{12}=0.8704$, $\Delta m^2_{12}=7.6\times10^{-5}$ eV$^{2}$, $|\Delta m^2_{32}|=2.4\times10^{-3}$ eV$^{2}$, $\delta_{\mathrm{CP}}=0$, and $\Delta m^2_{32}>0$.
\begin{table}[th]
\begin{center}
\caption{The expected number of $\nu_e$ events in T2K with $7.8\times 10^{21}$~POT for the neutrino beam mode and $7.8\times 10^{21}$~POT for the ant-neutrino~\cite{Abe:2014tzr}.} \label{ta:t2k:nue}
\begin{tabular}{l|c|ccccc}
\hline
\hline
                               &           & Signal                      & Signal                                                        & Beam CC                            & 
                                    Beam CC                                      & NC \\
Beam Mode            &  Total & $\nu_\mu \to \nu_e$ & $\overline{\nu}_\mu \to \overline{\nu}_e$  & $ \nu_e + \overline{\nu}_e$ & 
                                    $ \nu_\mu + \overline{\nu}_\mu$  & NC \\
\hline
neutrino beam        &  291.5  & 211.9  & 2.4    & 41.3 & 1.4 & 34.5 \\
anti-neutrino beam &  94.9 & 11.2 & 48.8 & 17.2 & 0.4 & 17.3 \\
\hline
\hline
\end{tabular}
\end{center}
\end{table}
\begin{table}[th]
\begin{center}
\caption{The expected number of $\nu_\mu$ events in T2K with $7.8\times 10^{21}$~POT for each beam operation mode~\cite{Abe:2014tzr}.} \label{ta:t2k:numu}
\begin{tabular}{l|c|cccc}
\hline
\hline
                               &           &  CCQE         & CC non-QE                     & CC $ \nu_e + \overline{\nu}_e$ &  \\
Beam Mode            &  Total & $\nu_\mu (\overline{\nu}_\mu) $ & $\nu_\mu (\overline{\nu}_\mu)$ & $\nu_\mu (\overline{\nu}_\mu) \to \nu_e (\overline{\nu}_e)$ & NC \\
\hline
neutrino beam        &  1,493  & 782 (48)   & 544 (40)    & 4 & 75 \\
anti-neutrino beam &     715  & 130 (263) & 151 (138)  & 0.5 & 33 \\
\hline
\hline
\end{tabular}
\end{center}
\end{table}

\subsubsection{CP sensitivity}\
\\

\begin{figure}
\begin{center}
\includegraphics[width=76mm]{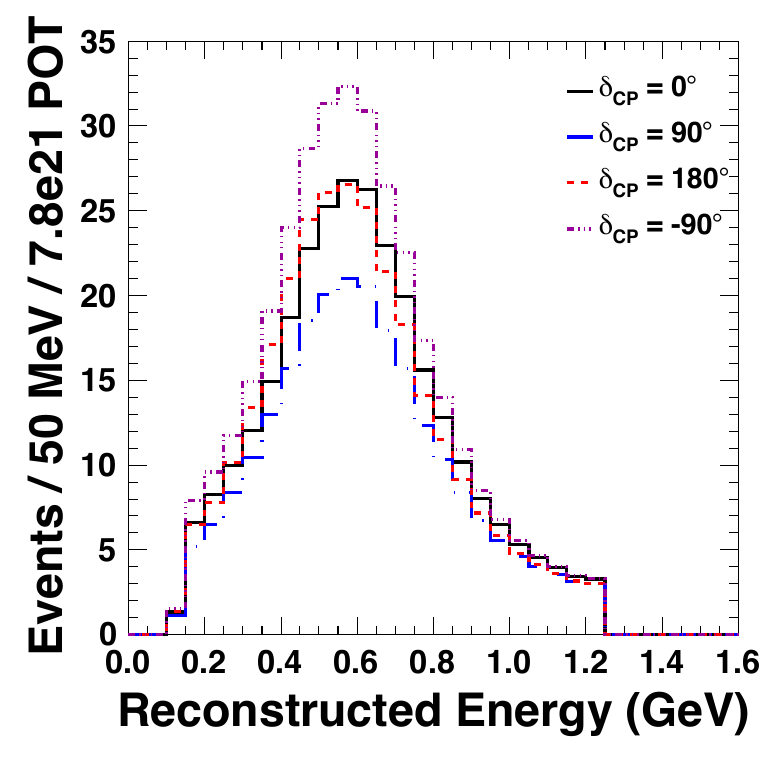}
\includegraphics[width=76mm]{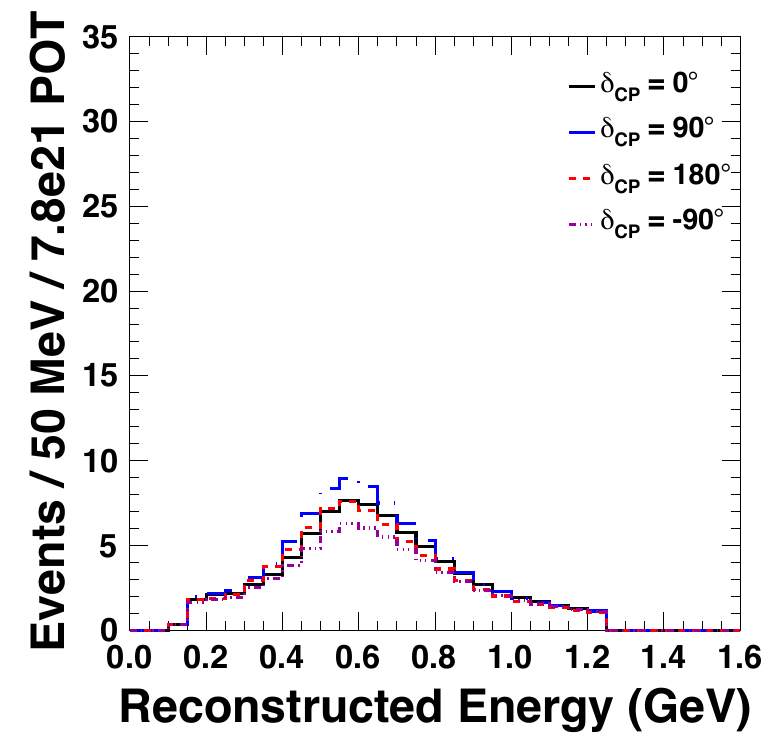}
\caption{The reconstruction energy of expected T2K electron neutrino appearance events with various $\delta_{\mathrm{CP}}$ parameters~\cite{Abe:2014tzr}. (Left) Neutrino beam operation with $7.8\times 10^{21}$~POT and (Right) Anti-neutrino beam operation with $7.8\times 10^{21}$~POT.} 
\label{fig:t2k:dcp}
\end{center}
\end{figure}

\begin{figure}
\begin{center}
\includegraphics[width=8cm]{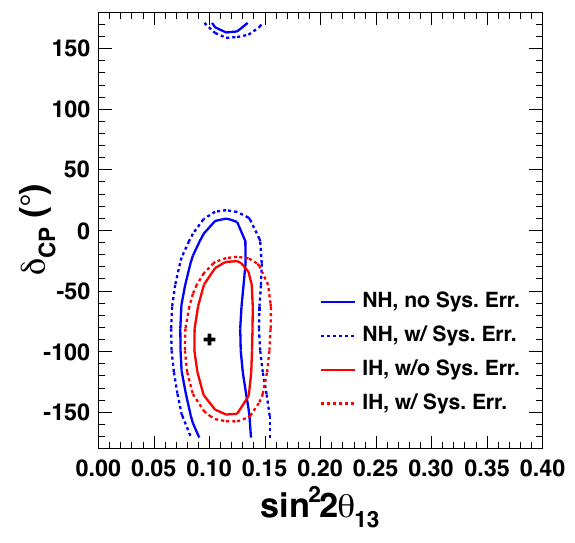}
\caption{The 90~\% contour region of the $\delta_{\mathrm{CP}}$ versus $\sin^2 2\theta_{13}$ plane in the expected T2K sensitivity with $7.8\times 10^{21}$~POT~\cite{Abe:2014tzr}. 
POT are assumed to be equally distributed to the neutrino beam mode ($3.9\times 10^{21}$~POT) and the anti-neutrino ($3.9\times 10^{21}$~POT).} 
\label{fig:t2k:dcp2}
\end{center}
\end{figure}

Since the electron neutrino appearance is sensitive to CP violation, the variation of the number of electron neutrino events with $\delta_{\mathrm{CP}}$ parameters is shown in Figure~\ref{fig:t2k:dcp}.
In maximum, we expect a 27~\% change, compared to no CP violation with $\delta_{\mathrm{CP}}=0$.
Hereafter, we assume the beam exposure to be 50~\% for the neutrino beam and 50~\% for the anti-neutrino.
We also assume $\sin^2 2 \theta_{13} = 0.10 \pm 0.005$ as the ultimate $\theta_{13}$ value from reactor experiments.
In the case of the maximum CP violation ($\delta_{\mathrm{CP}}=-90^\circ$), the T2K sensitivity for $\delta_{\mathrm{CP}}=-90^\circ$ is shown in Figure~\ref{fig:t2k:dcp2} with 90~\% C.L.

\begin{figure}
\begin{center}
\includegraphics[width=75mm]{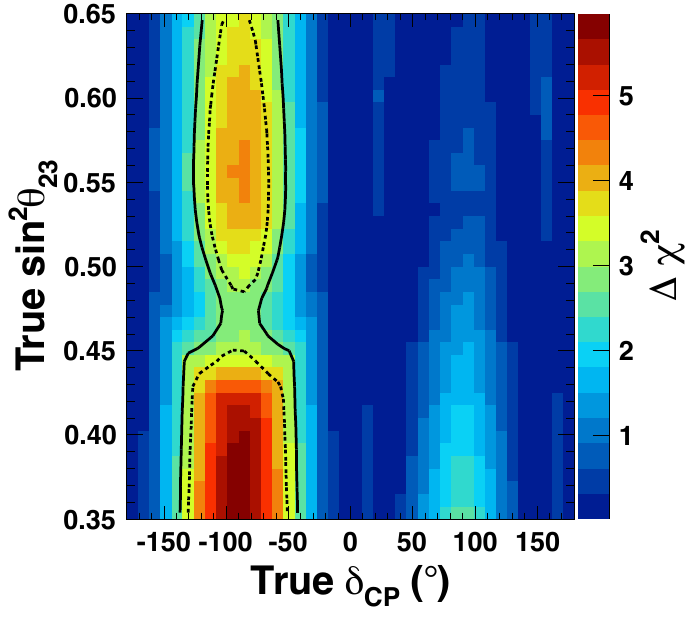}
\includegraphics[width=75mm]{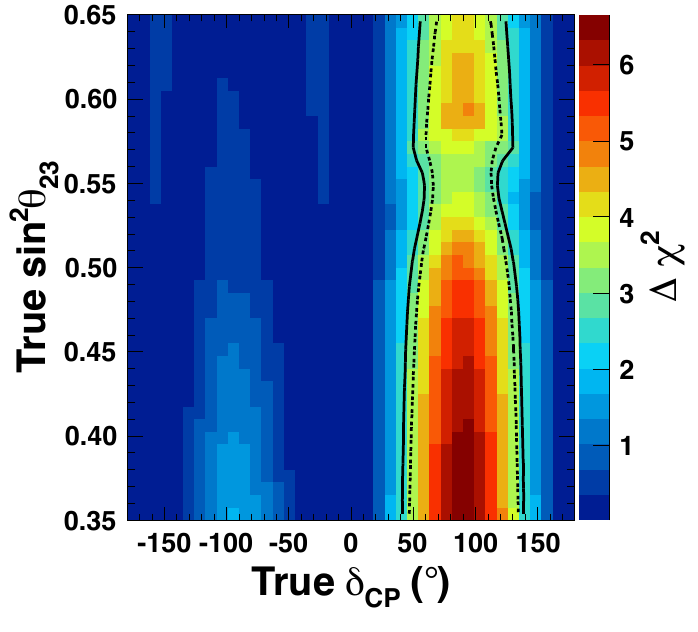}
\caption{T2K expected $\chi^2$ difference between the true point $(\delta_{\mathrm{CP}}, \sin^2 \theta_{23})$ and the point $\delta_{\mathrm{CP}}=0$~\cite{Abe:2014tzr}.
The map of $\chi^2$ difference shown in color is calculated assuming no systematic errors. The solid contours show the 90~\% C.L. sensitivity with statistical error only, while the dashed contours include the 2012 T2K systematic error. (Left) the contour in the case of Normal mass hierarchy, and (Right) in the case of Inverted mass hierarchy.
POT are assumed to be equally distributed to the neutrino beam mode ($3.9\times 10^{21}$~POT) and the anti-neutrino ($3.9\times 10^{21}$~POT).} 
\label{fig:t2k:cp23}
\end{center}
\end{figure}

In reality, the sensitivity to CP violation also depends on $\theta_{23}$.
In Figure~\ref{fig:t2k:cp23}, as the T2K sensititivy,  we show the $\chi^2$ difference between the true point with $(\delta_{\mathrm{CP}}, \sin^2 \theta_{23})$ and the hypothesis test point with $\delta_{\mathrm{CP}}=0$. 

\subsubsection{Precision measurements of neutrino oscillation parameters}\
\\

\begin{figure}
\begin{center}
\includegraphics[width=7.5cm]{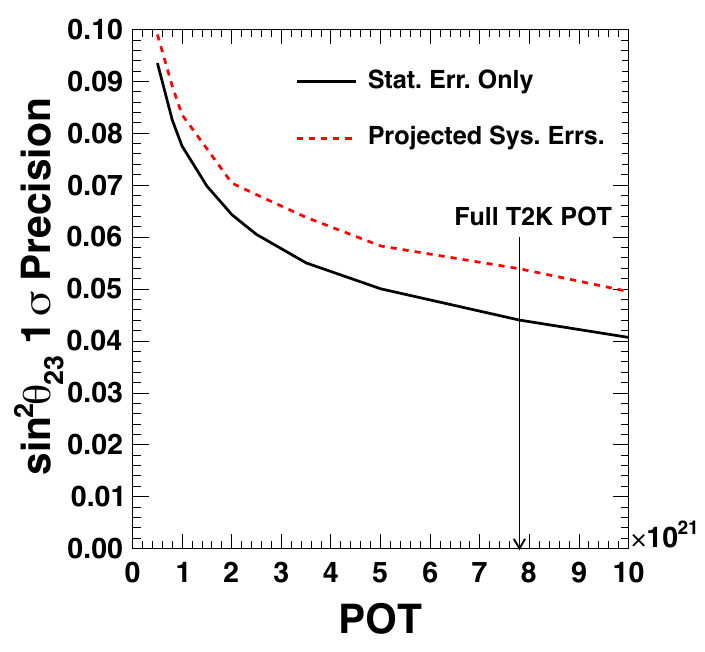}
\includegraphics[width=7.5cm]{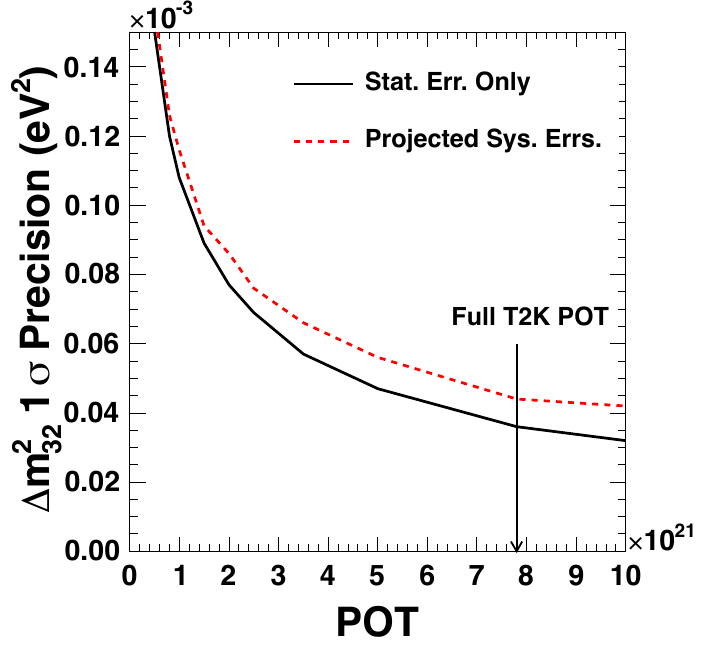}
\caption{T2K expected sensitivity to $\sin^2 \theta_{23}$ and $|\Delta m^2_{32}|$ as a function of POT for the normal mass hierarchy~\cite{Abe:2014tzr}. 
The solid lines are without systematic error and the dashed one with the systematic error from the T2K 2012 analysis. POT are assumed to be $3.9\times 10^{21}$~POT for the neutrino beam mode and $3.9\times 10^{21}$ for the anti-neutrino.} 
\label{fig:t2k:sm23}
\end{center}
\end{figure}

Since most of the T2K measurements are limited by statistics, more data improve the precision of the measurements.
Among the oscillation parameters, $\sin^2 \theta_{23}$ and $|\Delta m^2_{32}|$ are interesting because of their relatively larger uncertainty compared to other parameters.
Figure~\ref{fig:t2k:sm23} shows the expected precision of $\sin^2 \theta_{23}$ and $|\Delta m^2_{32}|$ as a function of POT for the normal mass hierarchy case. 
Hereafter, the total exposure in T2K is assumed to be $7.8\times 10^{21}$~POT which are equally distributed to the neutrino beam mode ($3.9\times 10^{21}$~POT) and the anti-neutrino ($3.9\times 10^{21}$~POT).
The statistical uncertainty of $\sin^2 \theta_{23}$ and $|\Delta m^2_{32}|$ is 0.045 and $0.04 \times 10^{-3}$~$\rm eV^2$, respectively, at the T2K full statistics. The precision of $\sin^2 \theta_{13}$ is influenced by the precision of other oscillation parameters including $\sin^2 \theta_{23}$ and 
$\delta_{\mathrm{CP}}$.

\begin{figure}
\begin{center}
\includegraphics[width=7.5cm]{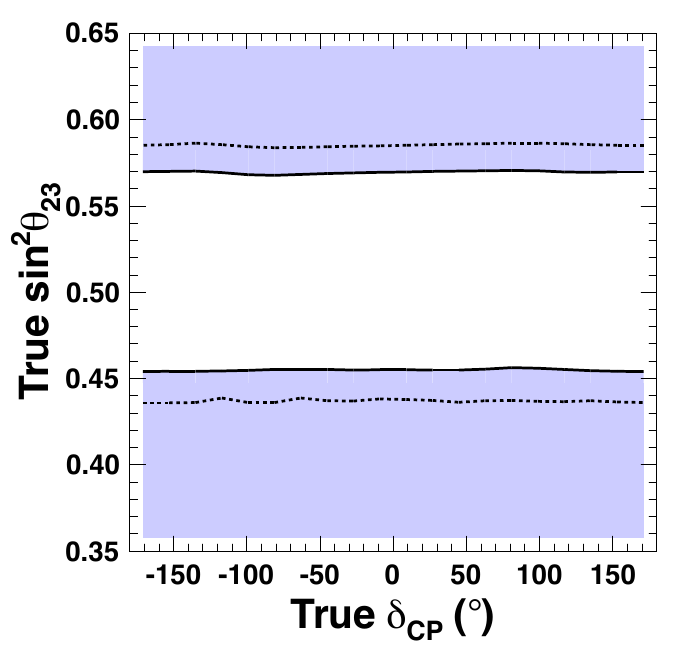}
\includegraphics[width=7.5cm]{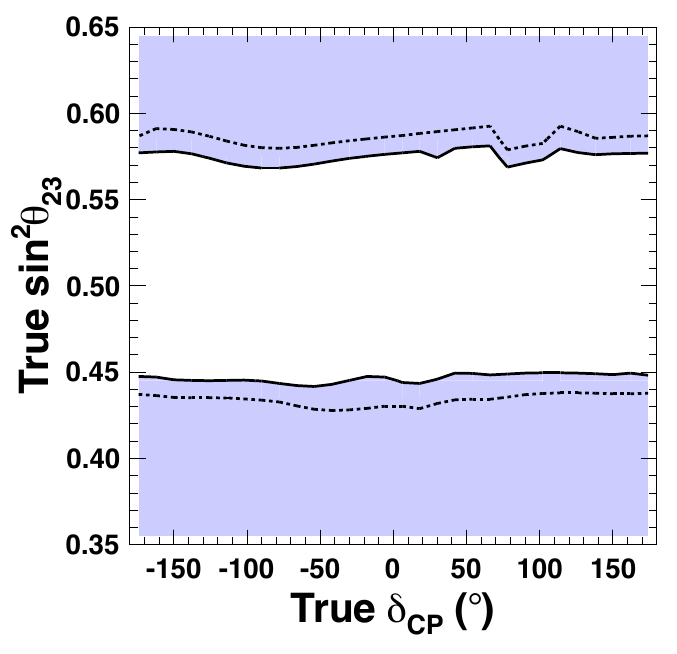}
\caption{(Left)The shaded region is where T2K has more than a 90~\% C.L. sensitivity to reject maximal mixing~\cite{Abe:2014tzr}. 
(Right) The shaded region is where T2K has more than a 90~\% C.L. sensitivity to reject one of the octants of $\theta_{23}$~\cite{Abe:2014tzr}. 
The mass hierarchy is considered unknown in the normal mass hierarchy case.
The shaded region is calculated assuming no systematic errors (statistical error only), and the dashed contours show the sensitivity including the systematic errors. POT are assumed to be $3.9\times 10^{21}$~POT for the neutrino beam mode and $3.9\times 10^{21}$ for the anti-neutrino.} 
\label{fig:t2k:s23}
\end{center}
\end{figure}

An interesting question about $\theta_{23}$ is which $\theta_{23}$ is exactly $45^\circ$ or not.
In the case of $\theta_{23} \neq 45^\circ$, which octant does the value of $\theta_{23}$ fit in, ($\theta_{23} > 45^\circ$ or  $\theta_{23} < 45^\circ$)?
Figure~\ref{fig:t2k:s23} shows the region where T2K can reject the maximum mixing $\theta_{23} = 45^\circ$ and the regions where T2K can 
reject one of the octants of $\theta_{23}$. The octant of $\theta_{23}$ is determined at 90~\% C.L. if $|\theta_{23} - 45^\circ | > 4^\circ$.

\subsubsection{Mass Hierarchy}\
\\

\begin{figure}
\begin{center}
\includegraphics[width=7.5cm]{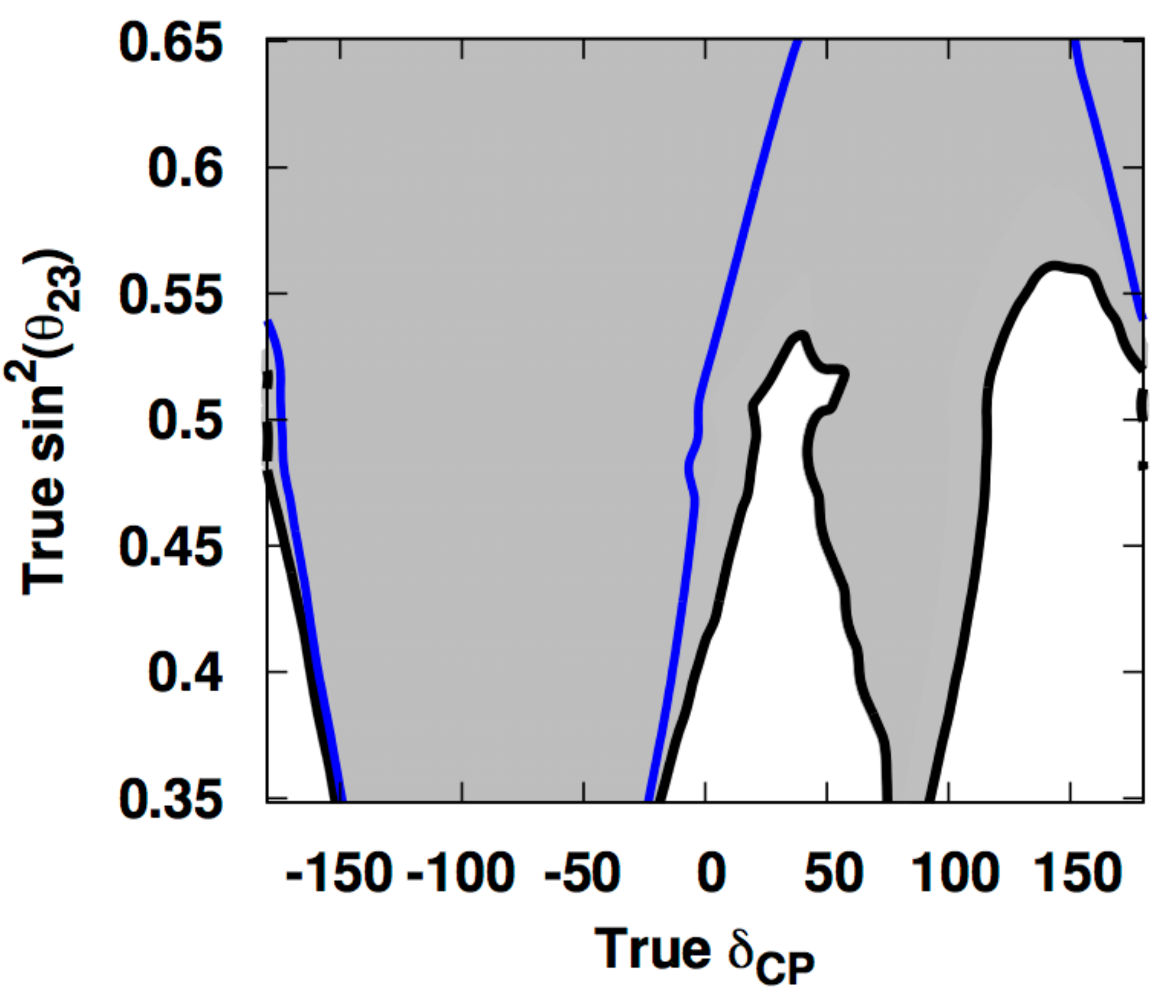}
\includegraphics[width=7.5cm]{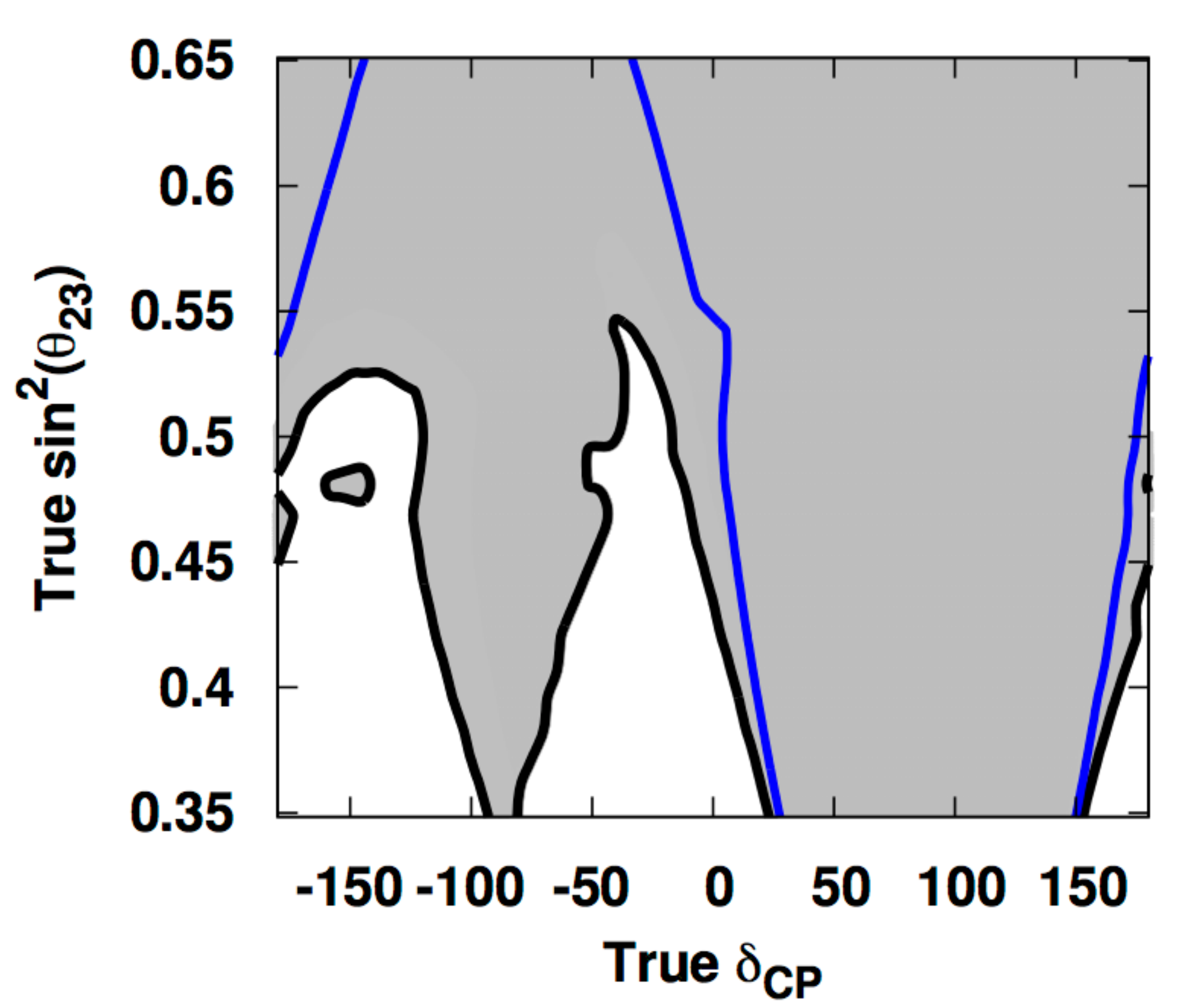}
\caption{The regions where the wrong mass hierarchy is expected to be rejected at 90~\% C.L. by the NOvA measurement (blue line) and the NOvA + T2K measurements (black shaded regions)~\cite{Abe:2014tzr}. (Left) the true hierarchy is normal and (right) it is inverted. The T2K POT are assumed to be $3.9\times 10^{21}$~POT for the neutrino beam mode and $3.9\times 10^{21}$ for the anti-neutrino. The NOvA POT ($3.6\times 10^{21}$) are also assumed to be distributed to the neutrino beam mode and the anti-neutrino equally.} 
\label{fig:t2knova:mh}
\end{center}
\end{figure}

Because of the relatively short baseline ($\sim$300~km) of T2K, the experiment is less sensitive to the mass hierarchy (more sensitive to CP). However, the measurement of T2K (sensitive to CP) can contribute to improving the mass hierarchy sensitivity of NOvA by helping untangle the two effects of CP and mass hierarchy in NOvA. Figure~\ref{fig:t2knova:mh} shows the 90~\% sensitivity region for mass hierarchy with the T2K and NOvA measurements. The sensitivity is really expanded by adding the T2K measurements, especially for $\delta_{\mathrm{CP}} \sim 90^\circ (-90^\circ)$ case in the normal (inverted) hierarchy case.
%

\section{Conclusions; Building the future}

We have presented the results and prospects of neutrino oscillation measurements by the present generation of experiments: MINOS/MINOS+, T2K and NOvA.
The phenomenology of neutrinos is being rapidly revealed by these experiments with a dramatical improvement of the precision of neutrino oscillation parameters in 10 years. In the standard neutrino oscillation scenario, all three mixing angles have been measured and found to be large enough to explore CP violation. 
Surprisingly, the new data from the accelerator experiments is beginning to become sensitive to CP violation, when coupled with the precise knowledge of  the mixing angles.

Upcoming results from on-going experiments, especially T2K and NOvA, will have large impact for the following reasons.
\begin{itemize}
\item The measurements of T2K and NOvA individually are the most sensitive to CP violation. By combining both, the sensitivity will be further improved.
\item  In order to explore CP violation, the precision of mixing angles is essential. In particular, the value of $\theta_{23}$ plays a key role in specifying  the complicated parameter space of delta-CP and the mass hierarchy. These experiments are the most sensitive to $\theta_{23}$.
\item The NOvA experiment has some sensitivity to the mass hierarchy. By combining with T2K, the sensitive region will be expanded.
\item The future neutrino experiment, Hyper-Kamiokande and DUNE, will have greatly expanded sensitivity to CP violation. The experiences of T2K and NOvA, together with the improvement of systematic uncertainties are key inputs to the future experiments.
\end{itemize}

Going beyond the standard neutrino oscillation scenario, unexpected phenomena may appear in the most sensitive experiments.
Thus, there is discovery potential for MINOS/MINOS+, T2K and NOvA at any time.

Success of on-going experiments is building the future.

\section*{Acknowledgements}

Fermilab is operated by Fermi Research Alliance, LLC under Contract No. DE-AC02-07CH11359 with the United States Department of Energy. 
J-PARC is operated by KEK and JAEA with MEXT in Japan. An author (TN) acknowledges the support of MEXT and JSPS in Japan with the Grant-in-Aid for Scientific Research A 24244030 and for Scientific Research on Innovative Areas 25105002 titled ``Unification and Development of the Neutrino Science Frontier''.
Fruitful discussions with members of the T2K, MINOS+, and NOvA collaborations are gratefully acknowledged.

\end{document}